\documentclass[aps,prb,reprint,twocolumn,showpacs,superscriptaddress]{revtex4-1}
\usepackage{amsmath, amssymb, amsthm}
\usepackage[breaklinks]{hyperref}
\usepackage{graphicx}
\usepackage{tcolorbox}
\usepackage{leftidx}
\usepackage{url}
\usepackage{enumitem}
\usepackage{physics}
\usepackage{tikz}
\usepackage[normalem]{ulem}
\usetikzlibrary{calc}
\usetikzlibrary{decorations.pathmorphing}
\usepackage{pifont}
\usepackage{multirow}
\usepackage{hhline}

\newcommand{\cmark}{{\color{blue} \ding{51}}}%
\newcommand{\xmark}{{\color{red} \ding{55}}}%
\newcommand{\spflip}{Z_2^{y}}%
\newcommand{\flip}{{\cal P}}%

\begin{document}

\title{Building Symmetry Enriched Topological Phases from a Bipartite Lattice Construction and Anyon Condensation}
\author{Jong Yeon Lee}
\affiliation{Department of Physics, Harvard University, Cambridge, Massachusetts 02138, USA}
\author{Ari M. Turner}
\affiliation{Department of Physics, Israel Institute of Technology, Haifa, 3200003, Israel}
\author{Ashvin Vishwanath}
\affiliation{Department of Physics, Harvard University, Cambridge, Massachusetts 02138, USA}

\date{\today}

\begin{abstract}
We introduce a construction of symmetry-enriched topological orders on bipartite lattices in which two $\mathbb{Z}_2$ spin liquids defined on each sublattice are combined, and then anyons are condensed to reduce the topological order. By choosing different anyon condensate structures, one can vary the fractionalization pattern of the resulting spin liquid, 
some of which cannot be readily constructed from parton based approaches. We demonstrate the construction for $i$) a spin-1/2 honeycomb lattice where we construct a featureless state as well as intermediate states with topological order, $ii$) a nonsymmorphic lattice, and $iii$) lattices with magnetic translation symmetry. At last, we discuss constraints on non-chiral topological orders in a bosonic system under magnetic field.


\end{abstract}

\maketitle

\tableofcontents

\section{Introduction}

Symmetry properties of the atoms making up a crystal can affect the statistical properties of the low-energy excitations in a gapped system, according to the theorem by Hastings and Oshikawa\cite{Oshikawa2000, Hastings2005} 
, which generalizes the work by Lieb, Schultz and Mattis in one dimension\cite{LSM_original, LSM}. This 
HOLSM theorem states that  
if the atoms in the unit cell of a crystal have a net spin which is a half-integer, then the ground state cannot
be gapped and symmetric 
without
having excitations that behave as anyons. Thus, aside from a gapless ground state, there are three alternatives:
breaking translational symmetry (e.g. singlet bond pairing), breaking the spin-rotation symmetry, or producing
a state which seems not to have any order, but actually has a hidden \emph{topological order} consisting of fluctuating gauge field lines
which can support anyon excitations.

Can this theorem be extended to lattices with an even number of spin-1/2s per unit cell when extra symmetries are added?
With a lattice that has two atoms per unit cell, each of spin-$\frac12$, one can form a state with spin-rotation and translational symmetries by creating a valence bond between the atoms in one unit cell.  However, rotation, reflection, and more complicated lattice symmetries can forbid this 
construction\cite{Sid2013,Po2017_LSM}. In a honeycomb lattice, one can choose to form valence bonds between pairs of atoms connected by hexagonal edges
pointing in any fixed direction without breaking translational symmetry, but this will break rotational symmetry.  There may
be another state with a more complicated pattern of resonating bonds, that has no broken symmetry or anyons, a so-called \emph{featureless state} suggested in Ref.~\onlinecite{Kimchi2013}.
In fact, such a state has been constructed numerically with the full crystal and spin symmetry group\cite{JianZaletel2016,YR2016}. Our goal in this paper is to understand analytic routes for states such as these.

Essentially, the difficulty of constructing a featureless state for lattice models with specific values of $U(1)$ charge per unit cell (e.g. spin-1/2 honeycomb lattice) arises from the fact that there is no known method to distribute $U(1)$ charge symmetrically without creating massive entanglement which tends to result in a topological order in many cases\cite{Z2_IGT_3}. On the other hand, it is well known how to construct a topological order in such models. Thus, it would be useful if we could access a featureless state from a state with topological order. Therefore, we attempt 
to access different phases such as a featureless state or a new type of topologically ordered phase starting from well-known states with topological orders.


We will describe a way of forming a state which does not break symmetries that may have topological order, and then collapse its topological order into a featureless state if possible.   
The idea is to decompose the lattice into sublattices and form \emph{spin liquid} states on the sublattices which people understand better. (Here, spin liquid (SL) refers to a paramagnetic phase of a spin system with topological order.) 
For example, the honeycomb lattice can be formed from two triangular lattices that have $\mathbb{Z}_2$ SLs on each of them.  
Then, we try to form condensates of some of the anyons to eliminate the topological order.  Similar constructions have been introduced for systems with on-site symmetries\cite{Wang2013_Layer,JianQi2014}, but our construction incorporates spatial symmetries that can permute anyons, enriching the resulting topologically ordered phase. 

To see whether it is possible to condense the anyons, one must
confirm that the anyons transform in a simple way (which means they can form a condensate without breaking any symmetry) or in a fractionalized \emph{projective} way under crystal symmetries and the other
symmetries. We outline a method for calculating the way anyons transform under the full symmetry group
starting from the smaller group of symmetries of the sublattices.

We will apply these ideas to several examples.  The main one is a construction of a state on the honeycomb lattice that has no
topological order or broken symmetry. We will also consider cases where the symmetries \emph{do} forbid a featureless state (as
has been shown previously) and give simple constructions of states with a minimal topological order.
The examples include a system with a glide-reflection or magnetic translation symmetry. We verify that the constructed topologically ordered phases have non-trivial anyon symmetry fractionalizations, which is closely related to the absence of a featureless state.

Let us make the statement of the HOLSM theorem more precise.  We will first define 
what we mean by a featureless state.
A two-dimensional gapped and symmetric system with local interactions can have an intrinsic topological order defined by the existence of {anyons}, which are gapped low-energy excitations with non-trivial braiding statistics.\cite{Wen1990, Wen2004} Further equipped with symmetries, an intrinsic topological order becomes a symmetry-enriched topological (SET) order. A SET order is characterized by symmetry fractionalization, meaning that anyons carry fractional quantum numbers of the symmetry group. A famous example is the $\nu=1/3$ fractional quantum Hall phase with $U(1)$ charge conservation symmetry, where quasi-particles carry $1/3$ of the electron charge.\cite{Stern2008}  Since fractional quantum numbers carried by anyons cannot change without a quantum phase transition, they have been used to distinguish between quantum phases with the same intrinsic topological order. 
\cite{EssinHermele_Z2_2013,Maissam_SymFrac_2014,Mesaros2013_SET,Hung2013_SET_Kmatrix,HungWen2013_SET,YMLAshvin2016_SET_Z2} 

When the state is gapped and fully symmetric without any topological order, we call it a featureless state. A subset of featureless states are called symmetry protected topological (SPT) phases,  which lack any notable physical property in the bulk but exhibit
interesting physics at the boundary\cite{Haldane1983,GuWen2009,Pollmann2010_EntSpectrum,Fidkowski2011_Fermion1D,Cirac2011_PEPS,Chen2011_Boson1D, Chen2012_BosonHigher, Chen2013_BosonHigher, HasanKane2010_TI,QiZhang2011_TSc}; however, their topological properties at the boundary disappear 
under the absence of symmetries.

The idea of the HOLSM theorem is that microscopic data can constrain the type of gapped phases that can be realized in a particular lattice model.  For example, one can eliminate featureless states (i.e. SPT phases) for a certain class of lattice models. 
The HOLSM theorem states that for a translationally invariant system with a non-integer electron filling or half-odd integer spin per unit cell, every gapped ground state must have either an intrinsic topological order or discrete symmetry breaking. Therefore, the theorem can be rephrased as a no-go theorem for the system to sustain a featureless ground state.  
This theorem can be made even more precise---one can show that in
the cases where the state does not break symmetry, and has topological order, the SET properties of it are also
constrained.
For a given 2D SET order, its symmetry enriched properties, i.e. symmetry fractionalization of anyons, must be consistent with the microscopic data such as a total spin per unit cell or on-site/lattice symmetries\cite{Maissam_SymFrac_2014, MZAV2015, HermeleChen2016, YangQi2015, PRX_Anomaly2015}.
It has been recently realized that the no-go theorem for a featureless state can be incorporated into this framework by viewing a featureless state as a totally trivial SET phase. \cite{PRX_LSM2016} 
 


However, the absence of a no-go theorem (so far) does not guarantee the existence of a featureless state, and one has to construct such a state explicitly. This has been done recently for the honeycomb lattice via a numerical method, so-called tensor network construction\cite{YR2016}. Our method will help to understand such states analytically, and can be generalized to other cases.  (In addition to the tensor network construction, there are also theoretical
 studies on $(2+1)$D system in the spin-1/2 honeycomb lattice showing that its field theory description does not contain any anomaly,  suggesting the possibility of the featureless state\cite{Cenke2017, Max_Ryan2017}.)  

The paper is organized as follows: We begin by outlining the approach of dividing a system up into sublattices in order to construct states with unbroken symmetry, but maybe with topological order. In Sec.~\ref{sec:review}, we provide a pedagogical review on symmetry fractionalization and how to fully characterize symmetry fractionalization class of the topological order 
In Sec.~\ref{sec:nonsymmorphic}, we describe the procedure for calculating the symmetry fractionalization class of a bipartite lattice, illustrating this for a lattice with nonsymmorphic group $pg$.
In the Sec.~\ref{sec:bipartite}, we use our method to construct a featureless state in the spin-1/2 honeycomb lattice. First, we discuss possible symmetry fractionalization classes of $\mathbb{Z}_2$ SLs in a triangular lattice with spin-1/2. By combining two triangular sublattices, we form a honeycomb lattice. 
In the Sec.~\ref{sec:condensation}, we discuss how to condense anyon bound states $W_1=e_A e_B$ and $W_2=m_A m_B$ in the previously constructed $\mathbb{Z}_2 \times \mathbb{Z}_2$ SLs on the honeycomb lattice to obtain a featureless ground state.  In Sec. \ref{sec:intermediate} we study the intermediate phases that are obtained by condensing just one of the bound states.  In particular, we note that the \emph{symmetry of the condensate} determines some aspects of the \emph{SET order of the intermediate phases.} Finally, in the Sec.~\ref{sec:application}, we illustrate applications of our construction for a lattice with magnetic translation symmetry. 
Detailed calculations of symmetry fractionalization classes will be discussed in the appendices.


\section{Review: Obstructions to Featureless States and Symmetry Fractionalization}\label{sec:review}

\subsection{Attempts to form Featureless States and their Downfall}
Is there a generic strategy to build a featureless state for a given physical model? 
Explicit constructions like AKLT-type wave functions \cite{AKLT_1D_1987,AKLT_2D_2012} are only possible in limited cases, and tensor-network generalized versions of AKLT-type wave functions in a virtual Hilbert space\cite{YR2015,YR2016,YR2017,JianZaletel2016} have the disadvantage that one has to do numerical calculations to check whether they are actually featureless, because a tensor network state that is formally symmetric may actually have a spontaneously broken symmetry\cite{YR2016}.
The most naive attempt would be to construct a symmetric superposition of all possible valence bond configurations. A famous example is a resonating valence bond (RVB) state in a spin-1/2 triangular lattice\cite{MS2001}, but the RVB state hosts a $\mathbb{Z}_2$ (intrinsic) topological order.
Indeed, this must be the case, thanks to the HOLSM theorem that a featureless state is prohibited.

However, a valence bond state is a good starting point in spite of the topological order, because the state at least preserves all the symmetries, and there is a method, ``condensation of anyons'', that can sometimes eliminate the topological order, which we will summarize briefly.  Cosider a $\mathbb{Z}_2$ spin liquid. There are two types of anyons, an $e$ spinon and an $m$ vison, which are mutual semions. One can condense $e$ spinons, causing $e$'s to no longer be well-defined excitations and causing $m$'s to be confined because they have a nontrivial statistics with excitations in the condensate\cite{FB2017, YR2017}. Thus, one can collapse the topological order by anyon condensation.  

In this way, one can try to construct a featureless state.  One can start from a spin liquid state.  Then, one tries to choose a type of anyon or bound states of anyons
to condense in such a way that the topological order is eliminated.  If the anyons that are chosen do not have \emph{fractional} quantum numbers, then it is possible to make a symmetric condensate, so that the symmetries remain unbroken at the same time as topological order is elminated.  


However, if anyons to be condensed carry fractional quantum numbers, one cannot get a featureless state in this way. For example, in a spin-1/2 triangular lattice (or any lattice with a half-integer spin per unit cell), the condensation of the anyons cannot work to produce a featureless state, because the HOLSM theorem forbids it.  Indeed, the spinon and vison
both carry fractional quantum numbers (we will explain what is meant by fractional quantum numbers below).  This implies that condensing one of them would break either the $SO(3)$ spin-rotation or lattice symmetry. Thus, the HOLSM theorem not only provides a no-go for a featureless state, but also constrains the symmetry fractionalization for anyons to be non-trivial, as pointed out in Ref.\,\onlinecite{PRX_LSM2016}.

In order to construct featureless states by this method, we will therefore need to understand what it means for anyons to transform in a fractional way.

\subsection{Fractionalized Symmetries}
A two-dimensional intrinsic topological phase is classified by its anyons, their fusion rules, and their braiding statistics. With a global symmetry $G$, the system can be further classified into a symmetry enriched topological (SET) phase, characterized by a \emph{symmetry fractionalization pattern}. First, let us define a symmetry fractionalization. In a generic SET phase, we assume that the action of a symmetry operator can be decomposed into the product of operators, each of which is supported in a finite region around an anyon excitation. For example, consider a state with $N$ anyon excitations where each anyon $a_i$ is well-separated from the others. For a symmetry element $g \in G$, let $R_g$ be the action of $g$ on the full Hilbert space. Then, $R_g$ can be decomposed as 
\begin{equation}
R_g \ket{\Psi} = \prod_{i=1}^N \Omega_g (a_i) \ket{\Psi}
\end{equation}
where $\Omega_g(a_i)$ is a fractionalized local operator acting on the neighborhood of $a_i$. This is called \emph{symmetry localization hypothesis}\cite{EssinHermele_Z2_2013,Maissam_SymFrac_2014,Fidkowski_Twist_2016}, a central assumption for the classification of SET phases. With this assumption, an individual anyon can realize a projective representation of $G$ instead of a linear representation. (We will not usually use distinct symbols like $R_g$ and $\Omega_g$ to distinguish between the full symmetry and the localized action on an anyon.)
However, two different projective representations may differ only by a gauge choice so that they are physically equivalent. Thus projective representations are organized into equivalence classes, and there may exist many different equivalence classes for a given $G$ and anyon contents.  In physics, an equivalence class for an individual anyon is called the fractionalization class. We define a symmetry fractionalization pattern to be the set of fractionalization classes for all distinct anyons.

We will illustrate symmetry fractionalization in a physical setting by going through an example of a toric code model with 2D translational symmetry, following the approach in Ref.~\onlinecite{EssinHermele_Z2_2013}. The 2D translation symmetries form a group $G = \{ T_1^{n_1} T_2^{n_2} | n_1, n_2 \in \mathbb{Z} \} \simeq \mathbb{Z} \times \mathbb{Z}$. $G$ is completely characterized by two generators $T_1$ and $T_2$, translations in $x$ and $y$ directions respectively, and a relation $T_1 T_2 T_1^{-1} T_2^{-1}$, which is identified as an identity element. For later usage, here we remark that any group can be fully specified by a set of generators $S$ and a set of relations $R$, where $R$ contains products of generators, each of which is identified as the identity element. This way of representing a given group is called the \emph{presentation}, and we write $G = \braket{S\,}{\, R}$; in this case, $G=\braket{T_1,T_2}{T_1T_2T_1^{-1}T_2^{-1}}$.

The toric code model realizes a $\mathbb{Z}_2$ topological order\cite{Kitaev2006}, the simplest non-trivial topological order. This topological phase has two distinct anyon excitations labeled by $e$ and $m$. Because of a given fusion rule $e \times e = 1$ and $m \times m = 1$, they can be created only in pairs. Consider the action of $T_1$ on a state with a pair of $e$ excitations at the points $r$ and $r'$. By the symmetry localization hypothesis, the action of $T_1$ can be decomposed into
\begin{eqnarray}
    T_1 \ket{\psi} &=& T^e_{1,r} T^e_{1,r'} \ket{\psi} 
\end{eqnarray}
where $T^e_{1,r}$ and $T^e_{1,r'}$ are fractionalized operators for $T_1$, acting non-trivially only near the neighborhood of $r$ and $r'$. Since the degrees of freedom that constitute the model transform linearly under the symmetry operators, any identity relation, such as $T_1 T_2 T_1^{-1} T_2^{-1}$, must act trivially on the state. In terms of fractionalized operators, this can be expressed as
\begin{eqnarray}\label{eq:constraint}
    \prod_{i \in \{r,r'\}} \qty[ T^e_{1,i} T^e_{2,i} (T^e_{1,i})^{-1} (T^e_{2,i})^{-1} ] = 1
\end{eqnarray}
Thus, for an individual fractionalized operator, $T^e_{1,i} T^e_{2,i} (T^e_{1,i})^{-1} (T^e_{2,i})^{-1}$ does not have to be the identity, and it may equal a non-trivial phase factor $\eta^e$. Since this non-trivial phase factor should be the same for all $e$ excitations, Eq.~\ref{eq:constraint} implies $\qty(\eta^e)^2 = 1$. Thus $\eta^e \in \{1,-1\} \simeq \mathbb{Z}_2$, and $T^e_{1,i}$ and $T^e_{2,i}$ do not have to be commutative. Physically, this corresponds to the situation where $e$ excitation experiences an emergent magnetic $\pi$-flux  per unit cell. 

The constraint on $\eta^e$ is called a compatibility condition, and is determined by the fusion rules of underlying anyons. For example, if a fusion rule is given by $e \times e \times e = 1$ instead of $e \times e = 1$, we would have $(\eta^e)^3 = 1$ because Eq.~\ref{eq:constraint} would become a product of three equivalent anyons. Then we get $\eta^e = e^{i 2\pi m/3} \in \mathbb{Z}_3$ for $m=0,1,2$. This phase factor $\eta^e$ forms an abelian group, called a \emph{coefficient group} ${\cal A}$. The same procedure can be repeated for the $m$ excitations, giving an additional phase factor $\eta^m$. Since Eq.~\ref{eq:constraint} is the only nontrivial relation among generators for the group $G$, we conclude that a $\mathbb{Z}_2$ topological order enriched by the translational symmetry group $G$ has four distinct SET phases, characterized by $(\eta^e, \eta^m) \in \mathbb{Z}_2 \times \mathbb{Z}_2$.

In general, symmetry fractionalization manifests itself by how products between group elements that are equal to the identity for the microscopic degrees of freedom (i.e., the relations) become non-trivial phase factors when applied to anyons. In some cases, there is an additional subtlety: the phase factors may become dependent upon some arbitrary choices (we call it a gauge choice).
The factorization of the symmetry into parts is ambiguous---a symmetry $\mathcal{O}$ acting on an anyon can be redefined as follows:
\begin{equation}\label{eq:gauge}
 \mathcal{O}|\psi_a\rangle  \mapsto e^{i\phi_{\cal O} (a)} \mathcal{O}|\psi_a\rangle
\end{equation}
for any state $\psi_a$ with anyon-charge equal to $a$. Here the phase $\phi_{\cal O}(a)$ can depend on the type of anyon.  Such transformations can change
the phase factors appearing in the relations between fractionalized operators, and in that case, one would regard the new set of phase factors as being equivalent to the old. 
Hence SET phases are defined by equivalence classes of projective representations of a global symmetry group $G$ with phase factors from a coefficient group $\cal A$, where
$\cal A$ is the set of phase factors consistent with the fusion rules.

How can we find all the equivalence classes for possible projective representations? A mathematical tool called \emph{group cohomology} allows us to do such a task. The second cohomology group $H^2(G,{\cal A})$ is an abelian group whose elements have a one-to-one correspondence with equivalence classes of projective representations. We pause to consolidate the physical intuition we developed via the example of the toric code: 

\vspace{0.05in}
\begin{tcolorbox}
{\noindent {\bf Intuition} Consider a group $G$ characterized by a set of free generators $S$ and a set of relations $R$. For a given abelian group ${\cal A}$, determined by fusion rules of underlying intrinsic topological order, SET phases with a symmetry $G$ are classified by the distinct ways to assign elements of ${\cal A}$ to every relation of $R$ consistently. 
}\end{tcolorbox}

The following mathematical lemma (3.16) of Ref.~\onlinecite{GCnote} gives a more precise description of the set of equivalence classes of projective representations.

\begin{tcolorbox}
{\noindent {\bf Lemma} Consider a discrete group $G$ with presentation $G = \expval{ S | R }$, where $S$ is a set of free generators and $R$ is a set of relations of generators in $S$ identified as the identity. For an abelian group ${\cal A}$, there is one-to-one correspondence between $H^2(G,{\cal A})$ and a quotient group $\text{Hom}_{\mathbb{Z}G}(R/R', {\cal A})/\text{Der}(S,{\cal A})$, where $R/R'$ is the abelianization of $R$ and $\text{Der}(S,{\cal A})$ is the \emph{derivation} function\cite{GCnote}.
}\end{tcolorbox}

\vspace{0.06in}

A group of homomorphisms $\text{Hom}_{\mathbb{Z}G}(R/R', {\cal A})$ represents all consistent assignments of elements of ${\cal A}$ into elements of abelian group of relations $R/R'$\footnote{$R'$ is a commutator subgroup of $R$. $R/R'$, a quotient group of $R$ by $R'$ gives an abelian group generated by $R$}, formalizing  the idea of `different ways' to assign elements of ${\cal A}$ to distinct relations of $R$. $H^2(G,{\cal A})$ is the second group cohomology of a group $G$ with an abelian coefficient group ${\cal A}$, whose elements directly correspond to distinct SET phases. $\text{Der}(S,{\cal A})$ represents a set of redundancies arising from the gauge choices made for projective representations, thus it makes sense to quotient out $\text{Hom}_{\mathbb{Z}G}(R/R', {\cal A})$ by $\text{Der}(S,{\cal A})$ to obtain equivalence classes for projective representations.

By using the lemma, it is possible to understand possible SET phases without sophisticated mathematics in terms of physically relevant expressions, the way each symmetry relation is assigned a non-trivial phase factor. 
For example, let's classify SET phases for $\mathbb{Z}_2$ topological order with a wallpaper group $p4m$, which is the symmetry group for a square lattice. The fusion rules $e \times e = m \times m = 1$ again imply that the coefficient group is ${\cal A} = \mathbb{Z}_2$. $G_{p4m}$ is fully  characterized by generators $S = \{ T_1, T_2, C_4, \sigma \}$ with (minimal) seven relations $R = \{r_1, r_2, \dots, r_7\}$. $T_{1/2}$ are translations, $C_4$ is a four-fold rotation, and $\sigma$ is a reflection with respect to the $y$-axis. The seven relations are
\begin{eqnarray}
r_1 &=& T_1 T_2 T_1^{-1} T_2^{-1} \nonumber \\
r_2 &=& \sigma^2 \nonumber \\
r_3 &=& (C_4)^4 \nonumber \\
r_4 &=& T_1 \sigma T_1^{-1} \sigma \nonumber \\
r_5 &=& T_2 \sigma T_2 \sigma \nonumber \\
r_6 &=& C_4 \sigma C_4 \sigma \nonumber \\
r_7 &=& T_1 C_4 T_2 C_4^{-1}.
\end{eqnarray}
Based on the toric code example, in this case, one might expect $2^7$ different fractionalization classes for each $e$ and $m$ respectively. However, this is wrong. While the first six relations are invariant under redefining each operator by a multiplicative factor $\tilde{O}_i = \eta_i O_i$ where $\eta_i \in {\cal A}$ and $O_i \in S$, $r_7$ changes its sign. If we redefine $\tilde{T}_1 = - T_1$, then $\tilde{r}_7 = - r_7$. Thus, the phase factor assignment on $r_7$ is gauge-dependent, and does not have any physical meaning. Mathematically, $r_7$ is what we quotient out with Der$(S,A)$, which encodes gauge redundancy. Thus, there are $2^6$ fractionalization classes for each $e$ and $m$, and we get total $2^{12}$ different SET phases in this case, if we have no additional symmetry.

So far, we have discussed only cases where $G$ does not permute anyons. For example, in Wen's plaquette model\cite{WenPlaquette2003} which realizes $\mathbb{Z}_2$ topological order, translation symmetry  permutes $e$ and $m$. In such a case, we need to discuss SET classification in a broader mathematical context, and the above approach lacks some of the sophistication necessary to enumerate all possible SET phases. In  Appendix\,\ref{AppendixTwisted}, we discuss the classification of SET phases with anyon-permuting symmetry in detail.

{\bf Caveat} In this discussion, we simplified the classification by considering only bosonic anyons (zero topological spin). However, for an anyon with non-zero topological spin, self-statistics can modify the phase factors with respect to the  rotation/reflection symmetries\cite{EssinHermele_Z2_2013,YML2014,Zaletel_TwistFactor}. Since we are interested in bosonic anyons that can be condensed, we will not have to consider this complication.

\section{Fractionalization of Symmetries in Combinations of Lattices}\label{sec:nonsymmorphic}

If one would like to make a featureless state on a certain crystal, or at least a state with the minimum amount of 
topological order, one can build the crystal out of sublattices whose symmetry is simpler.  For example, to
 make a featureless state on the honeycomb lattice, one can divide the honeycomb lattice into two triangular sublattices (called the $A$ and $B$ atoms). We start with simple resonating valence bond states on each sublattice (e.g., we start with a symmetric state of free fermions and then project the wave function
so that each site has exactly one fermion on it). This state forms a $\mathbb{Z}_2\times\mathbb{Z}_2$ spin liquid, because the spinon and vison excitations on each sublattice are distinct. The visons transform projectively under translations while spinons transform projectively under spin rotations, so neither of them can be condensed without breaking a symmetry.  However,
we can form bound states of the spinon excitations in the two sublattices and of the vison excitations in the two sublattices.
If we can show that these transform non-projectively, then they can be condensed without breaking symmetries, and this
would cause all the individual spinons and visons to become confined, removing the topological order. Such phase transitions triggered by condensation (or proliferation) of spinons has been well known for $\mathbb{Z}_2$ and $\mathbb{Z}_2 \times \mathbb{Z}_2$ topological orders without lattice symmetries. \cite{Tupitsyn2010, ColorCode_1,ColorCode_2, Zarei2016}. It is
clear that the bound states transform trivially under spin rotations and translations, because the projective minus-signs
for these symmetries are squared when two excitations are present.  However, the behavior of the excitations under other
symmetries may be more complicated because the symmetries can act differently on the two sublattices.

Once one knows the fractionalization of the anyons in one of the sublattices, it is possible to deduce how they transform under the full symmetries of the combined lattice, including symmetries that exchange the sublattices. In fact, one can prove that the symmetry fractionalization pattern of the combined lattice is \emph{uniquely} defined for a given symmetry fractionalization pattern of anyons in sublattices.

We will describe a general procedure for doing this here. First choose one of the sublattices, $L_1$.  For each of the other sublattices, $L_i$, choose a symmetry that maps $L_1$ to $L_i$; we will call this an \emph{identifying map}
 $\mathcal{S}_i:L_1\rightarrow L_i$. Each symmetry $U$ can be expressed in terms of symmetries of  $L_1$ because sublattices are enforced to be equivalent by the symmetry of the combined lattice.
 Suppose $U$ maps the lattices to one another according to $U:L_i\rightarrow L_j$.  Then
we will express $U$, when it acts on the $i^\mathrm{th}$ sublattice, as
\begin{equation} \label{eq:expandsymmetry}
U|_i=\mathcal{S}_jX\mathcal{S}_i^{-1}
\end{equation}
where $X$ is one of the symmetries restricted to $L_1$, which will usually depend on $i$. We emphasize that $\mathcal{S}_i$ is just a unidirectional map from a specific sublattice ($L_1$)
to one of the other sublattices. 

\begin{figure}[t]
    \vspace{0.15in}
  \centering
  \scalebox{0.9}{
  \begin{tikzpicture}
    \coordinate (Origin)   at (0,0);
    \coordinate (XAxisMin) at (-5,0);
    \coordinate (XAxisMax) at (5,0);
    \coordinate (YAxisMin) at (0,-5);
    \coordinate (YAxisMax) at (0,5);

	\node[left, black] at (-3.4,0.0) {{\color{blue} \large ${P}_x$}} ;

	\clip (-3.2,-2.9) rectangle (4.8,2.9); 


    \foreach \y in {-1,1}{
        \draw [->,black,ultra thick] (-5, 2*\y) -- (5, 2*\y);
    }

	\draw [->,blue,ultra thick] (-5, 0) -- (5, 0);

    \foreach \x in {-1,...,1}{
        \draw [->,black,ultra thick] (2.6*\x, -5) -- (2.6*\x, +5);  
    }
    
    \foreach \x in {-1,...,1}{
        \draw [black,dashed] (2.6*\x+1.3, -5) -- (2.6*\x+1.3, +5);  
    }
    
    \foreach \x in {-1,...,1}{
        \foreach \y in {-1,...,1}{
            \node[draw,circle,inner sep=2pt,fill] at (2.6*\x+0.2,2.0*\y-0.2) {};
            \node[draw,circle,inner sep=2pt] at (2.6*\x+0.2+1.3,2.0*\y+0.2) {};
        }
    }

    \foreach \x in {-1,...,1}{
        \foreach \y in {-1,...,1}{
            \pgfmathparse{int(round((\x+1)+(\y+1)*3+1))}
       		\let\theIntINeed\pgfmathresult     
            \node[left, black] at (2.6*\x+0.7,2.0*\y-0.4) {\theIntINeed};
        }    
    }

    \foreach \x in {-1,...,1}{
        \foreach \y in {-1,...,1}{
            \pgfmathparse{int(round((\x+1)+(\y+1)*3+1))}
       		\let\theIntINeed\pgfmathresult     
            \node[left, black] at (2.6*\x+0.8+1.3,-2.0*\y+0.4) {\theIntINeed$'$};
        }    
    }

  \end{tikzpicture}
  } 
  \caption{    \label{pg} Simplest lattice structure for non-symmorphic group \textit{pg}. The black and white dots form the A/B-sublattices respectively. The black lines are drawn to represent each unit cell. The dashed lines are separated by a half-lattice vector translation in the $x$-direction. The reflection symmetry $P_x$ we use in the text is defined with respect to the blue line. The sites are numbered to help picture the identifying map, $\mathcal{S}$, which is defined as the glide reflection symmetry $g\big|_A$ restricted to the A-sublattice. This maps the black dots labeled by $n$ to the white dots labeled by $n'$. Under $g\big|_B$, $n' \mapsto n+1$, meaning that $g\big|_B=T_x{\cal S}^{-1}$.  }
\end{figure}
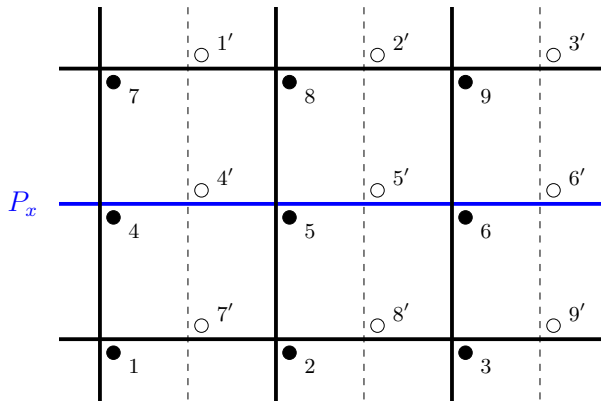

Before returning to the honeycomb lattice, let us apply this technique to a lattice with the symmetry group $G_{pg}$, see Fig.~\ref{pg}. This lattice has two translations $T_x$ and $T_y$, and glide-reflection symmetry $g = \tau_x \circ P_x$, where $\tau_x$ is half of the $T_x$ translation and $P_x$ is a reflection with respect to the $x$-axis.

Then, consider a spin-model with this symmetry, with a spin-$\frac12$ per site.  Like in the honeycomb lattice, there
are two spins per site, so this system is not forced to break symmetry or have topological order by \emph{translational} symmetry alone.  We will try to use our method to construct a featureless state: this system has a bipartite lattice structure, drawn as black and white dots in the figure, so we can construct a $\mathbb{Z}_2\times \mathbb{Z}_2$ spin liquid.  We will find that one set of anyons can be condensed, reducing this
to a $\mathbb{Z}_2$ spin liquid, but no more.  This is consistent with the result of Ref.~\onlinecite{Sid2013}, which shows that there are no featureless states with a half-integer spin in the reduced unit cell of a crystal with $pg$ symmetry.

When each sublattice is assumed to be in a $\mathbb{Z}_2$ spin-liquid phase, the system realizes $\mathbb{Z}_2 \times \mathbb{Z}_2$ topological order with four different anyons $e_A/m_A$ and $e_B/m_B$, where the $A$ and $B$ lattices are the
black and white dots respectively. How can we classify these anyons? The symmetry group $G_{pg}$ is characterized by two generators $x=g$ and $y=g^{-1}T_y$ together with the single relation $x^2y^2=1$. In group presentation language, $G_{pg}=\langle x,y|x^2y^2\rangle$. 

There is a complication in determining how anyons transform when lattices are permuted.
In the previous section, we explained that a symmetry fractionalization class can be represented by a set of phase factors representing the actions of symmetry relations. However, symmetries such as the glide-reflection $g$ exchange the $A$ and $B$ sublattices, and by extension exchange anyon types.
In the presence of a symmetry that exchanges anyon types, some of the phases that are ordinarily gauge invariant become
gauge dependent and cannot be used to distinguish SET phases\cite{Maissam_SymFrac_2014, Semion2016} (See Appendix\,\ref{AppendixTwisted} for a general framework).

We can see this for the defining relation of this lattice, $x^2y^2=gT_yg^{-1}T_y$ which is the identity up to a sign when it acts on an anyon; e.g., 
$gT_yg^{-1}T_y|m_{A}\rangle=s_{A}|m_{A}\rangle$.
This phase is not gauge invariant. Note that $g$ takes an $m_A$ type anyon to an $m_B$ type anyon and vice versa.
Hence if we apply the relation to $m_A$, the first $T_y$ (reading from the right) acts on an anyon of type $A$ and the second acts on one
of type $B$.  We can now pick different gauge choices for the two $T_y$'s according to Eq. (\ref{eq:gauge}), meaning that we multiply by $-1$ to the action of $T_y$ when it acts on $m_A$ but not $m_B$. This changes the sign of $s_A$.
However, there \emph{is} a gauge-invariant phase, because if we consider also an $m_B$ anyon and define
$gT_yg^{-1}T_y|m_B\rangle=S_B|m_B\rangle$, then $s_As_B$ \emph{is} gauge invariant, because changing the sign of $T_y^A$ affects both signs in the same way.  This is the only gauge-invariant quantity associated with the space group symmetry for the $m$'s; there is a similar invariant for the $e$'s (as well as parameters associated with time-reversal and spin-rotation symmetry).

Now each sublattice has symmetry $G_{p1}$, which is generated by just the translations $T_x$, $T_y$. (The lattice itself has rectangular symmetry, but the Hamiltonian is required only to have symmetries that are part of the symmetry group
of the full lattice.) The property of $m_A$ on the $A$-sublattice is defined by whether $T_x$ and $T_y$ commute or anticommute. Let us take
the $A$-sublattice as the reference lattice, and define $\mathcal{S}=g\big|_A$ (See Fig.~\ref{pg}), which maps it onto the $B$-sublattice. (The identifying map for the $A$-sublattice with itself can be taken as the identity.) Then, following
Eq. (\ref{eq:expandsymmetry}), we express $g$ and $T_y$ as
\begin{eqnarray}
g\left(|\psi_1\rangle\otimes|\psi_2\rangle\right)&=& T_x\mathcal{S}^{-1}|\psi_2\rangle\otimes \mathcal{S}|\psi_1\rangle \nonumber \\
T_y\left(|\psi_1\rangle\otimes|\psi_2\rangle\right)&=&T_y|\psi_1\rangle\otimes \mathcal{S}T_y^{-1}\mathcal{S}^{-1}|\psi_2\rangle.
\end{eqnarray}
since the full system is a tensor product of states $\psi_1$ and $\psi_2$ on the sublattices. (For example, the first equation means that $g|_B=T_x\mathcal{S}^{-1}$ and $g_A=\mathcal{S}$ which follow from the facts that $\mathcal{S}=g_A$ and $T_x=g^2$.) 
We can now apply the relation $gT_yg^{-1}T_y$ to the anyons to determine their fractionalization properties with respect to $G_{pg}$.
The action of $T_y$ on the $B$-sublattice could be represented simply by $T_y$ rather than by $\mathcal{S}T_y\mathcal{S}^{-1}$, but we write it the latter way, as required by Eq. (\ref{eq:expandsymmetry}), because then all the $\mathcal{S}$'s will cancel out when we calculate the relations between the group elements.  We find
\begin{equation}
gT_yg^{-1}T_y|\psi_1\rangle\otimes|\psi_2\rangle=(T_xT_y^{-1}T_x^{-1}T_y|\psi_1\rangle)\otimes|\psi_2\rangle.
\end{equation}
So if $T_xT_y^{-1}T_x^{-1}T_y=\eta_m$ for an $m_A$ anyon, then $gT_yg^{-1}T_y$ applies the sign $s_A=\eta_m$ to
an $m_A$ excitation, and leaves an $m_B$ invariant i.e., $s_B=1$. Thus the gauge-invariant parameter $s_As_B=\eta_m$.
The two anyon-types seem to behave differently because the way we chose to reduce the symmetries to the $A$-sublattice is effectively equivalent to making a choice of gauge for $g$ and $T_y$. 
In general, whenever one builds a state out of disconnected states on sublattices that are symmetric with one another, the fractionalization of the anyons in the full system (under all the symmetries) can be reduced to the fractionalization of one of the sublattices, just like this.

What are the implications for featureless states on the $pg$ lattice?
If one makes a resonating valence bond state on each of the $A$ and $B$ sublattices, the fact that
there is a spin-$\frac12$ per unit cell (in the sublattice) partly determines how the spinon and vison
transform.  $T_x$ and $T_y$ must anticommute for
the vison, $m$, while the spinon must have a half-integer spin.  Furthermore, we can choose a spin liquid such that
$T_x$ and $T_y$ commute for the spinon\cite{Wen2002, FW2016}.
Thus the product of $gT_yg^{-1}T_y$ for $e_A$ and $e_B$ is equal to 1.  This implies that a bound state formed
by combining these two particles together transforms trivially under $gT_yg^{-1}T_y$. (Note that, unlike for a single anyon, this relation
\emph{is} gauge invariant for the bound state because it does not change the topological type.)  Such a bound state also has an integer spin.  Hence the bound state has non-fractional transformation under the symmetries and can be condensed.  (We will explain below why when a certain anyon has non-fractional transformation properties then there is a condensate of this anyon that is \emph{invariant} so that it can be condensed.) This leaves only the excitations $e'=e_A$ (which is equivalent to $e_B$ because of the condensate)
and $m'=m_Am_B$ (because $m_A$, $m_B$ are now confined).

The remaining excitations have non-trivial quantum numbers---$e_A$ transforms projectively under $SO(3)$ while
$m_Am_B$ acquires a minus sign under $gT_yg^{-1}T_y$ (since $T_x$ and $T_y$ anticommute for $m_A$).
Thus, condensing $e'$ will break the $SO(3)$ symmetry and condensing $m'$ will break the glide-reflection symmetry. This result is consistent with  the extended version of the HOLSM theorem for nonsymmorphic space groups.\cite{Sid2013} In fact, the existence of the two symmetries $(i)$ $U(1) \subset SO(3)$ symmetry and $(ii)$ glide-reflection symmetry is essential in the proof of the extended version of the HOLSM theorem, and the fractionalization of these two symmetries is not just coincidence, as supported by Ref.~\onlinecite{PRX_LSM2016}.


A system with the same property, having an integer filling $\nu=1$ per unit cell and anyons with non-trivial fractionalization, was constructed previously for the lattice group $p4g$ in Ref.~\onlinecite{Sungbin2016}. They first constructed $\mathbb{Z}_4$ topological order via a parton-approach and condensed a pair of visons in the effective theory description to obtain $\mathbb{Z}_2$ topological order without breaking any symmetry.

\section{Bipartite Lattice Construction of a Featureless State on the Honeycomb Lattice}\label{sec:bipartite}


\subsection{Bipartite Lattice Construction}


Consider a honeycomb lattice $L$ and its triangular sublattices $L_A$ and $L_B$. Assume the two sublattices are decoupled from each other, and each sublattice realizes a ground state symmetric with respect to the underlying triangular lattice symmetry, spin-rotation, and time-reversal symmetry. The system is defined by the following Hamiltonian:
\begin{equation}
    H = H_A \otimes I_B + I_A \otimes H_B
\end{equation}
where $H_{A,B} = \sum_{ i,j \in L_{A,B}  } J_{ij} \vec{S_i} \cdot \vec{S_j}$ describes a symmetric Hamiltonian for each sublattice and $I_A$ and $I_B$ are identity operators acting on sublattice $L_A$ and $L_B$.  For Hamiltonians with the right type of frustration,
each sublattice has a symmetric and gapped ground state. (It may be helpful to include ``ring exchange'' interactions involving several sites simultaneously.) By the HOLSM theorem, such a ground state of a triangular lattice with a spin-1/2 per site is topologically non-trivial. Thus, it should be a $\mathbb{Z}_2$ spin liquid (or something more complicated). $\mathbb{Z}_2$ SL states with different symmetry fractionalization classes can be accessed by the PSG approach.\cite{FW2006,YML2016} Later on, we turn on the interaction between two sublattices so that excitations residing in different sublattices can interact with each other.

In the honeycomb lattice formed by combining the two sublattices, the anyons on each triangular sublattice belong to distinct classes; the order is described by $\mathbb{Z}_2 \times \mathbb{Z}_2$. 
We have two distinct spinons $e_A$ and $e_B$, and two distinct visons $m_A$ and $m_B$. 

Our goal is to show that there are spin liquid states for the two sublattices such that the bound states $W_1=e_Ae_B$ and $W_2=m_Am_B$ have trivial phase factors. Then condensing them will not break any symmetry and will lead to
a state without topological order.

\subsection{Parton Construction and Fractionalization on the Triangular Sublattice}
Although we can classify entire SET phases via an abstract mathematical machinery, group cohomology, the machinery does not tell us when a given SET phase can be realized using certain particles, e.g. spin-$\frac12$ particles on every site. To do this, we must  construct examples of the various phases. The most common approach for constructing states with topological order is the mean-field parton construction.\cite{Wen2002,Wen2002_2, FW2006, FW2010, FW2016, YML2016} In a mean-field parton construction, we explicitly break a fundamental degree of freedom, such as an electron or spin, into fractional entities called partons. Assuming certain mean-field configurations, we can write down a quadratic Hamiltonian for partons, in which the partons can move independently from one another. Then, this parton ansatz is projected onto the part of the Hilbert space corresponding to physical states. For a given Hamiltonian, one can obtain many different ansatzes starting from different mean-field configurations; thus, we should evaluate the energy of the projected state for the given Hamiltonian to determine the true ground state.

To be somewhat more precise, let's take an example. We first represent the spin operator, using Schwinger's mapping of spins to bosons, as $\vec{S}_i = \frac{1}{2} b^\dagger_{i,\alpha} \vec{\sigma}_{\alpha,\alpha'} b_{i,\alpha}$ where $\alpha = \uparrow, \downarrow$; physical states must satisfy the 
constraint $\sum_{\alpha} b^\dagger_\alpha b_\alpha = 2S$, given spin $S$ per site. 
Although this constraint seems very strong, a highly frustrated Hamiltonian can have a ground state that is approximately described by {starting} with
a quadratic Hamiltonian in the bosons, and then implementing the constraint by just projecting onto the physical states.  
For certain mean-field configurations, this state will have physical excitations corresponding to the partons in spite of the constraint. These excitations have half-integer spin and are the spinons.  This state has topological order due
to the fractional nature of these excitations, and must also have vison excitations (excitations that pick up a phase of $-1$ when they encircle a spinon).

\begin{table}[t]
    \begin{center}
      \begin{tabular}{| c | c | c | c |}
    \hline
    $\,\,\,$ Algebraic Identities $\,\,\,$& $\,\,\,$bosonic spinon $\,\,\,$& $\,\,\,\,\,\,$vison$\,\,\,\,\,\,$ \\ \hline\hline
    $T_1 T_2 T_1^{-1} T_2^{-1} = 1 $& $\eta_1 = (-1)^{p_1}$ & -1 \\ 
    \hline
    $\sigma^2 = 1$& $\eta_2 = (-1)^{p_2+p_3}$ & 1 \\
    \hline  
    $(C_6)^6 = 1$ & $\eta_3 = (-1)^{p_3}$ & -1 \\
    \hline
    $R^2 = (\sigma C_6)^2 = 1$ & $\,\,\,\, \eta_4 = (-1)^{p_2} \,\,\,\,$ & 1 \\
    \hline
    $\sigma T_1 \sigma T_2^{-1} = 1$ & $\alpha^e_1$ & $\alpha^m_1$ \\ 
    \hline
    $C_6^{-1} T_2^{-1} C_6 T_1  = 1$ & $\alpha^e_2$ & $\alpha^m_2$ \\ 
    \hline
    $\,\,\, C_6^{-1} T_2^{-1} T_1 C_6 T_2 = 1 \,\,\,$ & $\alpha^e_3$ & $\alpha^m_3$ \\ 
    \hline
    $T_1 {\cal T} T_1^{-1} {\cal T}^{-1} = 1$ & 1  & 1 \\ 
    \hline
    $T_2 {\cal T} T_2^{-1} {\cal T}^{-1} = 1$ & 1  & 1 \\ 
    \hline
    $\sigma {\cal T} \sigma {\cal T}^{-1} = 1$ & $\eta_2$  & 1 \\ 
    \hline
    $\,\,\, R {\cal T} R^{-1} {\cal T} = 1\,\,\,$ & $\eta_4$  & 1 \\ 
    \hline
    $ {\cal R}(2\pi) = 1 $ & -1 & 1 \\
    \hline
    $ \,\,\, {\cal R}(\theta) {\cal O} {\cal R}^{-1}(\theta) {\cal O}^{-1}  = 1 \,\,\,$ & 1 & 1 \\
    \hline
  \end{tabular}
  \caption{     \label{Triangle} This shows symmetry fractionalizations in triangular lattice. Here, ${\cal T}$ is time-reversal operator and ${\cal R}(\theta)$ is spin-rotation operator by angle $\theta$. ${\cal O}$ in the last row represents any operator in $G \times SO(3) \times \mathbb{Z}_2$, which means that spin-rotation commutes. Phase factors are slightly modified compared to Ref.~\onlinecite{YML2016} since we use a different coordinate system.  }
\end{center}
\vspace{-0.1in}
\end{table}

To further understand a mean-field ansatz for a topological order, we need to determine the projective transformations of the spinons and visons.  The projective representation of the spinons, as described above, is identical with the ``Projective Symmetry Group'' 
of the state constructed this way\cite{FW2006}.  The projective transformation for the visons can be determined by several approaches: (\emph{i}) Adding $Z_2$ gauge fields to the mean-field Hamiltonian and computing Berry phases\cite{Huh2011} (\emph{ii})
Requiring that the spin liquid does not support gapless edge states and deducing the consequence on visons\cite{YML2014, YML2016, Alicea2014} (\emph{iii}) Using flux-anomaly arguments to constrain possible fractionalization pattern for visons.\cite{HermeleChen2016, YangQi2015}.

When we construct a $\mathbb{Z}_2$ spin liquid in this way on each sublattice of the honeycomb lattice,  what are the possible choices for its fractionalization pattern? The
Hamiltonian has $SO(3)$ spin-rotation, $Z_2^T$ time-reversal, and triangular lattice symmetry $G_s$. The lattice space group $G_s$ is generated by two translations $T_{1,2}$, mirror reflection $\sigma$, and site-centered $\pi/3$ rotation $C_6$. We label each lattice site on a triangular lattice by its positional vector $r = m a_1 + n a_2$, where $a_1 = (1,0)$ and $a_2 = (1/2,\sqrt{3}/2)$ are Bravais lattice vectors corresponding to translation $T_1$ and $T_2$ respectively. 
The reflection generator $\sigma$ is defined in such a way that it exchanges $T_1 \leftrightarrow T_2$. The fractionalization class of the topological excitations in the parton ansatz hosting $\mathbb{Z}_2$ topological order is classified by projective representations of $G_s \times SO(3) \times Z_2^T$ with $\mathbb{Z}_2$ coefficients.   Supposing that we choose to use a state that was constructed by the mean-field parton construction in Ref.~\onlinecite{FW2006}, the possible fractionalization patterns are given in TABLE~\ref{Triangle}. 
 
The mean-field parton states come in eight different classes, labeled by $p_i=0,1$ with $i=1,2,3$. Each row represents a phase factor acquired by spinons and visons under a symmetry relation. There are seven relations defining the triangular lattice symmetry group, but only four of them are gauge-independent. 
In TABLE~\ref{Triangle}, rows 5-7 give phase factors labeled by $\alpha_i$. These are not meaningful because $\alpha_i$ are gauge-dependent factors and do not contribute to the classification of the fractionalization class. In the language introduced earlier, these are exactly the relations quotiented out by the derivation group $\text{Der}(S,{\cal A})$. However, we include them because they
will be helpful for understanding the honeycomb lattice. Although not meaningful individually, products of them on the two sublattices can form an invariant factor. 

Note that a similar table in Ref.~\onlinecite{YML2016} listed all gauge-dependent factors to be $1$. Such gauge choice is possible, but we have different choices for translations and reflection operators; in our choice, we have a constraint $\eta_1 \eta_2 \alpha_1 \alpha_2 = 1$. There is no such constraint on $\alpha_3$, so we can always take a gauge choice to set $\alpha_3 = 1$. 
Using the PSG solution of Ref.~\onlinecite{FW2006}, we can actually derive that for a certain gauge choice, $\alpha_1 = (-1)^{p_2+p_3}$, $\alpha_2 = (-1)^{p_1}$, and $\alpha_3 = 1$.

The $SO(3)$ symmetry is continuous, unlike the other symmetries considered so far. There are two possible projective representations of $SO(3)$, the integer spin representations, and the familiar half-integer representations where the $2\pi$ rotation results in a factor of $-1$.  There are no additional complications
arising from relationships with the discrete symmetries\cite{EssinHermele_Z2_2013}, on account of this result: Whenever a continuous symmetry commutes with all the discrete symmetries, it also commutes
in the \emph{projective} representation. This is proved by defining a phase factor $\eta(\theta) = {\cal R}(\theta) {\cal O} [{\cal R}(\theta)]^{-1} {\cal O}^{-1}$, which must be a continuous function of $\theta$. 
Since ${\cal R}(0)$ is the identity operator, $\eta(0) = 1$, and $\mathbb{Z}_2$ symmetry fractionalization tells that $\eta(\theta)$ can take only one of two values, $\pm 1$. As $\eta(\theta)$ is a continuous function, we can conclude that $\eta(\theta) = 1$ for all $\theta$.\cite{EssinHermele_Z2_2013}

\subsection{Honeycomb Lattice}

We are now prepared to determine the fractionalization pattern of excitations in the honeycomb lattice, and show that the 
pair excitations of the underlying triangular lattices  $W_1=e_Ae_B$ and $W_2=m_Am_B$ are never fractionalized. Therefore, they can be condensed to get out of the topological order without breaking the symmetry.
We will do this in two ways.  The first way uses the method of Sec.~\ref{sec:nonsymmorphic}, identifying the two sublattices with one another, so that any
transformation of the honeycomb lattice on the $A$ or $B$ excitations can be reduced to transformations of the triangular sublattice.  

\begin{figure}[t]
    \vspace{0.15in}
  \centering
  \scalebox{1.0}{
  \begin{tikzpicture}
    \coordinate (Origin)   at (0,0);
    \coordinate (XAxisMin) at (-5,0);
    \coordinate (XAxisMax) at (5,0);
    \coordinate (YAxisMin) at (0,-5);
    \coordinate (YAxisMax) at (0,5);

    \clip (-2.1,-3.1) rectangle (4.1,3.3); 
         
    \draw [black, thick, dotted](-1.73205081*1.2,-1.0*1.2) -- (1.73205081*2.3,1.0*2.3);
    \node[above, black] at (1.73205081*2.1,1.0*2.1+0.1) {\large $\boldsymbol{\sigma} $} ;
    
    \foreach \x in {-7,-6,...,7}{
      \foreach \y in {-7,-6,...,7}{
        \node[draw,circle,inner sep=2pt,fill] at (2*\x+1*\y,1.73205*\y) {};
      }
    }
   
    \foreach \x in {-7,-6,...,7}{
      \foreach \y in {-7,-6,...,7}{
        \node[draw, thick, circle,inner sep=2pt,fill=white] at (2*\x+1*\y+1,1.73205*\y-0.57735) {};
      }
    }

    \draw [->,red,ultra thick](0.1 , 0.0) -- node [pos=0.5, above, black] {$\color{red} T^A_1$} (1.9, 0.0);
    \draw [->,red,ultra thick](0.0 , 0.0) -- node [pos=0.95, left, black] {$\color{red} T^A_2$} (0.95, 1.73205*0.95);

    \draw [->,blue,ultra thick](1,-0.57735)-- node [pos=0.65, left, black] {$\color{blue} T^B_2$} (0.05,1.1547-0.0866);
              
    \draw [->,blue,ultra thick](1,-0.57735) -- node [pos=0.7, right, black] {$\color{blue} T^B_1$} (1.95,1.1547-0.0866);

   \node[draw,circle,inner sep=2pt,fill] at (0,0) {}; 
   \node[draw, thick, circle,inner sep=3pt] at (0,0) {}; 
   \node[below, black] at (0.1,-0.1) {${\cal O}_A$} ;
    
   \node[draw, thick, circle,inner sep=2pt,fill=white] at (1,-0.57735) {}; 
   \node[draw, thick, circle,inner sep=3pt] at (1,-0.57735) {}; 
   \node[below, black] at (1.1,-0.67735) {${\cal O}_B$} ;
 
  \end{tikzpicture}
  }  \vspace{-0.05in}
  \caption{    \label{HC} \small Black dots represents sublattice $A$. White dots represents sublattice $B$. ${\cal O}_A$ and ${\cal O}_B$ represent origins (reference points) for each sublattice. Dotted line represents the axis of reflection symmetry in honeycomb lattice. }
\end{figure}
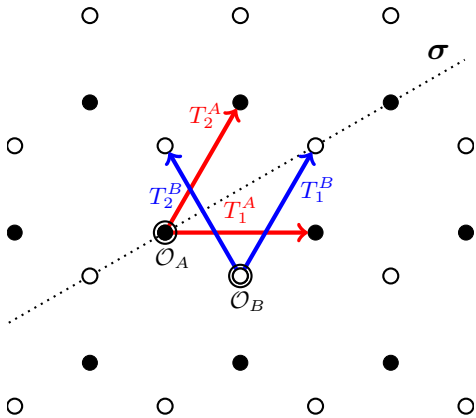

In the limit where two sublattices are decoupled, a wave function of the honeycomb lattice can be expressed as a tensor product of wave functions on each triangular sublattice. 
The total wave function is written as $\ket{\Phi} = \sum_i \ket{\psi^i_1}_A \otimes \ket{\psi^i_2}_B$, where $A$ and $B$ denote each triangular sublattice. A honeycomb lattice has the same symmetry group as a triangular lattice. We will distinguish honeycomb lattice symmetry operators using bold letters.
The symmetry group of the honeycomb lattice is again described by two translations $\boldsymbol{T}_{1,2}$, mirror reflection $\boldsymbol{\sigma}$, and six-fold rotation $\boldsymbol{C}_6$, but now, the rotation axis is at the center of hexagon not at the site. (It is not actually necessary to impose the order six symmetry of a sublattice about one of its sites since
this symmetry is broken when the two lattices are combined together.  However, since we already understand the relations and generators for the full symmetry group, we will assume the
additional symmetry to be present.) Here, we define translations of the
honeycomb lattice to coincide with those of triangular sublattice $L_A$. The two translations of $L_B$ are then defined in such a way that $\boldsymbol{C}_6 T_1^A \boldsymbol{C}_6^{-1} = T_1^B$ and $\boldsymbol{C}_6 T_2^A \boldsymbol{C}_6^{-1} = T_2^B$, as in Fig.~\ref{HC}.

\begin{table}[t]
    \begin{center}
    \begin{tabular}{| c | c | c | c |}
    \hline
    $\,\,\,$ Algebraic Identities $\,\,\,$& $\,\,\,$ $e_A$ (or $e_B$) $\,\,\,$& $\,\,\,\,\,\,\,$ $e_A e_B$ $\,\,\,\,\,\,\,$ \\ \hline\hline
    $\boldsymbol{T}_1 \boldsymbol{T}_2 \boldsymbol{T}_1^{-1} \boldsymbol{T}_2^{-1} = 1 $& $\bar{\eta}_1 = (-1)^{p_1}$ & 1 \\ 
    \hline
    $\boldsymbol{\sigma}^2 = 1$& $\bar{\eta}_2 = (-1)^{p_2+p_3}$ & 1 \\
    \hline  
    $(\boldsymbol{C}_6)^6 = 1$ & $\,\,\,${gauge-dependent} $\,\,$ & 1 \\
    \hline
    $\boldsymbol{R}^2 = (\boldsymbol{\sigma} \boldsymbol{C}_6)^2 = 1$ &  $\,\,\,${gauge-dependent} $\,\,$ & 1 \\
    \hline
    $ {\cal R}(2\pi) = 1 $ & -1 & 1 \\
    \hline
  \end{tabular}
  
  \vspace{0.08in}
  
  \begin{tabular}{| c | c | c | c |}
    \hline
    $\,\,\,$ Algebraic Identities $\,\,\,$& $\,\,\,$ $m_A$ (or $m_B$) $\,\,\,$& $\,\,\,\,\,$ $m_A m_B$ $\,\,\,\,\,$ \\ \hline\hline
    $\boldsymbol{T}_1 \boldsymbol{T}_2 \boldsymbol{T}_1^{-1} \boldsymbol{T}_2^{-1} = 1 $& -1 & 1 \\ 
    \hline
    $\boldsymbol{\sigma}^2 = 1$& 1 & 1 \\
    \hline  
    $(\boldsymbol{C}_6)^6 = 1$ & $\,\,\,${gauge-dependent} $\,\,$ & 1 \\
    \hline
    $\boldsymbol{R}^2 = (\boldsymbol{\sigma} \boldsymbol{C}_6)^2 = 1$ & $\,\,\,${gauge-dependent} $\,\,$ & 1 \\
    \hline
    $ {\cal R}(2\pi) = 1 $ & 1 & 1 \\
    \hline
  \end{tabular}
  \caption{     \label{Honeycomb} This shows some of symmetry fractionalizations for $\mathbb{Z}_2 \times \mathbb{Z}_2$ SL constructed from $\mathbb{Z}_2$ SL in triangular sublattice whose symmetry fractionalization is specified in TABLE~\ref{Triangle} in terms of $p_i$'s and $\alpha_i$'s. As symmetry group for honeycomb lattice has exactly the same structure of triangular lattice, the complete pattern would look like TABLE~\ref{Triangle}. However, unlike the TABLE~\ref{Triangle} where first four rows give gauge-invariant phase factors, here only first two rows give gauge-invariant phase factors. The other two rows give gauge-dependent phase factors (See Appendix.~\ref{AppendixTwisted}).  We used bold symbols to distinguish honeycomb lattice symmetries from triangular sublattice symmetries.}
\end{center}
\vspace{-0.1in}
\end{table}

Any symmetry operation of the honeycomb lattice can be decomposed into a tensor product of symmetry operators of individual sublattices and an identifying map ${\cal S}$ that maps the $A$ sublattice to the $B$ sublattice.
The identifying map ${\cal S} = \boldsymbol{C}_6 \big|_{A}$ is defined 
to take $T_1^A \rightarrow T_1^B$, $T_2^A \rightarrow T_2^B$ (See Fig.~\ref{HC}). This operation is equivalent to a $\pi/3$ counter-clockwise rotation with respect to the center of the hexagon for sublattice $A$. (Caution: ${\cal S}$ is one-directional map from sublattice $A$ to $B$. $\boldsymbol{C}_6|_B$ would be represented as $T_1 C_6^2 {\cal S}^{-1}$ if one analyzes carefully.)


We can then express all symmetry operators of the honeycomb lattice as follows:
{\small
\begin{eqnarray} \label{eq:sym_defining}
    \boldsymbol{T_1} : \Big( \ket{\psi_1}_A \otimes \ket{\psi_2}_B \Big) &\mapsto& \Big( T_1 \ket{\psi_1} \Big)_A \otimes \Big( \mathcal{S}T_2^{-1}T_1 \mathcal{S}^{-1} \ket{\psi_2} \Big)_B \nonumber \\ 
    \boldsymbol{T_2} : \Big( \ket{\psi_1}_A \otimes \ket{\psi_2}_B \Big) &\mapsto& \Big( T_2 \ket{\psi_1} \Big)_A \otimes \Big( \mathcal{S}T_1 \mathcal{S}^{-1}\ket{\psi_2} \Big)_B \nonumber \\
    \boldsymbol{\sigma} : \Big( \ket{\psi_1}_A \otimes \ket{\psi_2}_B \Big) &\mapsto& \Big( \sigma \ket{\psi_1} \Big)_A \otimes \Big( \mathcal{S}C_6^{-2} T_1^{-1} \sigma\mathcal{S}^{-1} \ket{\psi_2} \Big)_B \nonumber \\
\boldsymbol{C_6} : \Big( \ket{\psi_1}_A \otimes \ket{\psi_2}_B \Big) &\mapsto& \Big( T_1 C_6^2 \mathcal{S}^{-1}\ket{\psi_2} \Big)_A \otimes \Big( \mathcal{S}\ket{\psi_1} \Big)_B \nonumber  \\        
    && \mbox{}
\end{eqnarray}
}
The other operators, spin-rotation ${\cal R}^h(\theta)$ and time-reversal $\boldsymbol{\cal T}$, are simply represented by tensor products. Note, we have chosen an order of multiplication; for example, $\boldsymbol{T_1}$ on sublattice $B$ is given as $(T^B_2)^{-1}T^B_1$ rather than $T^B_1 (T^B_2)^{-1}$. At the level of linear representation, they are equivalent. However, for projective representations, the order matters since commutativity is not guaranteed. 
Fortunately, these two definitions differ only by phases, just another gauge choice (as in Eq.~\ref{eq:gauge}), and the physical gauge-invariant phase factors are unaffected.

 
 We now determine the symmetry fractionalization classes for bound states $W_1$ and $W_2$. For example,
\begin{eqnarray}
    &&\boldsymbol{T_1 T_2 T_1^{-1} T_2^{-1}} \big|_{e_A e_B} \equiv \nonumber \\
    && \Big( T_1 T_2 T_1^{-1} T_2^{-1} \big|_{e_A} \Big) \cdot \Big( T_2^{-1} T_1 T_2 T_1^{-1} \big|_{e_B} \Big) = (\eta_1)^2 \,\,\,
\end{eqnarray}
where $\eta_1$ is from TABLE~\ref{Triangle}. Since $(\eta_1)^2 = 1$, $\boldsymbol{T}_1 \boldsymbol{T}_2 \boldsymbol{T}_1^{-1} \boldsymbol{T}_2^{-1}$ acts linearly on the bound state (trivial fractionalization).  We  similarly compute all other symmetry relations characterizing the fractionalization class in Appendix \ref{Appendix1}, and find that bound states $W_1$ and $W_2$ always have trivial symmetry fractionalization classes. See TABLE~\ref{Honeycomb}, where the gauge-invariant phase factors characterizing a fractionalization class are listed for $e_A$, $e_B$ and $e_A e_B$. We can follow the same procedure for the $m$-particles.
 
Although the method of using identifying maps to reduce the problem to a single sublattice is systematic, one does not have to hew rigidly to this method.  For the honeycomb lattice, a more creative approach
 explains why the bound states always transform trivially.  We use symmetry to relate excitations on the two sublattices.  We show by symmetry that the $e_A$ and $e_B$ always change by the same phase $\pm 1$, so their bound state
$W_1$ changes by $(\pm 1)^2=1$.

The basic idea is to begin with a relation $\mathcal{R}(g_1,\dots,g_k)=1$ among the symmetries.  Suppose $e_A$ transforms projectively, so $\mathcal{R}|e_A\rangle=\eta_A|e_A\rangle$ for any state
containing an $e_A$ excitation.
Then take a symmetry $h$ that exchanges the $A$ and $B$ sublattices, and apply the relation $\mathcal{R}$ to $h^{-1}|e_B\rangle$,
giving $\mathcal{R}h^{-1}|e_B\rangle=\eta_A h^{-1}|e_B\rangle$ (since $h^{-1}|e_B\rangle$ is an excitation on the $A$-sublattice $\mathcal{R}$ picks up the factor of $e_A$). By multiplying by $h$, $h\mathcal{R}h^{-1}|e_B\rangle=\mathcal{R}(g_1^h,\dots,g_k^h)|e_B\rangle=\eta_A|e_B\rangle$, where $g_i^h$ is
short for $hg_ih^{-1}$. This shows that the B-excitations transform the same way as the A's under the \emph{corresponding} symmetries; this determines completely the transformation of all excitations
on the B-sublattice once one knows how the excitations on the A-sublattice, since we know how they transform under a complete set of relations. 
However this does not immediately imply
that the bound state $W_1$ transforms trivially because a given transformation acts the same way on both of the excitations in the bound state, but we have determined only how
the $e_B$ transforms under conjugated symmetries.  

For the honeycomb lattice, let us begin with the translation commutator, which illustrates the problem of getting conjugate relations:
\begin{equation}
\boldsymbol{T}_1\boldsymbol{T}_2\boldsymbol{T}_1^{-1}\boldsymbol{T}_2^{-1}|e_A\rangle=\bar{\eta}_{1A}|e_A\rangle\label{eq:a}
\end{equation}
where the subscript $A$ implies that $\bar{\eta}_{1A}$ is the phase for an anyon on sublattice $A$. 
Since $I=(\boldsymbol{C}_6)^3$ exchanges the two sublattices, we will use it to determine how $e_B$ transforms.
From the above equation (which is true for \emph{any} state of an $e_A$ excitation),
\begin{equation}
\boldsymbol{T}_1\boldsymbol{T}_2\boldsymbol{T}_1^{-1}\boldsymbol{T}_2^{-1}I^{-1}|e_B\rangle=\bar{\eta}_{1A}I^{-1}|e_B\rangle.
\end{equation}
Multiplying both sides by $I$ gives
\begin{equation}\label{eq:translation_commutator}
    \boldsymbol{T}_1^{I}     \boldsymbol{T}_2^{I}     (\boldsymbol{T}_1^{-1})^{I} (\boldsymbol{T}_2^{-1})^{I} \ket{e_B} = \bar{\eta}_{1A} |e_B\rangle
\end{equation}
where we define $g^h \equiv h g h^{-1}$ for $g,h \in G$.
  Notice that symmetry relations for the honeycomb lattice give the following expressions at the projective representation level:
\begin{equation}
    I\boldsymbol{T}_1 I^{-1}=\beta_1\boldsymbol{T}_1^{-1} \qquad I\boldsymbol{T}_2 I^{-1}=\beta_2\boldsymbol{T}_2^{-1}
\end{equation}
where
the $\beta$'s are phase factors that can depend on anyons acted on. Therefore, we can replace the Eq.~\ref{eq:translation_commutator} by
\begin{equation}
    \boldsymbol{T}_1^{-1} \boldsymbol{T}_2^{-1} \boldsymbol{T}_1 \boldsymbol{T}_2  \ket{e_B} = \bar{\eta}_{1A} |e_B\rangle \label{eq:b}
\end{equation}
for any state on the B sublattice, where the extra phase factors $\beta_1$ and $\beta_2$ are canceled. 

Now, this shows that the B-excitations transform the same way as the A-excitations under the \emph{corresponding} symmetries, that is Eq. \ref{eq:b} translates the B-excitations in the opposite direction relative to Eq. \ref{eq:a}.  To find how some relation transforms the bound state $e_Ae_B$ we have to know how the same relation acts on both excitations.
We can just replace $|e_B\rangle$ in Eq. \ref{eq:b} by
$\boldsymbol{T}_2^{-1}\boldsymbol{T}_1^{-1} |e_B\rangle$, which is also an $e_B$-type excitation. Then, after simplifying it,
\begin{equation}
\boldsymbol{T}_1\boldsymbol{T}_2\boldsymbol{T}_1^{-1}\boldsymbol{T}_2^{-1}|e_B\rangle=\bar{\eta}_{1B}|e_B\rangle = \bar{\eta}_{1A}|e_B\rangle.
\end{equation}
Hence $\bar{\eta}_{1B} = \bar{\eta}_{1A}$, and $W_1$ changes by $\bar{\eta}_{1A}^2=1$.

For the remaining products of symmetries, there is always an operator that exchanges the sublattices without changing the operators.
Two of the relations ($\boldsymbol{C}_6^6$, $\boldsymbol{R}^2$) are of the form $M^{2n}$ where $M$ is a transformation that exchanges $A$ and $B$ excitations, and one can use $M$ itself
to relate the phases of $e_A$ and $e_B$.Suppose for example $\boldsymbol{C}_6^6|e_A\rangle=\gamma_A|e_A\rangle$,
where $\gamma_A$ is not gauge-invariant.  Redefining $|e_B\rangle=C_6|e_A\rangle$, we get $\boldsymbol{C}_6^5|e_B\rangle=\gamma\boldsymbol{C}_6^{-1}|e_B\rangle$, which
implies $\boldsymbol{C}_6^6|e_B\rangle=\gamma_A|e_B\rangle$.  Likewise $e_A$ and $e_B$ transform with the same phase under
$\boldsymbol{R}^2$, so $W_1$ is invariant. The time reversal and $360^\circ$ rotations on the two sublattices are internal symmetries, so they are exchanged by any of the symmetries that switch $A\leftrightarrow B$.
Finally, one can deduce that $\boldsymbol{\sigma}^2$ gives the same phase when applied to both $e_A$ and $e_B$ by using the relation $I\boldsymbol{\sigma}I^{-1}\propto \boldsymbol{\sigma}$, since $I$ exchanges $A$ and $B$.

These two approaches both show that $W_1$ transforms trivially under all the relations, and the same arguments apply to $W_2$.
This implies that both bound states
they can be condensed without breaking any symmetry, so that the topological order disappears.  A more precise argument is
that, for any excitation that transforms linearly, there is some state that has \emph{trivial} quantum numbers, so it can be condensed.  The next section proves this explicitly.

We further confirmed our result by calculating twisted second group cohomology for these lattices. In Appendix.~\ref{AppendixTwisted}, the allowed fractionalization classes of $\mathbb{Z}_2\times\mathbb{Z}_2$ order where $\boldsymbol{C}_6$ exchanges anyons are enumerated using an algorithm associated with group cohomology. We find that we have indeed obtained all the possibilities, assuming the $\mathbb{Z}_2\times\mathbb{Z}_2$ order that is permuted by the symmetries in the way we have assumed.


\section{Anyon Condensation}\label{sec:condensation}

In the previous section, we confirmed that the symmetry fractionalizations of the anyons $W_1 = e_A e_B$ and $W_2 = m_A m_B$ are totally trivial in the absence of inter-sublattice interactions. Since the fractionalization class is a topological invariant, if we adiabatically turn on some interactions between sublattices, the symmetry fractionalization class of the bound states should remain the same, and we will assume this from now on. These bound states are bosonic excitations, since they are made of bosonic excitations with no mutual statistics, and there is no mutual statistics between $W_1$ and $W_2$. Thus, in principle, we can condense these excitations by tuning some interactions. Condensation of the two bound states will confine all other topological excitations and give a topologically trivial state. Such phase transitions triggered by condensation (or proliferation) of spinons has been well known for $\mathbb{Z}_2$ and $\mathbb{Z}_2 \times \mathbb{Z}_2$ topological orders without lattice symmetries. \cite{Tupitsyn2010, ColorCode_1,ColorCode_2, Zarei2016}. 

There are several interesting aspects to the condensation of anyons, so we will describe it more explicitly.
Let us focus on condensing the $e_Ae_B$ anyons.  The goal is to start with the SL wave functions, and introduce anyons into it so that the parity of the number of anyons in any given region fluctuates strongly. The anyons cannot be correlated in bound pairs, or else they will not cause the confinement of the anyons that have non-trivial braidings with them. We will also want to create a condensate of a low density so that the state is not completely distorted.  We will discuss qualitative i) How to choose a Hamiltonian that favors a condensate ii) Why knowing that the anyons transform non-projectively implies that there is a symmetric condensate (i.e., a state that does not transform \emph{at all}).  

\begin{figure}
\includegraphics[width=.45\textwidth]{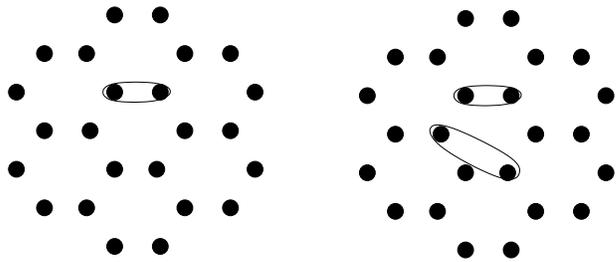}
\caption{ \label{honeycombpuzzle} Forms of the wave-function of an $e_Ae_B$ pair. (Left) A simple idea for how to create the $e_Ae_B$ bound state---the oval represents a singlet of excitations on the two lattice sites. However, the wave-function is odd under reflection in the bond, so there is no way to create a completely symmetric state starting from this one. (Right) Instead, if one creates four excitations in a configuration without any symmetry, the configuration can be superimposed with all the symmetric transformations of itself without any cancellation occurring.  Note that there are three anyons on B sites and one on an A site, so the net charge is correct. Alternatively, one could add some non-topological excitation to the singlet in the left figure, at a position that breaks the reflection symmetries.}
\end{figure}

First, choose a local configuration with the charge $e_Ae_B$ (see Fig.~\ref{honeycombpuzzle}), and let $W^\dagger$ be an operator that creates this configuration at a certain position of the lattice.  Then the condensed wave function can be represented schematically as 
\begin{equation}
    \ket{\Psi_C} \approx e^{\kappa \sum_S SWS^\dagger}|\mathrm{spin}\ \mathrm{liquid}\rangle
    \label{eq:symmetrize}
\end{equation}
where $\kappa$ is small to keep the density of anyons low. The sum $\sum_S$ is over all the symmetries of the lattice, to create a symmetric wave function.
In general,
\begin{equation}
    S_1 \sum_{S\in G} [SW^\dagger S^\dagger] S_1^\dagger=\sum_S e^{i\phi(S,S_1)}(S_1S)W^\dagger(S_1S)^\dagger
\end{equation} 
where $\phi$ is a phase factor associated with a projective symmetry representation for the anyons, and $G$ is the total symmetry group. There is no symmetry fractionalization in this case, which means that one can change the ``gauge" of the symmetries so that all the $\phi$'s are zero. Then the expression is the same as $\sum_S SW^\dagger S^\dagger$; i.e., the condensate is symmetric, hence condensing it does not break the symmetry.  (In the next section, we will see that condensates that change sign under some of the symmetries
actually are allowed as well.)
The exponential is not to be interpreted completely literally: one should expand it in a Taylor series, ignore each odd order term (because it is not possible to create an odd number of anyons), and then create the even numbers of anyons in the even-order terms by connecting strings between arbitrary pairs.  The wave function is independent of how the anyons are paired because they do not have any mutual statistics.  In addition, one can omit terms where two anyons overlap since we have not defined how to create two anyons at the same point. 

There are several issues which we can elaborate on.  First of all, a wave function like this, although it is globally symmetric, may actually represent a  cat state formed by superimposing a set of states each separately breaking the symmetry (e.g., dimer-orderings rotated in all directions).\cite{Kimchi2013, JianZaletel2016, YR2016}  This is caused by overlaps between anyons which can lead to correlations, so it can be avoided by making $\kappa$  small enough that the anyons are far from one another.

Secondly, what Hamiltonian can stabilize this state? To ensure the state is symmetric as just discussed, we need $\kappa$ to be small.
It is difficult to construct a Hamiltonian that produces a Bose condensate with a small density when the excitations being condensed have $\mathbb{Z}_2$ quantum numbers, unlike in ordinary Bose condensation. Suppose one takes a Hamiltonian $H=- \sum_{<ij>} \lambda (W_i^\dagger W_j^\dagger+ \text{h.c.})+\sum_i \epsilon n_i$.  The first term both moves single anyons and creates pairs of anyons and the second term suppresses anyons energetically.    As $\epsilon$ is lowered, anyons will first appear only in virtual pairs, which is not a true condensate.   The $\lambda$ term must be strong enough to move anyons apart from one another, but then it will also be strong enough to create a high density of anyons. 
However, we can take a Hamiltonian that includes a long-range repulsion between anyons to keep the density low, $\sum_i \epsilon n_i- \sum_{ij} \lambda_{ij} (W_i^\dagger W_j^\dagger+\text{h.c.})-\sum_{ij}t_{ij} W_i^\dagger W_j+U\sum_{ij}n_in_j$. The term $U_{ij}$ is a repulsive interaction between anyons that is large below some distance $L$.  This allows one to diagonalize the Hamiltonian in a subspace consisting of
anyons spaced by at least this distance.  The term $\lambda_{ij}$ creates anyons in pairs and allows them to move apart.  This should be nonzero for pairs of sites separated by slightly more than $L$, because otherwise, it will have no effect on account of the repulsion. The other terms
$t_{ij}$ and $\epsilon$ describe the motion and the energy cost of an anyon. When $\epsilon,t_{ij},\lambda_{ij}\ll U_{ij}$, this system should undergo an ordinary Ising transition where the anyons condense as $\epsilon$ decreases, except with the anyons kept at a low density.


Finally, there is an issue related to symmetry.  If $W^\dagger$ is not chosen right, and it turns out to be odd under a reflection or a $180^\circ$ rotation, the wave-function $SWS^\dagger$ would cancel. Consider trying to make a symmetric state out of a pair of $e_A$ and $e_B$ anyons.  Under both spin and orbital symmetries, the state must be symmetric.  As in atomic physics there is a connection between these symmetries: because the spin part of the wave-function is antisymmetric, the orbital part must also be antisymmetric (since the $e_A$ and $e_B$ act as bosons relative to one another).  However this allows only lattice versions of orbitals of the form $p_x$ that break rotational symmetry or $p_x+ip_y$ that break time-reversal symmetry.

Explicitly, let $W^\dagger= c_{1\downarrow}^\dagger c_{2\uparrow}^\dagger-c_{1\uparrow}^\dagger c_{2\downarrow}^\dagger$ (see Fig.~\ref{honeycombpuzzle}) where $1$ and $2$ represent two adjacent sites. If we symmetrize over the symmetries of this hexagon, $\sum_{i=1}^2\sum_{j=1}^6 \boldsymbol{\sigma}^i \boldsymbol{C_6}^j W^\dagger \boldsymbol{C_6}^{-j} \boldsymbol{\sigma}^i $ we get zero; the reflection across the bond changes the sign of the wave function.
However, this is not due to a topological property of the state. The argument that the angular momentum must be odd does not apply when there are more than two anyons.  To see this in a general way, let $W^\dagger$ create any group of anyons that does \emph{not} have any symmetry (see Fig.~\ref{honeycombpuzzle}).  Then the different terms of the symmetrized operator $\sum SW^\dagger S^\dagger$ are all distinct so they do not cancel. Thus, the proposed wave function indeed gives rise to a symmetric state where anyon bound states are condensed. The resulting wave function where both $e_A e_B$ and $m_A m_B$ bound states are condensed will be totally symmetric and topologically trivial, thus featureless.

Actually, rather than changing $W^\dagger$ to an asymmetric $W^\dagger$ to ensure that the wave-function does not cancel, we could use the original $W^\dagger$ to construct a state that is symmetric under all transformations up to minus signs.  We will see in the next section that condensing such a state does not break any crystal symmetry either. The argument we have just given is still useful though because it can be used to show that both symmetric and antisymmetric condensates exist, and being able to create symmetric as well as antisymmetric condensates leads to a greater variety in the wave-functions we can construct. 
We will see that there are four types of wave-function (depending on which crystal symmetries reverse the sign), and that the intermediate states (with only a $W_1$ or only a $W_2$ condensate) have different symmetry fractionalization patterns for each case.

\section{Intermediate Phases: Anyon Condensate and Gauge Fixing}
\label{sec:intermediate}

In the previous section, we showed that one can construct a symmetric wave function by condensing bound states of $W_1 = e_A e_B$ or $W_2 = m_A m_B$ (or both). Initially, the state was described by $\mathbb{Z}_2 \times \mathbb{Z}_2$ topological order with the generating set of anyons $\expval{e_A,m_A} \times \expval{e_B,m_B}$. The generating set can be expressed in more relevant form, $\expval{W_1, m_A} \times \expval{e_B, W_2}$. Condensation of bosonic anyons confines all other anyons with non-trivial mutual statistics with the condensed ones. Thus, if $W_1$ is condensed, the system will be in $\mathbb{Z}_2$ topological order generated by $\expval{e_B, W_2}$, and if $W_2$ is condensed, the system will be in $\mathbb{Z}_2$ topological order generated by $\expval{W_1, m_A}$. Now, we can ask what SET phases are realized by these intermediate $\mathbb{Z}_2$ topological orders. To answer this, we should investigate the symmetry fractionalization classes of the anyons after condensation.  \\

\subsection{Condensation and Gauge-Fixing}


First, let's generalize our discussion from Sec. \ref{sec:bipartite} on symmetry fractionalization to the setting where symmetry actions permute anyons. When this happens, we may guess intuitively that there will be fewer classes of symmetry fractionalization types.  The symmetry fractionalization breaks down when anyons are permuted; 
in general, the symmetry cannot be performed on a single anyon in a local way as it involves the annihilation
of a single anyon and creation of another anyon with a different type.  Nevertheless, in place of a local operator, we can take an operator that has two parts: 1) a non-local anyon permutation (for example, an operator that moves the original anyon off to ``infinity'' and brings the appropriate new anyon from ``infinity'') 2) a local unitary transformation that performs the proper symmetry and it seems reasonable to use the same formal approach of factoring a symmetry into symmetries acting on individual anyons.

One intuitively expects that some of the symmetry fractionalization classes will collapse in this case since the symmetries map between two different sets---this is somewhat like the question of classifying states by whether they are even or odd under a certain operator: the operator has to map the state back into the same space for this to be defined.  For example, consider the orbital state of an electron whose spin state is up.  One can define the parity of the orbital state under reflection but not under reflection combined with flipping the spin, since one cannot define the relative phase of spin-up and spin-down states. In the context of symmetry fractionalization, $\mathbb{Z}_2$ symmetries behave in a similar way.  If the symmetry does not permute anyons, then in $\mathbb{Z}_2$ topological order we have four symmetry fractionalization classes (based on how each of $e$ and $m$ transform).   However, when the $\mathbb{Z}_2$ symmetry does permute
$e\leftrightarrow m$, these four distinct symmetry fractionalization classes coallesce into one\cite{Maissam_SymFrac_2014}. Conceptually, we can ascribe such behavior to the fact that the symmetry localization hypothesis, which is the key assumption for symmetry fractionalization, breaks down when anyons are permuted.


In our case, the phase factor associated with $\boldsymbol{C}_6$ rotation symmetry (i.e., the value of $\boldsymbol{C}_6^6=\lambda=\pm 1$) is not gauge-invariant for a single anyon in the original $\mathbb{Z}_2 \times \mathbb{Z}_2$ topological phase. This is because $\boldsymbol{C}_6$ exchanges $e_A \leftrightarrow e_B$. We have a gauge-freedom to multiply $\boldsymbol{C}_6$ by an additional phase factor that \emph{depends on the type of anyon the symmetry is applied to}.  If $\boldsymbol{C}_6\rightarrow \boldsymbol{C}_6 e^{i\phi(a)}$ where $a$ is the anyon type before the symmetry is applied, then $(\boldsymbol{C}_6)^6\rightarrow e^{3i(\phi(e_A)+\phi(e_B))}$ which can cancel $\boldsymbol{C}_6^6$ (by taking $e^{i\phi(e_A)}=1,e^{i\phi(e_B)}=\lambda$).  (This does not contradict our claim above that the bound states $e_Ae_B$ and $m_Am_B$ have well-defined phase factors $\boldsymbol{C}_6^6$ (that are equal to 1), since $\boldsymbol{C}_6$ maps the anyon bound states into themselves.)

Now, we can ask the following question: if the anyon bound state $e_A e_B$ is condensed, what would we get for symmetry fractionalization classes of $e_A$? In the $e_A e_B$ condensed phase, $e_A$ would be identified with $e_B$. Thus, we can just call both $e_A$ and $e_B$ the same name, the $e$-particle, and the action of $\boldsymbol{C}_6$ would map the $e$-particle into itself, and therefore $\boldsymbol{C}_6^6$ is well defined in this phase.  A similar question can be asked for the symmetry $\boldsymbol{R}=\boldsymbol{\sigma} \boldsymbol{C}_6$; is $\boldsymbol{R}^2$
equal to $\pm 1$ for $e_A$ in the intermediate phase? We will focus on $\boldsymbol{C}_6$ for now.

In a more precise statement, when $e_A e_B$ is condensed, the action of $\boldsymbol{C}_6$ rotation symmetry can be realized locally. Consider an original ground state wave function with single $e_A$ excitation. (There must be another $e_A$ excitation as it cannot be created alone) When we apply $\boldsymbol{C}_6$ rotation symmetry, $e_A$ at position $r$ would be mapped into $e_B$ at position $\boldsymbol{C}_6(r)$. This cannot be achieved by a local operator since the local operator cannot destroy a single anyon and create another anyon with a different type. However, in a condensate wave function, the wave function consists of all possible superposition of $e_A e_B$ excitations, including vacuum. Locally, it has the form: $\ket{\Psi^{\text{cond}}_r} \sim \ket{0} + \alpha \ket{e_A e_B}$. When we excite $e_A$, then it would be $\ket{e_A} + \alpha \ket{e_B}$, which is the superposition of $e_A$ and $e_B$ excitations. Similarly, when we excite $e_B$, it will still be the superposition of $e_A$ and $e_B$. Thus, macroscopically $e_A$ and $e_B$ excitations are indistinguishable, although they are located in different sublattices. Then, the action of $\boldsymbol{C}_6$ can be realized locally since we do not need to destroy $e_A$ and create $e_B$; we can realize the action of $\boldsymbol{C}_6$ by just moving $e_A$ and $e_B$ by local operators as in Ref.~\onlinecite{EssinHermele_Z2_2013}. To summarize, the symmetry localization hypothesis is restored for $e$-particles under the condition that $e_A e_B$ is condensed.

A relevant example has been studied in Ref.~\onlinecite{Semion2016}. In this work, a double-semion (or twisted $\mathbb{Z}_2$ topological order) model is considered, which contains four different anyons $\{1, s, s', b=ss'\}$. Here, the symmetry fractionalization of time-reversal symmetry is ill-defined because time-reversal permutes $s$ and $s'$. However, when a bosonic excitation $b = ss'$ is condensed at the boundary, $s$ and $s'$ become indistinguishable, and we can define the action of time-reversal symmetry on the semion and anti-semion properly. Therefore, the symmetry fractionalization of time-reversal symmetry becomes well-defined at the boundary, either ${\cal T}^2 = 1$ or ${\cal T}^2 = -1$ depending on the amplitude of the bosonic condensate.

\subsection{Honeycomb Lattice}

Looking back to our case of the $\mathbb{Z}_2 \times \mathbb{Z}_2$ SL in the honeycomb lattice, the action of the $\pi/3$ rotation $\boldsymbol{C}_6$ is not well-defined on a single $e_A$ or $e_B$ before the condensation of the bosonic particle $W_1 = e_A e_B$. However, once $W_1$ is condensed, we can  define the action of $\boldsymbol{C}_6$ on the $e$-particle properly, as $e_A$ is identified with $e_B$. The subtlety is, while the amplitude of bosonic anyons are spatially uniform in the double-semion model, here the amplitudes of bosonic anyons can have some spatial structure, which is symmetric up to a gauge transformation.

To facilitate the discussion, we define inversion symmetry $I=\boldsymbol{C}_6^3$, which is a $180^\circ$ rotation. Therefore, we want to figure out the action of  $I^2$ on each anyon after the condensation of $W_1$.The condensate wave-function can have either even or odd parity under a $I$. In fact, Eq.~\ref{eq:symmetrize} describes only the first possibility; to  encompass the second possibility, the equation should be modified into
\begin{equation} \label{eq:condensate_phase}
	\psi_{c}^\dagger=\sum_{S \in G_1} SW^\dagger S^\dagger + \epsilon \sum_{S \in G_2} SW^\dagger S^\dagger
\end{equation} where $G_1 \subset G$ is the subset which does not exchange sublattices, and $G_2 \subset G$ is the subset which exchanges sublattices ($G_1 \cup G_2 = G$). This wave function has a parity $\epsilon = \pm 1$ with respect to $I$, because the composition with the inversion symmetry $I$ exchanges $G_1$ and $G_2$.  

Such a state may seem to break rotational symmetry; however, the state is defined as $e^{\kappa \psi_c^\dagger}|G\rangle$, where the exponential is projected onto the even-order terms, which all have an even parity. 

Actually, this wave-function is not well-defined yet, unless we give a precise definition to the operators $S$.  They are actually fractionalized symmetries with a gauge ambiguity, because they act on $W^\dagger$ which creates anyons. In particular, although we know that the square of $I$ is fixed (and equal 1) for bound states, the sign of $I$ itself is ambiguous.  However, there is a convention that allows one to fix the sign: we first consider the actions of 
$I^A$ and $I^B$ on single anyons $e_A,e_B$ respectively. We choose the relative sign between $I_A$ and $I_B$  so that $I^2=I_B I_A=1$; this leaves only an ambiguity where the
sign of both $I_A$ and $I_B$ are changed.  Now when one transforms a bound state, one has to use both $I_A$ and $I_B$, so the transformation is well-defined  (it does not change if one flips the sign of both symmetries).  We will call this standard
convention for $I$, $I_0$, and we will assume that $I_0$ is used in the previous equation.

Although we have a convenient convention for fixing the phase of $I$, this does not mean that $I$ is an actual symmetry of the system, at least after there is a condensate.  We will show that the physical $I$, which is the symmetry of the condensate wave function, would have the action on $e_A$ such that $I^2 \big|_{e_A} = \epsilon$.
To prove this, we have to consider how the $\psi_c$ transforms under $I$ for each choice of $I$'s gauge.  Before condensation, one can choose the relative phase between $e_A$ and $e_B$ freely; we have two different gauge choices: $I^{(\alpha)}_BI^{(\alpha)}_A=\alpha \in \{1,-1\}$. The choice we made to define $\psi_c^\dagger$ is $I_0=I^{+1}$. We have $I^{(\alpha)}=(\alpha)^{N_A} I_0$, where $N_A$ is the number of $e_A$ excitations. (When $\alpha=-1$, this factor switches the sign of $I_A$ without switching the sign of $I_B$;  one could switch the sign of $I_B$ instead since only the relative sign of $I_A$ and $I_B$ matters.)  

As we defined, $\psi_c^\dagger$ has a parity $\epsilon$ under $I_0$:
\begin{eqnarray} \label{eq:condensate_transform}
	\psi_{c}^\dagger &=& \sum_{S \in G_1} I_0^\dagger(I_0S) W^\dagger  (I_0S)^\dagger I_0\nonumber \\
    && + \epsilon \sum_{S \in G_2} I_0^\dagger(I_0S) (W^\dagger ) (I_0S)^\dagger I_0\nonumber \\ 
   &=& \sum_{\tilde{S} \in G_2} I_0^\dagger\tilde{S}  W^\dagger  \tilde{S}^\dagger I_0 + \epsilon \sum_{\tilde{S} \in G_1} I_0^\dagger\tilde{S} W^\dagger \tilde{S}^\dagger I_0 \nonumber \\
   &=& \epsilon I_0^\dagger \psi_c^\dagger I_0.
\end{eqnarray}
The second identity is from the fact that $I_0: G_1 \leftrightarrow G_2$.
What we know is that $I_\alpha I_0^\dagger W^\dagger (I_\alpha I_0^\dagger)^\dagger = \alpha W^\dagger$ (because  $I_\alpha I_0^\dagger=(\alpha)^{N_A}$). Thus,
\begin{equation}
I_\alpha \psi_c^\dagger I^\dagger_\alpha = \epsilon I_\alpha I_0^\dagger \psi_c^\dagger I_0 I_\alpha^\dagger = \epsilon \alpha \cdot \psi_c^\dagger
\end{equation}
Therefore, the $\psi_c^\dagger$ is symmetric under $I_\alpha$ only if $\alpha = \epsilon$. To have a symmetric condensate wave function which does not break the inversion symmetry even locally, we require $\psi^\dagger_c$ to transform trivially under $I$. Since the condensate wave function is symmetric only under $I_\epsilon$ instead of another $I$ with a different gauge choice, $I_\epsilon$ is the symmetry for the system, and the gauge freedom is frozen. Now the action of $I^2$ on $e_A$ is well-defined; it is fixed to be $I_\epsilon^2$, acquiring a phase factor $\epsilon$ under $I^2$. (This argument could be made more precise by constructing a \emph{local operator} that rotates a specific $e_A$ excitation $180^\circ$, by using the condensate to convert the $e_A$ into an $e_B$.  Given a local realization of $I$ for a specific initial state, one can
calculate $I^2$ in general, as in Ref.~\onlinecite{EssinHermele_Z2_2013}.)


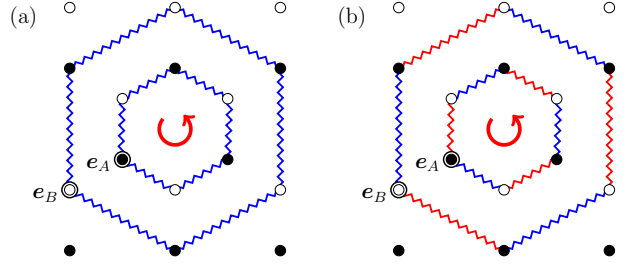
\begin{figure}[t]
    \vspace{0.15in}
  \centering
  \scalebox{0.7}{
  \begin{tikzpicture}
    \coordinate (Origin)   at (0,0);
    \coordinate (XAxisMin) at (-10,0);
    \coordinate (XAxisMax) at (10,0);
    \coordinate (YAxisMin) at (0,-10);
    \coordinate (YAxisMax) at (0,10);
    \node[left, black] at (-1.5,2.7) {\large (a)} ;

    \clip (-1.7,-2.0) rectangle (3.6,3.0); 
    
    \draw[decoration = {zigzag,segment length = 2mm, amplitude = 0.5mm}, decorate, color = blue, line width=1pt] 
    (0,0) -- (1,-0.61735);
    \draw[decoration = {zigzag,segment length = 2mm, amplitude = 0.5mm}, decorate, color = blue, line width=1pt] 
    (1,-0.57735) -- (2,0);
    \draw[decoration = {zigzag,segment length = 2mm, amplitude = 0.5mm}, decorate, color = blue, line width=1pt] 
    (2,0) -- (2.04,1.73205-0.57735);
    \draw[decoration = {zigzag,segment length = 2mm, amplitude = 0.5mm}, decorate, color = blue, line width=1pt] 
    (2,1.73205-0.57735) -- (1,1.73205);
    \draw[decoration = {zigzag,segment length = 2mm, amplitude = 0.5mm}, decorate, color = blue, line width=1pt] 
    (1,1.73205) -- (0,1.73205-0.57735);
    \draw[decoration = {zigzag,segment length = 2mm, amplitude = 0.5mm}, decorate, color = blue, line width=1pt] 
    (-0.03,0) -- (0,1.73205-0.57735);
 
    \draw [->,line width=2.0pt,color=red] (1,0.57735) ++(140:3mm) arc (-220:40:3mm) --++(130:1mm);

    \draw[decoration = {zigzag,segment length = 2mm, amplitude = 0.5mm}, decorate, color = blue, line width=1pt] 
    (-1,-0.57735) -- (1,-1.73205);
    \draw[decoration = {zigzag,segment length = 2mm, amplitude = 0.5mm}, decorate, color = blue, line width=1pt] 
    (1,-1.73205) -- (3,-0.57735);
    \draw[decoration = {zigzag,segment length = 2mm, amplitude = 0.5mm}, decorate, color = blue, line width=1pt] 
    (3,-0.57735) -- (3,1.73205);
    \draw[decoration = {zigzag,segment length = 2mm, amplitude = 0.5mm}, decorate, color = blue, line width=1pt] 
    (3,1.73205) -- (1,1.73205*2-0.57735);
    \draw[decoration = {zigzag,segment length = 2mm, amplitude = 0.5mm}, decorate, color = blue, line width=1pt] 
    (1,1.73205*2-0.57735) -- (-1,1.73205);
    \draw[decoration = {zigzag,segment length = 2mm, amplitude = 0.5mm}, decorate, color = blue, line width=1pt] 
    (-1,1.73205) -- (-1,-0.57735);
    

    \node[draw, thick, circle,inner sep=3pt,fill=white] at (0,0) {}; 
    \node[draw, thick, circle,inner sep=3pt,fill=white] at (-1,-0.57735) {}; 
    \foreach \x in {-7,-6,...,7}{
      \foreach \y in {-7,-6,...,7}{
        \node[draw,circle,inner sep=2pt,fill] at (2*\x+1*\y,1.73205*\y) {};
      }
    }
    \foreach \x in {-7,-6,...,7}{
      \foreach \y in {-7,-6,...,7}{
        \node[draw,circle,inner sep=2pt,fill=white] at (2*\x+1*\y+1,1.73205*\y-0.57735) {};
      }
    }

    \node[left, black] at (-0.1,-0.1) {\large $\boldsymbol{e}_A$} ;
    \node[left, black] at (-0.1-1,-0.1-0.57735) {\large $\boldsymbol{e}_B$} ;

  \end{tikzpicture}
  }
    \scalebox{0.7}{
  \begin{tikzpicture}
    \coordinate (Origin)   at (0,0);
    \coordinate (XAxisMin) at (-10,0);
    \coordinate (XAxisMax) at (10,0);
    \coordinate (YAxisMin) at (0,-10);
    \coordinate (YAxisMax) at (0,10);
    \node[left, black] at (-1.5,2.7) {\large (b)} ;

	\clip (-1.7,-2.0) rectangle (3.6,3.0); 
        \draw[decoration = {zigzag,segment length = 2mm, amplitude = 0.5mm}, decorate, color = blue, line width=1pt] 
    (0,0) -- (1,-0.61735);
    \draw[decoration = {zigzag,segment length = 2mm, amplitude = 0.5mm}, decorate, color = red, line width=1pt] 
    (1,-0.57735) -- (2,0);
    \draw[decoration = {zigzag,segment length = 2mm, amplitude = 0.5mm}, decorate, color = blue, line width=1pt] 
    (2,0) -- (2.04,1.73205-0.57735);
    \draw[decoration = {zigzag,segment length = 2mm, amplitude = 0.5mm}, decorate, color = red, line width=1pt] 
    (2,1.73205-0.57735) -- (1,1.73205);
    \draw[decoration = {zigzag,segment length = 2mm, amplitude = 0.5mm}, decorate, color = blue, line width=1pt] 
    (1,1.73205) -- (0,1.73205-0.57735);
    \draw[decoration = {zigzag,segment length = 2mm, amplitude = 0.5mm}, decorate, color = red, line width=1pt] 
    (-0.03,0) -- (0,1.73205-0.57735);
 
    \draw [->,line width=2.0pt,color=red] (1,0.57735) ++(140:3mm) arc (-220:40:3mm) --++(130:1mm);

    \draw[decoration = {zigzag,segment length = 2mm, amplitude = 0.5mm}, decorate, color = red, line width=1pt] 
    (-1,-0.57735) -- (1,-1.73205);
    \draw[decoration = {zigzag,segment length = 2mm, amplitude = 0.5mm}, decorate, color = blue, line width=1pt] 
    (1,-1.73205) -- (3,-0.57735);
    \draw[decoration = {zigzag,segment length = 2mm, amplitude = 0.5mm}, decorate, color = red, line width=1pt] 
    (3,-0.57735) -- (3,1.73205);
    \draw[decoration = {zigzag,segment length = 2mm, amplitude = 0.5mm}, decorate, color = blue, line width=1pt] 
    (3,1.73205) -- (1,1.73205*2-0.57735);
    \draw[decoration = {zigzag,segment length = 2mm, amplitude = 0.5mm}, decorate, color = red, line width=1pt] 
    (1,1.73205*2-0.57735) -- (-1,1.73205);
    \draw[decoration = {zigzag,segment length = 2mm, amplitude = 0.5mm}, decorate, color = blue, line width=1pt] 
    (-1,1.73205) -- (-1,-0.57735);
    

    \node[draw, thick, circle,inner sep=3pt,fill=white] at (0,0) {}; 
    \node[draw, thick, circle,inner sep=3pt,fill=white] at (-1,-0.57735) {}; 
    \foreach \x in {-7,-6,...,7}{
      \foreach \y in {-7,-6,...,7}{
        \node[draw,circle,inner sep=2pt,fill] at (2*\x+1*\y,1.73205*\y) {};
      }
    }
    \foreach \x in {-7,-6,...,7}{
      \foreach \y in {-7,-6,...,7}{
        \node[draw,circle,inner sep=2pt,fill=white] at (2*\x+1*\y+1,1.73205*\y-0.57735) {};
      }
    }

    \node[left, black] at (-0.1,-0.1) {\large $\boldsymbol{e}_A$} ;
    \node[left, black] at (-0.1-1,-0.1-0.57735) {\large $\boldsymbol{e}_B$} ;

  \end{tikzpicture}
  }

  \caption{    \label{Condensate} \small The figure represents the mean-field configurations $Q_{ij}$ that have the two different symmetries: (a) $\boldsymbol{C}_6^6 \big|_e = 1$ (b) $\boldsymbol{C}_6^6 \big|_{e}= -1$ on $e$-particles. The six-fold rotation exchanges two anyons $e_A$ and $e_B$, thus its action is related to the mean-field amplitudes for anyon bound states $e_A e_B$. Blue lines represent positive $Q_{ij}$, while red lines represent negative $Q_{ij}$. The right figure is symmetric under $\boldsymbol{C}_6$ only after associated gauge transformation. }
\end{figure}

In fact, we have been suppressing an additional parity in the definition of the condensate $\psi_c^\dagger$.  There are two independent symmetry operators which exchange two sublattices, the inversion symmetry $I$ and the reflection symmetry ${R} = \boldsymbol{\sigma} \boldsymbol{C}_6$.  The condensate wave-function can
have either parity under each of $R_0$ and $I_0$ (the standardized forms of $R,I$ that satisfy $R_0^2=1,I_0^2=1$).  To display these parities explicitly, note that while $I$ is orientation preserving, $R$ is not. Therefore, we can subdivide the group $G$ into orientation preserving symmetries $G_o$ and orientation reversing $G_r$, and divide each of them further depending on whether a symmetry exchanges sublattices or not (index 1,2). Thus, at full generality, Eq.~\ref{eq:condensate_phase} becomes
\begin{eqnarray}
	&\psi_{condensate}^\dagger=\sum_{S \in G_{o,1}} SWS^\dagger + \epsilon_1 \sum_{S \in G_{o,2}} SWS^\dagger \nonumber \\
&\,\, + \epsilon_2 \qty[\epsilon_1 \sum_{S \in G_{r,1}} SWS^\dagger +  \sum_{S \in G_{r,2}} SWS^\dagger ]
\end{eqnarray}
characterized by $\epsilon_1, \epsilon_2 \in \{1,-1\}$. Following the same reasoning, we can conclude that the condensate wave function is symmetric under $I_{\epsilon_1}$ and $R_{\epsilon_2}$, meaning that $e_A \equiv e_B$ acquires phase factors $\epsilon_1$ and $\epsilon_2$ respectively under $I^2$ and $R^2$ after the condensation.

This result can also be understood more explicitly by using a PSG representation of the wave function, as illustrated in Fig. \ref{Condensate}.  The original 
decoupled states on the two triangular sublattices is described by a quadratic Hamiltonian with only next-nearest neighbor interactions.
A condensate of bound states can be induced by adding nearest neighbor couplings as well, leading to the full Hamiltonian:
\begin{equation}
    H_{MF} = - J \sum_{ij} Q_{ij} \epsilon_{\alpha \beta} b_{i \alpha} b_{j \beta} + \text{H.c.} + \dots
\end{equation}
There are two possibilities, depending on whether the $Q$'s associated with the sides of a hexagon all have the same signs or alternating signs.   $\boldsymbol{C}_6$ symmetry allows only these possibilities.  In the latter case, the Hamiltonian is not immediately symmetric
under $\boldsymbol{C}_6$ symmetry, but one can define
$\boldsymbol{C}_6b_{i\alpha}\boldsymbol{C}_6^{-1}=\pm b_{\boldsymbol{C}_6(i)\alpha}$ where the sign is $(+)$ if $i$ is in the $A$-sublattice and $(-)$ if it is in the $B$-sublattice, and this is a symmetry.  The $b$ operators create the emergent spinons;  applying $\boldsymbol{C}_6$ six times gives a minus sign because
the spinon passes through the $B$-sublattice an odd number of times. The idea also works for visons in the exact same manner except they reside on the dual lattice.

\subsection{Intermediate Phases}
In the following, we describe the two different intermediate $\mathbb{Z}_2$ topological phases accessed by the condensation of either $W_1$ or $W_2$ from the $\mathbb{Z}_2 \times \mathbb{Z}_2$ topological order.

\vspace{0.05in}
\noindent {\bf Scenario 1} Consider a phase with condensation of the bound state $W_1$. In this phase, the system has $\mathbb{Z}_2$ topological order with anyonic excitations $e_A$ and $W_2$ which have a mutual phase of $-1$.
There is only one type of $e$ excitation because we can identify $e_A \equiv e_B$ since they can be transformed into one another using local operators in a phase where $e_A e_B$ is condensed.  On the other hand,
unpaired $m_A$ and $m_B$ are confined, and only the bound state $W_2$ has a finite energy. We will call $e_A \equiv e_B$ a spinon as it carries spin-1/2 and $m_A m_B$ a vison as it has mutual $\pi$ statistics with spinon. The topological properties of the system are fully characterized by symmetry fractionalization classes of spinons and visons. 

Using the flux-anomaly argument\cite{YangQi2015, HermeleChen2016}, we can prove that vison excitations must have a totally trivial fractionalization class if spinons carry spin-1/2. This agrees with what we have found about the vison, $m_A m_B$. What about the symmetry fractionalization class of spinons? They are listed in TABLE~\ref{Honeycomb}, except that the phases that are marked ``gauge-dependent'' now become gauge-invariant and can be made to have either sign by choosing
a condensate wave-function with the appropriate symmetry, as described in the previous section.

Although we have found four possibilities for $(\bold{C}_6)^6$ and $R^2$, some of these seem likely to be harder to achieve using local interactions.  Condensation of $W_1$ can be achieved in a parton approach by having some non-zero symmetric mean-fields between sublattice $A$ and $B$, and the fields can presumably be adjusted to give all four possibilities.  However, assuming non-zero mean-fields for \emph{nearest-neighbors} constrains the symmetric mean-field patterns and thus the fractionalization of $(\bold{C}_6)^6$, as we can see from the results of Ref.~\onlinecite{FW2010}.  

In the triangular lattice with non-zero nearest neighbor amplitudes, there are two $\mathbb{Z}_2$ SL solutions $\ket{\Psi^T_{p}}$ characterized by $p = p_1 = p_2 = 1- p_3$ (mod $2$) in TABLE~\ref{Triangle}. Starting from $\ket{\Psi^T_{p}} \otimes \ket{\Psi^T_{p}}$ living in honeycomb lattice, we give non-zero mean-fields between sublattices to obtain $\mathbb{Z}_2$ spin liquids in the honeycomb lattice, while the mean-fields within sublattices are kept the same. Our discussion above gives these possibilities for the symmetry fractionalization class of the spinon (See TABLE~\ref{Honeycomb})
\begin{equation} \label{symFrac_H}
    (\bar{\eta}_1, \bar{\eta}_2, \bar{\eta}_3, \bar{\eta}_4) = \big( (-1)^p, -1, \epsilon_1, \epsilon_2 \big)
\end{equation}
However, the more conventional mean-field parton approach in Ref. \cite{FW2010} found that there are only two symmetric mean-field configurations $\ket{\Psi^H_{q}}$ ($q=1,2$) with
 \emph{non-zero nearest neighbor} and \emph{next nearest neighbor} mean-fields. These two solutions are characterized by 
\begin{equation}
    (\bar{\eta}_1, \bar{\eta}_2, \bar{\eta}_3, \bar{\eta}_4) = \big( (-1)^q, -1, (-1)^{q+1}, -1 \big).
\end{equation}
These are the special cases of Eq. (\ref{symFrac_H}) found by letting $p=q$ and fixing $\epsilon_1=(-1)^{p+1}$, $\epsilon_2=-1$.

\vspace{0.1in}
\noindent {\bf Scenario 2} Consider a phase where the bound state $W_2$ is condensed. Although a spinon condensation has been discussed extensively in mean-field parton approaches, a vison condensation in a $\mathbb{Z}_2$ topological order has only been discussed in an effective theory of Ising gauge theory.\cite{Z2_IGT_1,Z2_IGT_2,Z2_IGT_3,Z2_IGT_4,Z2_IGT_5} Similarly, we map our problem into two copies of Ising gauge theory where spinons are gapped, and then assume the existence of an interaction which prefers condensation of $W_2$, implying that the expectation value of the $W_2$ field is non-zero. In this phase, $e_A$ and $e_B$ are confined, and the system has a $\mathbb{Z}_2$ topological order with anyonic excitations generated by $W_1 = e_A e_B$ and $m_A \equiv m_B$. $m_A$ and $m_B$ are identified because $W_2$ is condensed. Here, we call $W_1$ the spinon and $m_A \equiv m_B$ the vison.

A $\mathbb{Z}_2$ SL in this scenario is exotic. In this phase, the `spinon' excitation, which is $e_A e_B$, carries an integer spin instead of a half-integer one because the $e_A e_B$ bound state has a totally trivial fractionalization class. In the study of $\mathbb{Z}_2$ SLs, the most extensively used approach is the mean-field parton approach, and in the case of spin-1/2 per site that has been studied via the parton approach, the spinon  always carries a half-odd integer spin. Although it is possible to modify the parton-approach to one based on spinons with integer spin for a system with spin-1 per site, for the system with spin-1/2 per site, such cases has been thought to be difficult\cite{Bieri2012}. Since the `vison' excitation also carries an integer (zero) spin, this new phase of $\mathbb{Z}_2$ SL does not have topological excitations with a non-trivial projective representation under $SO(3)$ rotation. Moreover, its vison excitations can have a non-trivial fractionalization class inherited from the symmetry of the $W_2$ condensate wave function, as can be shown following the same arguments we applied to spinons, which is not allowed if the spinon excitations carry spin-1/2.\cite{YangQi2015}. (The procedure for forming a vison condensate is exactly the same as the one for forming a spinon condensate--there is no difference between excitations that are centered on sites or plaquettes since we can symmetrize both kinds of states under the whole symmetry group.)

\section{The HOLSM-type constraints and minimal topological orders}\label{sec:application}

Here, we demonstrate applications of our bipartite lattice construction to other systems where featureless states are guaranteed not to exist by HOLSM-type theorems. There are extensions of the HOLSM theorem which provide no-go theorems on featureless states for systems with ``nonsymmorphic" symmetries\cite{Sid2013} (such as glide-reflections) or even modified symmetries (magnetic translations\cite{YML2017}). As discussed earlier in the introduction, no-go theorems for featureless states imply gapped symmetric states must be topological and their symmetry fractionalization classes are constrained in a certain way. 

In Sec.~\ref{sec:nonsymmorphic}, we discussed a system with the nonsymmorphic lattice group $pg$. The system had two sites with spin-1/2 per unit cell, a net spin-1 per unit cell. However, the topological order we constructed could not be further reduced unless it breaks spin-rotation or glide-reflection symmetry. In fact, no other construction can succeed on account of an extension of the HOLSM theorem\cite{Sid2013}, which states that if the $U(1)$ charge density per unit cell $\nu \not\equiv 0 \mod{{\cal S}}$ where $\cal S$ is a certain number measuring how non-symmorphic the group is, there does not exist a featureless state. One can deduce a no-go theorem for a spin-system from this by regarding $S_z$ as the charge density. We then find that the sum of the \emph{magnitudes} of the spins in a unit cell must satisfy $\sum{S_i}\equiv 0 \mod{{\cal S}}$ in order for a featureless state to exist.
We observed that this is indeed the case; the $\mathbb{Z}_2$ topological order we constructed has both anyons with non-trivial fractionalization of spin-rotation symmetry and anyons with non-trivial fractionalization of glide-refection symmetry.


We expect topological orders are constrained in a similar way for any system that has a no-go theorem ruling out featureless states.
We will provide another example of this, a bosonic system with magnetic translation symmetries, using HOLSM-type arguments to constrain the possible fractionalization patterns, and then constructing states that have these fractionalization patterns using the method of sublattices.

\subsection{HOLSM-type Constraints for Lattices in Magnetic Fields} \label{sec:magnetic}
We will now consider bosons in a lattice with a \emph{magnetic field}.  We will focus on the case of a system with half-integer filling and half-a-unit of flux, since these favor
the type of $\mathbb{Z}_2$ order we have been considering as opposed to chiral states like the Laughlin state.  

When a magnetic field is present, new types of fractionalization are possible.  When anyons go around unit cells they can acquire phases in two ways, either from their fractionalized symmetries or from their fractional charge encircling the magnetic flux, and these two effects can metamorphose into one another in an interesting way. Let us first describe the system using spin variables. Consider a two-dimensional lattice with spin-$1/2$ per unit cell, together with a magnetic translation symmetry characterized by
\begin{equation}
   \boldsymbol{T}_1 \boldsymbol{T}_2 \boldsymbol{T}_1^{-1} \boldsymbol{T}_2^{-1} = \prod_i e^{i \pi S^z_i  }
\end{equation}
where the bold symbol implies the operators are magnetic translations with $\pi$-flux. 
We will impose spin-rotation symmetry $U(1)_z$ about the z-axis, spin-flip symmetry about the $y$-axis, and time-reversal
symmetry.
If we use the Holstein-Primakoff transformation to map this to a model of itinerant bosons, this is equivalent to a (hardcore) bosonic system at half-filling under magnetic $\pi$-flux. Under this mapping, the symmetries are
charge conservation symmetry, time reversal symmetry, and particle-hole symmetry, in addition to the magnetic translations.  For example, the Hofstadter model applied to hard-core bosons (at half-filling and with $\pi$-flux) has all these symmetries.
The boson density $n_b$ is related to the spin-$z$ components by the following relation:  $S_z = n_b - \frac12$ showing that
the spin-flip symmetry implies $\expval{S_z} = 0$ or equivalently $n_b = 1/2$ (See Appendix\,\ref{Magnetic}).

A spin-system has a physical time-reversal symmetry given by the product ${\cal T} = e^{i \pi S^y} {\cal K}$ where ${\cal K}$ is an antiunitary operator. Since a flux of $\pi$ stays the same under ${\cal T} \in Z_2^{\cal T}$ and maps into $-\pi \equiv \pi$ under spin-flip symmetry $e^{i \pi S^y} \in \spflip$, the Hamiltonian has both symmetries $Z_2^{\cal T}$ and $\spflip$. 
A simple example of a Hamiltonian that realizes these symmetry properties (although more frustration may be necessary to produce a gapped spin liquid) is an anisotropic
Heisenberg model where the couplings $K_{ij}$ depend on the links:
\begin{equation}
    H = \sum J S^z_i S^z_j + K_{ij} \qty( S^x_i S^x_j + S^y_i S^y_j );
\end{equation}
here $K_{ij} = -K$ for vertical links on the even $x$-coordinates while $K_{ij} = K$ for the rest. We have magnetic translation symmetry corresponding to the $\pi$-flux since $\prod_{\text{unit cell}} K_{ij} = - \abs{K}^4$ for any plaquette. 
The $SO(3)$ spin-rotation symmetry is broken down to $U(1) \rtimes \spflip $ symmetry.

First, let us consider just the $U(1)$ symmetry and translational symmetry, ignoring the others. For a bosonic model at half-filling under $\pi$-flux, a featureless state is prohibited by theorems proven recently\cite{YML2017}. 
Consider $U(1)$ on-site symmetry along with magnetic translation symmetry. According to Ref.~\onlinecite{YML2017}, if there exists an SPT phase, the system has a Hall conductance that satisfies the relation
\begin{equation} \label{eq:constraint_YML_IQH}
    \rho \equiv \sigma_{xy} \cdot \phi  \quad \mod 1
\end{equation}
where $\sigma_{xy}$ is the Hall conductivity divided by $e^2/h$, $\phi$ is the number of flux quanta per unit-cell (which is $1/2$ for $\pi$-flux), and $\rho$ is the particle number per unit-cell. Regarded as a bosonic system, our system has $\rho_b=1/2$. Eq.~\ref{eq:constraint_YML_IQH} implies that the Hall conductivity must be odd-integer multiples of $e^2/h$. However, a bosonic quantum Hall system can have only a Hall conductivity that is an even-integer multiple of $e^2/h$ (See Ref.~\onlinecite{YML2012}), implying that a featureless state cannot exist. Eq. ~\ref{eq:constraint_YML_IQH} can be applied more generally as well, for example, for half-filling, what values of the magnetic field can lead to a featureless state?  Eq.~\ref{eq:constraint_YML_IQH}
shows that the flux must be $\phi=1/(4n)$ per unit cell, and in this case the Hall conductivity would be $\sigma_{xy} \equiv 2n \mod 4n$. These are like integer quantum Hall states of fermions, because the number of particles per unit flux is an integer, $2n$, but they can be stabilized only with the help of interactions.  (Without interactions, a bosonic system cannot have a gap!) A bosonic integer quantum Hall (BIQH) phase with $\sigma_{xy}=2$ at $n=1$ has been reported in the numerical study of lattice models\cite{BIQH_1,BIQH_2}.

Thus, for a gapped and $U(1)$ symmetric ground state under $\pi$-flux, an intrinsic topological order is inevitable. What kinds of SET orders would be allowed in this situation? There are many well-known chiral topological orders with non-zero Hall conductivity for various filling and flux density (similar to states in the fractional quantum Hall effect), which we will list briefly at the end of Sec.~\ref{sec:generic_case}. However, we would like to focus on cases where there is an additional time-reversal symmetry besides the $U(1)$ symmetry. For either $\rho=1/2$ or $\phi=1/2$, we would have a time-reversal symmetry, which suppress a finite Hall conductivity. Thus, we will consider non-chiral topological orders, i.e. ones with zero Hall conductivity. For just the $U(1)$ symmetry group, but at half-filling and $\pi$-flux, we may expect toric-code type $\mathbb{Z}_2$ topological order. This is indeed possible, as we will show later on by the anyon condensation from a $\mathbb{Z}_2 \times \mathbb{Z}_2$ SL in bipartite lattice construction.



\subsection{Topological orders consistent with Spin-Flip and Time-Reversal Symmetry}

We have seen that the existence of $U(1)$ symmetry together with $\pi$-flux and half-filling prohibits a featureless state, so a gapped and symmetric ground state should have a topological order, which can be a toric-code type. We did not yet consider additional symmetries such as time-reversal ${\cal T}$ or spin-flip $e^{i\pi S^y}$. In fact, these additional symmetries impose further constraints on the topological orders that can be realized (See TABLE~\ref{table:summaryTO}).

We will consider four different cases: $G_1 = U(1) \rtimes \spflip$, $G_2 = U(1) \rtimes Z_2^{{\cal T}_1}$, $G_3 = U(1) \times Z_2^{{\cal T}_2}$, and $G_4 = (U(1) \rtimes \spflip) \times Z_2^{{\cal T}_2}$. We will represent the generator of $\spflip$, the spin-flip transformation, by $\flip$ from now on. $Z_2^{{\cal T}_1}$ and $Z_2^{{\cal T}_2}$ are two different realizations of time-reversal symmetry. There are two possibilities because time-reversal symmetry can be implemented differently depending on the microscopic data. The first case is ${\cal T}_1 = {\cal K}$, an anti-unitary operator which just involves complex conjugation. Here, the total symmetry group $G = U(1) \rtimes Z^T_2$; while $\cal T$ commutes with the charge operator ${\cal T} Q {\cal T}^{-1} = Q$, it does not commute with an element $e^{i Q \theta} \in U(1)$ since ${\cal T} e^{i Q \theta} {\cal T}^{-1} = e^{-i Q \theta}$. The second case is  ${\cal T}_2 = {e^{i\pi S^y}} {\cal K}$ where ${e^{i\pi S^y}}$ flips the sign of $U(1)$ charge.  In this case, the total symmetry group $G = U(1) \times {Z}^{{\cal T}_2}_2$ since ${\cal T}_2$ commutes with symmetry elements of $U(1)$.

To proceed, we need to illustrate the most important constraint: for a system with $\pi$-flux at half filling, $U(1)$ symmetry enforces the following: $(i)$ there must be an anyon carrying a half-odd integer charge (or spin) and $(ii)$ this anyon should be a background anyon. The idea of a background anyon is originated from the fact that the fractionalized translation symmetry can be understood as if a background anyon is placed on each unit cell. These constraints can be seen using the second theorem  in Ref.~\onlinecite{YML2017}. 
The filling-enforced constraint for a generic phase has the following form:
\begin{equation} \label{eq:constraint_YML_FQHE}
	\rho \equiv \sigma_{xy} \phi +  \frac{\theta_{F,a}}{2\pi} \quad \mod 1
\end{equation}
where $\theta_{F,a}$ is the mutual statistics between a fluxon $F$ and the background anyon $a$. Also, the Hall conductivity satisfies the condition, $\sigma_{xy} = \theta_{F,F}/2\pi \mod 1$. For a time-reversal symmetric system, $\sigma_{xy} = 0$. Then, since the fluxon is equivalent to the $U(1)$ $2\pi$-symmetry defect, Eq.~\ref{eq:constraint_YML_FQHE} implies the existence of a background anyon carrying a half-odd integer $U(1)$ charge. Now, we are ready to analyze constraints on non-chiral topological orders. We will assume that symmetries do not permute anyons, except time-reversal. One can also understand the following examples in a slightly different perspective (See Appendix.~\ref{appendix:Fluxon}).

\vspace{0.15in}
\noindent {\bf Case 1: $G = U(1) \rtimes \spflip$}. Assume there exists a toric-code type $\mathbb{Z}_2$ topological order, which is consistent with all symmetries. Since spinons carry $S_z = \pm 1/2$, they would see a magnetic flux of $\pm \pi/2$ due to the magnetic translation symmetry. Due to the $\mathbb{Z}_2$ fractionalization, there may or may not exist an overall additional factor $\eta = \pm 1$. 
Consider the following symmetry fractionalization for $e$-particle with $S_z$:
\begin{equation}
\boldsymbol{T}_1 \boldsymbol{T}_2 \boldsymbol{T}_1^{-1} \boldsymbol{T}_2^{-1} \big|_{e,S_z} = \eta  e^{i \pi S_z}
\end{equation}
Under the conjugation of $\flip \in \spflip$, the LHS is transformed to act on the $e$-particle with an opposite spin $-S_z$\footnote{To be more precise, we define $e$-particle to be an unique excitation with a $S_z = 1/2$, whose creation operator denoted by $b^\dagger_e$. Then, $e$-particle with a $S_z = -1/2$ would be expressed as $b^\dagger_e b_{c}$ where $b_c$ is the annihilation operator for a local physical excitation carrying $S_z = 1$. Thus, $\spflip: b^\dagger_e \mapsto b^\dagger_e b_c$}, while the RHS does not change since it is just a number ($S_z$ should be evaluated). This implies that it is impossible to write down an (mean-field) effective spinon Hamiltonian symmetric under $\spflip$. Therefore, $\spflip$ symmetry is inconsistent with magnetic translation symmetry and $\mathbb{Z}_2$ topological order cannot be realized. \footnote{Y. C. He, \emph{Private Communication}}. Similarly, it can be shown that $\mathbb{Z}_2$ double-semion order does not have a consistent fractionalization pattern.

\vspace{0.15in}
\noindent {\bf Case 2: $G = U(1) \rtimes Z_2^{ {\cal T}_1 }$}. We can follow the reasoning similar to the above one. Then, we can show that the $\mathbb{Z}_2$ toric-code  order is inconsistent. However, $\mathbb{Z} _2$ double-semion order is not prohibited. Double-semion order has anyons $\{1, s, \bar{s}, b\}$, where time-reversal symmetry exchanges ${\cal T}: s \leftrightarrow \bar{s}$. (Because their topological spins are exchanged by complex conjugation.) Here, $s$ and $\bar{s}$ carry $S_z = \pm 1/2$. Let's assume the following fractionalization pattern for the magnetic translation symmetry:
\begin{eqnarray} \label{eq:T1_DS}
     \boldsymbol{T}_1 \boldsymbol{T}_2 \boldsymbol{T}_1^{-1} \boldsymbol{T}_2^{-1} \big|_{s, S_z}&=&  \eta e^{i\pi S^z} \nonumber \\
     \boldsymbol{T}_1 \boldsymbol{T}_2 \boldsymbol{T}_1^{-1} \boldsymbol{T}_2^{-1} \big|_{\bar{s}, S_z} &=& - \eta e^{i\pi S^z},
\end{eqnarray}
where $\eta = \pm 1$ is a phase factor from $\mathbb{Z}_2$ fractionalization.
This fractionalization pattern is consistent under the time-reversal ${\cal T}_1$ since ${\cal T}_1:\, i \mapsto -i$ as well as $s \leftrightarrow \bar{s}$. Furthermore, it is consistent with the Eq.~\ref{eq:constraint_YML_FQHE} since this fractionalization pattern corresponds to the background anyon being either an $s$ or $\bar{s}$, which carry a half-odd integer $U(1)$ charge. (A trivial anyon or a bosonic anyon $b$ would not carry a half-odd integer $U(1)$ charge.)

This is interesting, because it was shown that double-semion topological order is not allowed with $U(1)$ and time-reversal symmetry in the absence of magnetic flux in the unit cell\cite{MZAV2015}. Let us interpret Ref.~\onlinecite{MZAV2015} in our framework; when there is no magnetic flux, if $s$ and $\bar{s}$ have different phase factors under translational symmetry like Eq.~\ref{eq:T1_DS}, it is inconsistent under time-reversal; thus, they should have the same phase factor under $T_1 T_2 T_1^{-1} T_2^{-1}$. Intuitively speaking, this implies that either the trivial anyon or the bosonic anyon $b= s\bar{s}$ is sitting on each unit cell, neither of which can screen the spin-1/2 projective representation per unit cell. Therefore, it is impossible to host double-semion order without symmetry breaking here.  However, magnetic translation symmetry invalidates such a picture, opening up a possibility for the double-semion model to be realized.

\vspace{0.15in}
\noindent {\bf Case 3: $G = U(1) \times Z_2^{ {\cal T}_2 }$}.  
Unlike the previous cases, the toric code order can have a consistent fractionalization pattern. Since $Z_2^{ {\cal T}_2 }$ symmetry flips spins as well as conjugate complex numbers, inconsistency does not arise. For the double-semion order, the both semions must have the fractionalization pattern
\begin{eqnarray}
     \boldsymbol{T}_1 \boldsymbol{T}_2 \boldsymbol{T}_1^{-1} \boldsymbol{T}_2^{-1} \big|_{s (\bar{s}), S_z}&=& \eta e^{i\pi S^z},
\end{eqnarray}
because ${\cal T}_2 : s \leftrightarrow \bar{s}, S_z \mapsto - S_z$. However, this implies that the background anyon is either a trivial anyon or a bosonic anyon $b=s \bar{s}$. Since $s$ and $\bar{s}$ carry $U(1)$ charge opposite to each other, $b=s \bar{s}$ must carry an integer charge. This is not allowed due to the Eq.~\ref{eq:constraint_YML_FQHE}, which tells that the background anyon should carry a spin-1/2 (a half $U(1)$ charge). Therefore, this further consideration shows that the double-semion $\mathbb{Z}_2$ order is inconsistent.

\vspace{0.15in}
\noindent {\bf Case 4: $G = \qty(U(1) \rtimes \spflip) \times Z_2^{ {\cal T}_2 }$}  
In this case, both the toric code and double-semion orders are ruled out. While the $\spflip$ symmetry is inconsistent with the toric code order, the $Z_2^{ {\cal T}_2 }$ symmetry is inconsistent with the double-semion order unless there is a spontaneous symmetry breaking.
Consequently, we can ask what is the minimal topological order with $(U(1) \rtimes Z_2) \times Z_2^{\mathcal T}$ on a lattice with $\pi$-flux. The answer lies in our bipartite lattice construction!

\subsection{Bipartite Lattice Construction of a $\mathbb{Z}_2\times\mathbb{Z}_2$ Spin Liquid under $\pi$-flux}
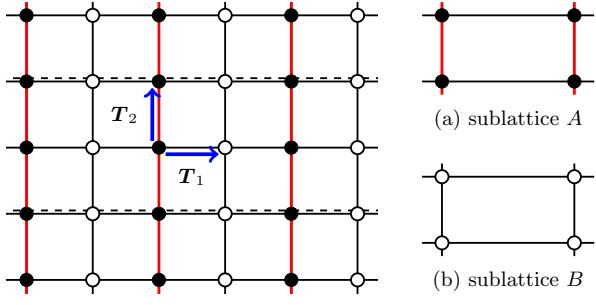
\begin{figure}[t]
  \centering
  \scalebox{0.88}{
  \begin{tikzpicture}
    \clip (-2.3,-2.2) rectangle (3.3,2.2); 
    \draw[style=help lines, thick, black] (-6,-6) grid[step=1] (6,6);

   \foreach \i in {-4,...,4}{
        \draw[red,very thick] (2*\i,-6) -- (2*\i,6);}

   \foreach \i in {-4,...,4}{
        \draw[black,dashed,thick] (-6,2*\i-0.95) -- (6,2*\i-0.95);}

    \foreach \x in {-7,-6,...,7}{
      \foreach \y in {-7,-6,...,7}{
        \node[draw,circle,inner sep=2pt,fill] at (2*\x,\y) {};
      }
    }
   
    \foreach \x in {-7,-6,...,7}{
      \foreach \y in {-7,-6,...,7}{
        \node[draw, thick, circle,inner sep=2pt,fill=white] at (2*\x+1,\y+1) {};
      }
    }

    \draw [->,blue,ultra thick](0.1,-0.1) -- (0.9,-0.1);
    \node [below] at (0.5,-0.2) {$\boldsymbol{T}_1$};    
    
    \draw [->,blue,ultra thick](-0.1,0.1) -- (-0.1,0.9);
    \node [left] at (-0.2,0.5) {$\boldsymbol{T}_2$};
    
  \end{tikzpicture}

    \hspace{0.18in}
  
  \begin{tikzpicture}
    \clip (-0.3,-2.2) rectangle (2.3,2.2); 
       
    \draw[black,thick] (-0.3, 1) -- (2.3, 1);
    \draw[black,thick] (-0.3, 2) -- (2.3, 2);
    \draw[red,very thick] (0, 0.8) -- (0, 2.2);    
    \draw[red,very thick] (2, 0.8) -- (2, 2.2);    
        
    \node[draw,circle,inner sep=2pt,fill] at (0,1) {};
    \node[draw,circle,inner sep=2pt,fill] at (2,1) {};
    \node[draw,circle,inner sep=2pt,fill] at (0,2) {};
    \node[draw,circle,inner sep=2pt,fill] at (2,2) {};
        
    \node [below] at (1, 0.7) {(a) sublattice $A$};

    \draw[black,thick] (-0.3, -1-0.44) -- (2.3, -1-0.44);
    \draw[black,thick] (-0.3, 0-0.44) -- (2.3, 0-0.44);
    \draw[black,thick] (0, -1.2-0.44) -- (0, 0.2-0.44);    
    \draw[black,thick] (2, -1.2-0.44) -- (2, 0.2-0.44);    
        
    \node[draw, thick, circle, inner sep=2pt, fill=white] at (0,-1-0.44) {};
    \node[draw, thick, circle, inner sep=2pt, fill=white] at (2,-1-0.44) {};
    \node[draw, thick, circle, inner sep=2pt, fill=white] at (0,0-0.44) {};
    \node[draw, thick, circle, inner sep=2pt, fill=white] at (2,0-0.44) {};
        
    \node [below] at (1, -1.3-0.44) {(b) sublattice $B$};

  \end{tikzpicture}
  }
  
  \vspace{0.1in}
  
  \caption{ \label{magnetic}Black dots represent sublattice $A$. White dots represent sublattice $B$. In this figure, black lines represent hopping terms with the positive sign, and red lines represent hopping terms with the negative sign. Due to the alternating signs, if we translate the system in $x$-direction, we should perform gauge transformation at sites on dashed lines (even rows) to go back to the original Hamiltonian. Figure (a) and (b) show unit cells for sublattice $A$ and $B$, which are different by gauge transformation. }
\end{figure}

Here, we show that there exists a $\mathbb{Z}_2 \times \mathbb{Z}_2$ topological order consistent with the symmetry group $\qty(U(1) \rtimes \spflip) \times Z_2^{ {\cal T}_2 }$ under $\pi$-flux. Combine two rectangular sublattices to form a lattice with a smaller unit cell. (Fig.~\ref{magnetic}) Although we consider rectangular lattices for graphical convenience, we do not consider crystal symmetries other than translations. As we can see in Fig.~\ref{magnetic}, there is one essential subtlety: Hamiltonians for the sublattices are equivalent only up to the following unitary transformation 
\begin{equation}
    {\cal G} = \prod_{\text{all sites}} e^{i \pi y_i S^z_i }.
\end{equation} 
This choice is required to construct a system with $\pi$-flux per unit cell. If we use this symmetry combined with pure translation in the $x$-direction as
an ``identifying map'' between the two sublattices, we find that $U(1) \rtimes \spflip$ symmetry acts differently on two sublattices: for $R \in U(1) \rtimes \spflip$, if $U[R]$ is the symmetry action on the sublattice $A$, then ${\cal G} \cdot U[R] \cdot {\cal G}^{-1}$ is the symmetry action on the sublattice $B$. The $U(1)$ part of the symmetry stays the same under the conjugation by ${\cal G}$, but $\flip = \prod e^{i \pi S_i^y}$ changes to
\begin{equation}\label{Z2_different}
    \tilde{\flip} = \prod_i e^{i \pi y_i S^z_i } e^{i \pi S_i^y} e^{-i \pi y_i S^z_i } = \prod_i (-1)^{2S \cdot y_i}  e^{i \pi S_i^y}
\end{equation}
where $S$ is the spin of the state acted upon. For an operator carrying an integer spin, $\tilde{\flip} = \flip$ but for an operator carrying a half-integer spin, $\tilde{\flip} \neq \flip$. This can also be understood as following: Let a ground state of sublattice $A$ be $\ket{\psi_0}$. Then, a ground state of sublattice $B$ would be ${\cal G}\ket{\psi_0}$. Thus, if the ground state of sublattice $A$ is symmetric under the operator $O$, then the ground state of sublattice $B$ is symmetric under the operator ${\cal G} O {\cal G}^{-1}$. Similarly, given a spinon-excited state $b^\dagger \ket{\psi_0}$ of sublattice $A$, a spinon-excited state of sublattice $B$ is ${\cal G} \cdot b^\dagger \ket{\psi_0}$. A ground state of the combined system is given by
\begin{equation}
    \ket{\psi}_{tot} = \ket{\psi_0} \otimes \qty( {\cal G} \ket{\psi_0} )
\end{equation}
In the Fig.~\ref{magnetic}, if $H_{int} \sim S^z_A  S^z_B + (S^x_A S^x_B + S^y_A S^y_B)$ for black links, then $H_{int} \sim S^z_A S^z_B - (S^x_A S^x_B + S^y_A S^y_B)$ for red links. Thus, two sublattices are differ by the gauge transformation ${\cal G}$.

Once two sublattices are combined, we need to introduce an interaction between sublattices. Spin interaction (hopping for bosons) between sublattices should be uniform so that they are consistent with $\pi$-flux per unit cell. Let them all be positive, as in Fig.~\ref{magnetic}. The system is not symmetric under naive translation ${T_1}$. Rather, the system is symmetric under the following magnetic translations:
\begin{equation}
    \boldsymbol{T}_1 = \qty[   \prod_{\text{all sites}} e^{i \pi y_i S^z_i}\, ] \cdot  T_1 \qquad \boldsymbol{T}_2 = T_2
    \label{eq:stripes}
\end{equation}
where we use bold symbols to distinguish magnetic translations with pure translations. Now, $\boldsymbol{T}_1, \boldsymbol{T}_2$ are proper symmetries of the system. They satisfy the relation
\begin{equation}\label{Z_2different}
    \boldsymbol{T}_1 \boldsymbol{T}_2 \boldsymbol{T}_1^{-1} \boldsymbol{T}_2^{-1}  =  \prod_{\text{all sites}} e^{i \pi S^z_i}
\end{equation}
which defines a system with $\pi$-flux per unit cell.

Now, we connect magnetic translation symmetries to sublattice translation symmetries. Let $\tilde{T}_1$ and $\tilde{T}_2$ be two translation operators for both sublattice, and ${\cal G}$ be the aforementioned gauge transformation. Although ${\cal G}$ itself is not the symmetry of the system, it will act as an identifying map in the Sec.~\ref{sec:bipartite}. For a certain gauge choice, we can represent the action of $\boldsymbol{T}_1$ and $\boldsymbol{T}_2$ in the combined lattice as the following:
\begin{eqnarray} \label{eq:magnetic_definition}
     &&\boldsymbol{T}_1: \ket{\psi_1}_A \otimes \ket{\psi_2}_B \mapsto \Big( \tilde{T}_1 {\cal G}^{-1} \ket{\psi_2} \Big)_A  \otimes \Big( {\cal G} \ket{\psi_1} \Big)_B \nonumber \\
     &&\boldsymbol{T}_2: \ket{\psi_1}_A \otimes \ket{\psi_2}_B \mapsto \Big( \tilde{T}_2 \ket{\psi_1} \Big)_A \otimes  \Big(  \tilde{T}_2 \ket{\psi_2} \Big)_B .
     \nonumber \\ &&\mbox{}
\end{eqnarray}
For the first line, we should act ${\cal G}^{-1}$ on $\ket{\psi_2}$ while ${\cal G}$ on $\ket{\psi_1}$ to have $\boldsymbol{T}_1^2 = \tilde{T}_1 \otimes {\tilde{T}}_1$. This is consistent with Eq.~\ref{eq:stripes} because $\boldsymbol{T}_1^2 = {\cal G}^2 T_1^2 = T_1^2 = \tilde{T}_1 \otimes \tilde{T}_1$. As ${\cal G}^2$ is just an identity, and ${\cal G}$ is not a symmetry of the system, we should not consider fractionalization associated with ${\cal G}^2$. Based on Eq.~\ref{eq:magnetic_definition}, we can derive the followings:
\begin{eqnarray}\label{eq:result_magnetic}
&&T_1 T_2 T_1^{-1} T_2^{-1} \big|_{e_A} = \boldsymbol{T}_1^2 \boldsymbol{T}_2 \boldsymbol{T}_1^{-2} \boldsymbol{T}_2^{-1} \big|_{e_A}= \eta^e_S \nonumber \\
&&\boldsymbol{T}_1 \boldsymbol{T}_2 \boldsymbol{T}_1^{-1} \boldsymbol{T}_2^{-1} \big|_{e_A e_B} = \eta_S e^{i\pi(S^z_{e_B} - S^z_{e_A})}  = - \eta_S e^{i\pi S^z_{e_A e_B}}. \nonumber \\
&&\mbox{}
\end{eqnarray}
because $e^{2\pi i S^z_{e_A}} = -1$. $\eta_S$ is the phase factor associated with the translation symmetry fractionalization of an $e_A$ particle in the sublattice $A$, which is the same for $B$.

Let us compare Eq.~\ref{eq:result_magnetic} with results from generic consideration. In general, the commutator of the two magnetic translations with respect to an $e_A$ or $e_B$ spinon excitation with $S_z$ is defined as
\begin{equation}\label{eq:magnetic_fractionalization}
    \boldsymbol{T}_1 \boldsymbol{T}_2 \boldsymbol{T}_1^{-1} \boldsymbol{T}_2^{-1} \Big|_{e_A/e_B}= \eta_{A/B} \cdot e^{i \pi S^z},
\end{equation}
which includes signs $\eta_{A/B}$ from fractionalization as well as the phase factor due to the magnetic flux. Using Eq.~\ref{eq:magnetic_fractionalization}, one obtains
\begin{eqnarray}
&& \boldsymbol{T}_1^2 \boldsymbol{T}_2 \boldsymbol{T}_1^{-2} \boldsymbol{T}_2^{-1} \big|_{e_A} = \nonumber \\ &&\boldsymbol{T}_1 \cdot \boldsymbol{T}_1 \boldsymbol{T}_2 \boldsymbol{T}_1^{-1} \boldsymbol{T}_2^{-1} \Big|_{e_B} \cdot \boldsymbol{T}_2  \boldsymbol{T}_1^{-1} \boldsymbol{T}_2^{-1} \Big|_{e_A} \nonumber \\
&& \quad \quad \quad = \eta_A \eta_B \cdot  e^{2\pi i S^z}
\end{eqnarray}
Since $2\pi i S^z = -1$ for any spinon, we obtain that 
\begin{equation}
\eta_A\eta_B= - \eta^e_S.
\end{equation}
Now to see whether we can form a condensate of $e_Ae_B$ bound states, we combine the fractionalization of the two types of spinons:
\begin{equation}
\boldsymbol{T}_1 \boldsymbol{T}_2 \boldsymbol{T}_1^{-1} \boldsymbol{T}_2^{-1} \Big|_{e_Ae_B}= \eta_A \eta_B \cdot e^{i \pi S^z_{e_A e_B}} 
\end{equation}
which again agrees with Eq.~\ref{eq:result_magnetic} since $\eta_A \eta_B = - \eta_S$. To preserve $U(1)$ symmetry, the bound state must have $S^z_{e_A e_B}=0$, and thus for translational
symmetry to be preserved, $\eta^e_S$ must be $-1$.
 This makes sense since after a condensate is created, the $e_A=e_B$ excitation has
a spin of $1/2$ so it must pick up a minus sign when it goes around two unit cells (i.e. the unit cell of the sublattice).  Now in the uncoupled sublattices, no effects of the magnetic field are felt by the unfractionalized degrees of freedom (because the unit cell encloses a whole flux quantum).  Hence, the spin liquid state on the sublattice must be chosen so that the ``emergent" magnetic field felt by the spinon agrees with the actual field. Such a spin liquid with the emergent $\pi$-flux exists as a physical mean-field solution\cite{Wen2002,Wen2002_2}, which validates our construction. The spin singlet state is also trivially fractionalized under time-reversal symmetry, ${\cal T}^2=1$, hence this symmetry does not have to be broken either.

However, we have to examine other symmetries as well, since our on-site symmetry is no longer continuous unlike the previous example of the spin-1/2 honeycomb lattice. Indeed, $SO(3)$ spin-rotation symmetry is broken down into $U(1)_z \rtimes \spflip$ and we must examine other symmetry relations, $\flip^2=1$, $\boldsymbol{T}_1 \flip = \flip \boldsymbol{T}_1$ and $\boldsymbol{T}_2 \flip = \flip \boldsymbol{T}_2$. Examining the relation $\boldsymbol{T}_2 \flip = \flip \boldsymbol{T}_2$, we can notice that this symmetry is fractionalized for the bound-state $e_A e_B$. For $e_A$ excitation, $\boldsymbol{T}_2 \flip = \flip \boldsymbol{T}_2$ holds everywhere, but for $e_B$ excitation, the action of $\flip$ on sublattice $B$ depends on the location of $e_B$ (even-row or odd-row). As $\boldsymbol{T}_2$ translates the location by one lattice site in $y$-direction, Eq.~\ref{Z2_different} implies that $\boldsymbol{T}_2 \flip = -\flip \boldsymbol{T}_2$ for an $e_B$ excitation. 
(This seems to break the translation symmetry, but one can check that it is actually \emph{enforced} by the magnetic traslation symmetry, on account of the Berry phases from the magnetic field.) 
Thus, for the bound state $e_A e_B$, this symmetry fractionalizes, and the condensation of $e_A e_B$ must break either $\spflip$ symmetry or enlarge the unit-cell in $T_2$ direction. This is consistent with the previous analysis (See TABLE~\ref{table:summaryTO}), where the $\spflip$ symmetry is not consistent with both of the toric code and double-semion orders. By breaking the $\spflip$ symmetry, we can access the $\mathbb{Z}_2$ topological order without enlarging the unit-cell.

As for visons, we know that $\eta^m_S = -1$ because they see spin-1/2 per unit cell as $\pi$-flux. By following the procedure as above, we can show that vison bound state $m_A m_B$ carries non-trivial translation symmetry fractionalization, $\eta^m_A \eta^m_B = + \eta_S^m = -1$. This is because visons do not carry a $U(1)$ quantum number affected by magnetic translation symmetry. Thus, condensation of $m_A m_B$ bound states would break translation symmetry and enlarge the unit cell. We can conclude that $\mathbb{Z}_2 \times \mathbb{Z}_2$ topological phase cannot be reduced further into a simpler topological phase without symmetry breaking.

{\bf Remark} One may wonder--since our construction started from two decoupled $\mathbb{Z}_2$ SLs and $\pi$-flux is manifested only by hoppings between sublattices, does $\pi$-flux play any role in the system? It does as long as hopping terms between sublattices are non-zero. Consider the anyon bound state $e_A e_B$, whose $S_z$ quantum number can be $0$ or $\pm 1$. Then, under the magnetic translation around the unit cell, a $S_z = \pm 1$ bound state acquires a phase factor of $-1$ relative to a spin-$0$ bound state due to the $\pi$-flux. Thus, we would observe non-trivial band dispersion for anyon bound states, which is doubly repeated due to $\pi$-flux. This relative phase is invariant under a different choice of $Z_2$ SL in sublattice. Moreover, unlike a single anyon excitation whose translation symmetry fractionalization is gauge-dependent due to the fact that the anyon-type changes under the symmetry, this phase factor is gauge-invariant, and is a measurable quantity of the system.

\begin{widetext}

{\centering

\begin{table}[t]
    \begin{center}
    
    \begingroup
	\setlength{\tabcolsep}{7pt} 
	\renewcommand{\arraystretch}{1.1} 

    \begin{tabular}{| c | c | c | c | c | }
    \hline
    Microscopic &  Symmetry Group & $\mathbb{Z}_2$ Toric Code & $\mathbb{Z}_2$ Double-Semion & $\mathbb{Z}_2 \times \mathbb{Z}_2$ Toric Code  \\ \hline\hline

    \multirow{5}{*}{ \shortstack{ $\rho = 1/2$ \\ $\phi = 1/2$ }  } & $U(1)$ & \cmark & \cmark & \cmark  \\
    \hhline{~----} & $U(1) \rtimes \spflip$ & \xmark & \xmark & \cmark   \\
   	\hhline{~----} & $U(1) \rtimes Z^{{\cal T}_1}_2$ & \xmark & \cmark & \cmark  \\
    \hhline{~----} & $U(1) \times Z^{{\cal T}_2}_2$ & \cmark & \xmark & \cmark   \\
    \hhline{~----} & $(U(1) \rtimes \spflip) \times Z^{{\cal T}_2}_2$ & \xmark & \xmark & \cmark   \\
    \hline 
    
    
    \multirow{2}{*}{ \shortstack{ $\rho = 1/2$ \\ $\phi = 1/n$ } } & $U(1)$ & \cmark & \cmark & \cmark  \\
    \hhline{~----} & $U(1) \times Z^{{\cal T}_2}_2$ & \cmark & \xmark & \cmark \\
   	\hline \hline
   
   	Microscopic &  Symmetry Group & {$\mathbb{Z}_n$ Topo. Order} & $U(1)_n \times U(1)_{-n}$ ($\mathbb{Z}_n$ Twisted)  & $\mathbb{Z}_n \times \mathbb{Z}_n$ Topo.  Order \\ \hline\hline
    
    \multirow{2}{*}{ \shortstack{ $\rho = 1/n$ \\ $\phi = 1/2$  } } & $U(1)$ & \cmark & \cmark & \cmark \\
    \hhline{~----} & $U(1) \rtimes Z^{{\cal T}_1}_2$ & \xmark (even $n$), \cmark (odd $n$) & \cmark (even $n$), \xmark (odd $n$)  & \cmark \\
   	\hline

	\end{tabular}
 
 \caption{ \label{table:summaryTO} Summary of non-chiral topological phases for the bosonic systems with different microscopic data and on-site symmetry group where a featureless state is prohibited by the LSM-type theorem. States with a check mark are not ruled out by our arguments, although we have not constructed all of them explicitly. When $G=U(1)$, there does not exist much constraint as long as there exists anyon with fractional charge to satisfy Eq.~\ref{eq:constraint_YML_FQHE}. }
\end{center}
\endgroup

\end{table}
}
 
\end{widetext}

\subsection{Smaller Fractions for the Flux or Filling} \label{sec:generic_case}

In this subsection, we analyze a constraint on the non-chiral topological order that can be realized in a bosonic system with magnetic translation symmetry in detail, at generic filling and flux condition. Consider a bosonic system with $U(1)$ symmetry and magnetic flux $\Phi_0$ per unit cell such that we can define conserved charge $Q$ and associated magnetic translation symmetry,
\begin{equation} \label{eq:magnetic_translation}
 \boldsymbol{T}_1 \boldsymbol{T}_2 \boldsymbol{T}_1^{-1} \boldsymbol{T}_2^{-1}  = \exp(i Q \Phi_0) \in U(1) 
\end{equation}
where $Q$ measures total $U(1)$ symmetry charge of the region where magnetic translation symmetry acts on. $Q$ can be either electric charge or spin $S_z$ components depending on the system of interest. Assume we have a time reversal symmetry ${\cal T}$. As discussed earlier, there are two ways to realize time-reversal symmetry, ${\cal T}_1 = {\cal K}$ 
and ${\cal T}_2 = {\cal P} {\cal K}$ where ${\cal P}$ is an unitary operator reversing the sign of $U(1)$ charge (for a spin model, ${\cal P} = e^{i\pi S^y}$). 

For the first case with $G = U(1) \rtimes Z_2^{T_1}$, $\expval{Q}$ can take an arbitrary value. However, the $U(1)$-flux is reversed under ${\cal T}_1$: ${\cal T}_1 \cdot  \exp(i Q \Phi_0)   \cdot  {\cal T}_1^{-1} = \exp(-i Q \Phi_0)$. Thus, Eq.~\ref{eq:magnetic_translation}, the relation defining the magnetic translation symmetry, is invariant under ${\cal T}_1$ only if $Q \cdot \Phi \equiv 0 \mod \pi$. For a generic system made of particles with integer charge, this holds only if $\Phi = 0,\pi$. For the second case with $G = U(1) \times Z_2^{T_2}$, the $U(1)$ charge density $\expval{Q} = N_0/2 \in \mathbb{Z}/2$ since ${\cal T}_2: Q \mapsto N_0-Q$ for some integer $N_0 \in \mathbb{Z}$. Thus, it can have either an integer or half-odd integer filling. (We will focus on a half-filling case as an integer filling does not have a no-go theorem for a featureless state by the Eq.~\ref{eq:constraint_YML_IQH}.) 
Since any element of $U(1)$ and ${\cal T}_2$ commute, $\exp(i Q \Phi_0)$ is invariant under ${\cal T}_2$. As a consequence, there is no restriction on the magnetic flux $\Phi_0$. We will explain possible constraints for each case in the following. For the summary, see the TABLE~\ref{table:summaryTO}.

\vspace{0.15in}

\noindent {\bf Case I: $\rho=1/2$ and $\phi = 1/n$}. At half-filling with a generic flux $\Phi_0 = 2\pi/n$ (flux density $\phi=1/n$), we can have a time-reversal ${\cal T}_2$ on top of $U(1)$ symmetry, thus $G = U(1)$ or $U(1) \times Z_2^{{\cal T}_2}$. In this case, the Eq.\ref{eq:constraint_YML_IQH} implies that a featureless state does not exist unless $n = 0 \mod 4$. Even if $n\equiv 0 \mod 4$, a state without topological order must have a nonzero Hall coefficient, and thus must break time reversal symmetry.

Using the same argument we discussed above, $\mathbb{Z}_2$ toric code order has a consistent symmetry fractionalization pattern with the given microscopic data, but $\mathbb{Z}_2$ double semion order does not. However, an actual parton construction for the toric code is not apparent, and seemingly a daunting task. To make a start on this issue, we can extend the approach used in the Sec.~\ref{sec:magnetic} to construct a $\mathbb{Z}_2^{\otimes n}$ topological order ($n$ copies of toric code), and then condense anyons to access $\mathbb{Z}_2$ topological order. For a given rectangular lattice, we take a sublattice consisting of every $n^\mathrm{th}$ site in the $x$-direction (similar to Fig.~\ref{magnetic}). We can prepare a $\mathbb{Z}_2$ SL state $\ket{\psi_0}$ on the first sublattice with the Hamiltonian $H_0$, and then copy it to the other sublattices
by using powers of the gauge transformation
\begin{equation}
    {\cal G} = \prod_r e^{i y_r \Phi_0  S^z_r}.
\end{equation}
Thus, the $n^\mathrm{th}$ sublattice has the Hamiltonian ${\cal G}^{n-1} H_0 {\cal G}^{-(n-1)}$. Then, we obtain a lattice with magnetic translation symmetry, $\boldsymbol{T}_1$ and $\boldsymbol{T}_2$, where $\boldsymbol{T}_1$ is a pure $x$-translation (which permutes sublattices) combined with an associated gauge transformation ${\cal G}$, and $\boldsymbol{T}_2$ is a pure $y$-translation. Note that $\boldsymbol{T}_1^n$ becomes equivalent to the translation operator of each sublattice in the $x$-direction and the $\boldsymbol{T}_2$ is equivalent to the original translation operator of each sublattice in $y$-direction. Therefore, the Hamiltonian for the system is given as
\begin{equation}
    H = H_0 \otimes {\cal G} H_0 {\cal G}^{-1}  \otimes \dots \otimes {\cal G}^{n-1} H_0 {\cal G}^{-(n-1)}
\end{equation}
Then, the ground state of $H$ is given by
\begin{equation}
    \ket{\psi} = \ket{\psi_0} \otimes {\cal G} \ket{\psi_0} \otimes \dots \otimes {\cal G}^{n-1} \ket{\psi_0}
\end{equation}
which is a $\mathbb{Z}_2^{\otimes n}$ topological phase. We can add  hopping terms  between the sublattices to couple them.
Now, condensing pairs of anyons in this case is more complicated than in the cases we have considered so far. If one tries to condense bound states formed from a pair of $e$'s on two sublattices, for example, one breaks the translational symmetry because these $e$'s are transformed into a different topological type by the translation of the lattice along $x$.
However, it seems likely to be possible to \emph{simultaneously} introduce several condensates, e.g. a condensate of the
$e_1e_2$ bound state, the $e_2e_3$ bound state,\dots, and the $e_ne_1$ bound state (where $e_i$ represents an $e_i$ excitation
on the $i^\mathrm{th}$ sublattice) which would reduce the topological order to the $\mathbb{Z}_2$ toric code order.

\vspace{0.15in}

\noindent {\bf Case II: $\rho=1/n$ and $\phi=1/2$}. At a generic filling $\rho = 1/n$ with $\pi$-flux ($\phi=1/2$), we can have a time-reversal ${\cal T}_1$ on top of $U(1)$, thus $G = U(1)$ or $U(1) \rtimes Z_2^{{\cal T}_1}$. In this case, featureless states are impossible since $\sigma_{xy} \in 2\mathbb{N}$ can never satisfy the Eq.~\ref{eq:constraint_YML_IQH}. Eq.~\ref{eq:constraint_YML_FQHE} requires the existence of an anyon carrying  $1/n$-charge, which has the braiding statistics $2\pi/n \mod 1$ with the $2\pi$ $U(1)$ symmetry defect. To accommodate such excitations, we need at least $\mathbb{Z}_n$ fractionalization.

For a non-chiral topological order, the simplest example is the topological order described by $\mathbb{Z}_n$ gauge theory, whose anyon contents are $\{e^{l_1} m^{l_2} | 0\leq l_1,l_2 < n\}$, where $\theta_{e,e} = \theta_{m,m} = 0$ and $\theta_{e,m} = 2\pi/n$. Also, without loss of generality, we can set $e$-particle to carry a $U(1)$-charge $\frac{1}{n}$ and $m$-particle to be $2\pi$-flux $F$ in order to satisfy Eq.~\ref{eq:constraint_YML_FQHE}.  
There is no obstruction for this $\mathbb{Z}_n$ topological order for $G=U(1)$, but for $U(1) \rtimes Z_2^{{\cal T}_1}$, we can show that there is no consistent symmetry fractionalization pattern that can be assigned to $e$-particles. First, ${\cal T}_1: e \mapsto e$ because $e^{l_1}$ and $e^{l_2}$ carry different $U(1)$ charges even upto a physical unit charge unless $l_1 = l_2$. Then, since $\theta_{{\cal T}_1 e, {\cal T}_1 m} = - \theta_{e,m}$, we can deduce ${\cal T}_1: m \mapsto m^{n-1}$. 
If $n$ is odd, for $n = 2l +1$, a consistent symmetry fractionalization pattern can be written as follows:
\begin{eqnarray}
\boldsymbol{T}_1 \boldsymbol{T}_2 \boldsymbol{T}_1^{-1} \boldsymbol{T}_2^{-1} \big|_{e} &=& \eta^l \cdot e^{i\pi S_z} = -1 \nonumber \\
\boldsymbol{T}_1 \boldsymbol{T}_2 \boldsymbol{T}_1^{-1} \boldsymbol{T}_2^{-1} \big|_{m} &=& \eta \cdot e^{i \pi S_z} = e^{2\pi i/n},
\end{eqnarray}
where $\eta = e^{2\pi i/n}$ is the $n^\mathrm{th}$ root unity--a phase factor coming from $\mathbb{Z}_n$ fractionalization. As ${\cal T}: m \mapsto m^{n-1}$, both equations are symmetric under ${\cal T}_1$. This pattern is equivalent to saying there is a background anyon $e \cdot m^l$. As this background anyon carries charge $\frac{1}{n}$, it is consistent. On the other hand, if $n$ is even, there is no pattern of fractionalization of the translation symmetries that is consistent with time-reversal.  
Combining $n$ e-particles together gives a single fundamental particle, so it must pick up a minus sign under transport around the unit cell.  Thus one $e$ must acquire a phase that is an $n^\mathrm{th}$ root of $-1$, which is complex so it breaks time-reversal symmetry.

For even $n$, as we did in the previous section, we can show the existence of $\mathbb{Z}_n \times \mathbb{Z}_n$ topological order explicitly through our bipartite lattice construction. Since the flux density $\phi=1/2$, we can combine two sublattices, each of which does not see any external flux and realizes the $\mathbb{Z}_n$ topological order. Thus, $\mathbb{Z}_n \times \mathbb{Z}_n$ topological order can be constructed easily.

Another possible non-chiral topological order is a $\mathbb{Z}_n$ twisted topological order, a $\mathbb{Z}_n$ analogue of $\mathbb{Z}_2$ double-semion order. There are more than one way to ``twist'' for $n>2$, but the most representative one is $U(1)_n \times U(1)_{-n}$ non-chiral fractional quantum Hall state described by a 
$K$-matrix $\bigl( \begin{smallmatrix}n & 0\\ 0 & -n\end{smallmatrix}\bigr)$. Since we are considering bosonic topological orders, such a state can only exist for even $n$. The state contains a set of anyons $\{s^{l_1} \bar{s}^{l_2}\, |\,0\leq l_1,l_2 < n \}$, where $\theta_{s,s} = - \theta_{\bar{s},\bar{s}} = 2\pi/n$ and $\theta_{s,\bar{s}} = 0$. A similar analysis shows that there exists a consistent symmetry fractionalization pattern, where $s$ ($\bar{s}$) carries $U(1)$-charge $\frac{1}{n}$ ($-\frac{1}{n}$) and $2\pi$-flux $F = s \bar{s}$. Here, ${\cal T}_1 : s \mapsto \bar{s}^{n-1} + c$, where $c$ is a local physical excitation carrying charge-1 (should be there to conserve charge). Then,
\begin{eqnarray}
\boldsymbol{T}_1 \boldsymbol{T}_2 \boldsymbol{T}_1^{-1} \boldsymbol{T}_2^{-1} \big|_{s} &=& e^{i\pi  S_z} \rightarrow  e^{i\pi/n} \nonumber\\ 
\boldsymbol{T}_1 \boldsymbol{T}_2 \boldsymbol{T}_1^{-1} \boldsymbol{T}_2^{-1} \big|_{\bar{s}} &=& \eta \cdot e^{i \pi S_z} \rightarrow e^{i\pi/n}
\end{eqnarray}
gives a consistent pattern, which can be understood as a background anyon $s$. However, the explicit construction of such a phase is unclear unlike the $\mathbb{Z}_n$ gauge theory from a parton construction.

{\bf Remark} When the system does not have a time-reversal symmetry (or breaks it spontaneously), a chiral topological order with a finite Hall conductivity can be realized. Consider a symmetry group $U(1)$ with the particle density $\rho$ and flux density $\phi$. At the half-filling under $\pi$-flux ($\rho = 1/2$, $\phi=1/2$), we have the ratio between particle density and flux density $\nu = 1$. In this case, a $\nu=1$ Moore-Read state\cite{MooreRead1991} is known to exist.
In general, there are many explicit parton constructions available\cite{Wen1991} in the lattice model for a generic $\rho$ and $\phi$, and some are shown even numerically\cite{Cooper2001,Hormozi2012}. For example, when $\rho = 1/3$ and $\phi=1/2$, there is an explicit parton construction of the chiral topological order with non-abelian statistics\cite{Wen1991}. When $\rho= 1/(4m)$ and $\phi=1/2$ with $m \in \mathbb{N}$, the ratio between the boson and flux density $\nu=1/2m$, and we expect Laughlin wavefunction for fractional quantum Hall state. For $n = 4$, it has been shown numerically that the physical system realizes $\nu=1/2$ bosonic fractional quantum Hall state\cite{Laughlin1983}, i.e. chiral spin liquid with non-zero magnetization in the spin-model language\cite{Kalmeyer1989}.

\section{Conclusion and Discussion}\label{sc6}

In this paper, we proposed a framework to construct a symmetry enriched topological order in a bipartite lattice. 
First, we constructed the featureless ground state in the spin-1/2 honeycomb lattice. we took $\mathbb{Z}_2$ SL wave functions of each sublattice, combined them, and condensed some of the anyons to obtain the featureless state. As an intermediate phase, we obtained an exotic type of $\mathbb{Z}_2$ SLs, where topological excitations do not carry any half-odd integer spin. 
We want to emphasize that in the $\mathbb{Z}_2 \times \mathbb{Z}_2$ topological phase, single anyons $e_A$ or $e_B$ do not have well-defined symmetry fractionalization with respect to the lattice symmetry that permutes anyons. However, the structure of the condensate of the anyon bound state $e_A e_B$ fixes the gauge-choice for a single $e_A$, determining the symmetry fractionalization of $e_A=e_B$ in the resulting $\mathbb{Z}_2$ topological order.

Next, we applied the proposed framework to understand the connection between extensions of the HOLSM theorem and the allowed gapped and symmetric phases for a given symmetry. While the HOLSM theorem places constraints on systems with a half-odd integer spin per unit cell, its extensions can put further constraints based on other microscopic data, such as nonsymmorphicity of lattice or the $U(1)$-flux per unit cell, i.e. replacement of translation symmetry by  magnetic translation symmetry.
We explored cases where the extensions of HOLSM theorem prevents the existence of a featureless state, and listed SET phases consistent with given symmetry information. In particular, we constructed SET phases with certain symmetry fractionalization patterns that are closely related to the HOLSM-type constraints.

One interesting question that can be addressed in a future is the explicit construction of $\mathbb{Z}_2$ double-semion topological order for the system at a half-filling with $\pi$-flux with $U(1) \rtimes Z_2^{{\cal T}_1}$ symmetry. Although it is proven to be impossible under the absence of $\pi$-flux, we showed that the constraint is circumvented when there is $\pi$-flux. Naively, we can combine two sublattices; one sublattice realizes a $\mathbb{Z}_2$ double semion order with broken time-reversal symmetry, and the other realizes a gauge transformed verion of it like in Sec.~\ref{sec:application} due to the $\pi$-flux. It might be able to recover time-reversal symmetry after condensing appropriate anyon bound states.

Our approach may be further generalized to give a conceptual route for establishing the existence of other featureless ground states, such as the one for a square lattice with spin-1 per site. In this case, the existence of a featureless state is not prohibited by the HOLSM theorem, and the featureless state was shown numerically using the tensor network state approach. In Ref.~\onlinecite{JianZaletel2016}, a generalized AKLT-type construction was used with four virtual spin-1/2's per site, from which a virtual symmetric Hilbert space was constructed. Then, the virtual state was projected to the physical Hilbert space with spin-1 per site, and the resulting state was verified to be topologically trivial and gapped. 
Simiarly, in our construction, we can think of spin-1 per site as two spin-1/2 per site in a virtual space, later projected into a spin-1 subspace. In a virtual space, we can form two copies of $\mathbb{Z}_2$ SLs in a square lattice with spin-1/2 per site, and consider bound states $e_A e_B$ and $m_A m_B$. Since phase factors for symmetry fractionalization class of bound states are just products of phase factors for equivalent anyons, we obtain totally trivial symmetry fractionalization classes for bound states. Then, we can condense $e_A e_B$ and $m_A m_B$ to obtain a featureless state in the virtual space---followed by the projection, we would obtain a featureless state in the physical system. Of course, we first have to verify whether the state survive as a featureless state after projection and not as a ``cat'' state. Then, It would be interesting to compare a featureless state obtained by our construction to the featureless states in Ref.~\onlinecite{YR2016,JianZaletel2016}, and investigate which class of SPT phases they belong to\cite{YR2017}.


Finally, we remark that this bipartite construction provides an interesting toy model for a system where spatial symmetries can exchange anyons. A few models have been suggested, such as the Wen's plaquette model\cite{WenPlaquette2003, Cho2012} or the bilayer toric code\cite{Maissam_SymFrac_2014}, but their precise symmetry fractionalization properties remain to be studied. Here we provide an explicit model where spatial symmetry exchanges anyons and detailed calculations of the twisted second group cohomology.  One interesting question would be to address the issue regarding the condensate of multiple anyons superposed in a symmetric way. This has a potential application to other multi-partite lattices, such as a kagome lattice which is made of three triangular sublattices. 
Since we can construct new SET orders with additional tunable features from known ones, a generalization of this approach would be a fruitful direction. A three-dimensional generalization of this construction also seems to be feasible, where loop-like excitations carry fractionalized quantum numbers. Recently, an anyon condensation approach was used to derive fracton topological orders in three dimensions\cite{Ma2017_Fracton}, and it would be interesting to extend this analysis to include symmetries as discussed here. These questions are left for future work. 

\acknowledgments

Authors would like to thank Graham Ellis, Andrew Essin, Zhaoxi Charles Xiong, Shubhayu Chatterjee, Yang Qi, Yi-Zhuang You, and Meng Cheng for helpful discussion. In particular, we thank Yin-Chen He and Han Ma for the discussion on double-semion order. Jong Yeon Lee and Ashvin Vishwanath were supported by the ARO MURI on topological insulators, grant W911NF-12-1-0961 and by a Simons Investigator award. Ari Turner was partially supported by a grant from the Simons Foundation and worked on this project while at the Aspen Center for Physics which is supported by National Science Foundation grant PHY-1607611.

\appendix

\section{Symmetry Fractionalization of Bound States in the Honeycomb Lattice}\label{Appendix1}
Based on Eq.~\ref{eq:sym_defining}, symmetry fractionalization phase factors of bound states (and individual anyons) in the honeycomb lattice can be calculated as the following. For $\boldsymbol{T_1 T_2 T_1^{-1} T_2^{-1}}$, 
\begin{eqnarray}
    &&\boldsymbol{T_1 T_2 T_1^{-1} T_2^{-1}} = \nonumber \\
    && \qty( T_1 T_2 T_1^{-1} T_2^{-1} ) \otimes \qty( T_2^{-1} T_1 T_1 T_1^{-1} T_2 T_1^{-1} ) =\nonumber \\
    && \qty( T_1 T_2 T_1^{-1} T_2^{-1} ) \otimes \qty( T_2^{-1} T_1 T_2 T_1^{-1} ) = \qty( \eta_1 )^2 = 1 \quad \,\,\,\,
\end{eqnarray}
Similarly, as $\boldsymbol{\sigma}^2 = \sigma^2  \otimes \qty( C_6^{-2} T_1^{-1} \sigma C_6^{-2} T_1^{-1} \sigma  )$ for $\boldsymbol{\sigma}^2$,
\begin{eqnarray}
     &&  C_6^{-2} T_1^{-1} \sigma C_6^{-2} T_1^{-1} \sigma   = \nonumber \\
     &&   C_6^{-2} [ \alpha_1 \sigma^{-1} T_2^{-1}] C_6^{-2} T_1^{-1} \sigma =\nonumber \\
     && \alpha_1 \cdot  C_6^{-2} \sigma^{-1} [\alpha_3 C_6^{-1}T_2^{-1} T_1 ] C_6^{-1} T_1^{-1} \sigma  =\nonumber \\
     && (\alpha_1 \alpha_3 ) \cdot  C_6^{-2} \sigma^{-1} C_6^{-1}T_2^{-1} [ \alpha_2 C_6^{-1} T_2 ] T_1^{-1} \sigma  =\nonumber \\
    && (\alpha_1 \alpha_2  \alpha_3 ) \cdot  C_6^{-2} \sigma^{-1} C_6^{-1} [\alpha_3 C_6^{-1} T_2^{-1} T_1 ] T_2  T_1^{-1} \sigma  =\nonumber \\
    && (\alpha_1 \alpha_2) \cdot C_6^{-2} \sigma^{-1} C_6^{-2} [ T_2^{-1} T_1 T_2  T_1^{-1} ] \sigma = \nonumber \\
    &&  (\alpha_1 \alpha_2 \eta_1 ) \cdot C_6^{-2} \sigma^{-1} C_6^{-2} \sigma = (\alpha_1 \alpha_2 \eta_1) 
\end{eqnarray}
where we used the fact that 
\begin{eqnarray}
    && C_6^{-2} \sigma^{-1} C_6^{-2} \sigma  = C_6^{-1} \qty( C_6^{-1} \sigma^{-1} C_6^{-1} \sigma^{-1}) \sigma C_6^{-1} \sigma  \nonumber \\
    && = \eta_4 \cdot  C_6^{-1} \sigma C_6^{-1} \sigma  = \eta_4 \cdot  C_6^{-1} \sigma^{-1} C_6^{-1} \sigma^{-1} = 1 \quad 
\end{eqnarray}
since $\sigma^{-1} = \eta_2 \sigma$ and $\qty( \eta_i )^2 = 1$. Thus,
\begin{equation*}
    \boldsymbol{\sigma}^2 = \sigma^2  \otimes \qty( C_6^{-2} T_1^{-1} \sigma C_6^{-2} T_1^{-1} \sigma  ) = \eta_2 \cdot (\alpha_1 \alpha_2 \eta_1) = 1
\end{equation*}
The last equality seems suspicious, but we can prove that $\eta_1 \eta_2 \alpha_1 \alpha_2 = 1$ indeed. Proof is given as the following:
\begin{eqnarray}
    && T_2 = \eta_4 \cdot T_2 \cdot \eta_4 = (\sigma C_6 \sigma C_6)\cdot  T_2 \cdot (\sigma C_6 \sigma C_6)^{-1} \,\,\,\,\,\,\,
\end{eqnarray}
From definitions, $C_6 T_1 C_6^{-1} = \alpha_2 T_2$ and $\sigma T_2 \sigma^{-1} = \alpha_1 \eta_2 T_1$, which gives $(\sigma C_6) T_1 (\sigma C_6)^{-1} = \alpha_1 \alpha_2 \eta_2 \cdot T_1 $. Furthermore, we have $\sigma C_6 T_2 C_6^{-1} \sigma^{-1}= \alpha_3 T_2^{-1} T_1$. Then,
\begin{eqnarray}
    && T_2 = \eta_4 \cdot T_2 \cdot \eta_4 = (\sigma C_6 \sigma C_6)\cdot  T_2 \cdot (\sigma C_6 \sigma C_6)^{-1} = \nonumber \\
    && (\sigma C_6) \alpha_3 T_2^{-1} T_1 (\sigma C_6) = (\eta_1 \alpha_3) \cdot (\sigma C_6)  T_1 T_2^{-1} (\sigma C_6)^{-1} \nonumber \\
    && = (\eta_1 \alpha_3) (\alpha_1 \alpha_2 \eta_2 \cdot T_1) \, (\alpha_3 T_2^{-1} T_1 )^{-1} = (\alpha_1 \alpha_2 \eta_1 \eta_2) \cdot T_2   \nonumber \\
    \mbox{}
\end{eqnarray}
Thus we conclude that $\alpha_1 \alpha_2 \eta_1 \eta_2 = 1$, implying this combination of phase factors becomes gauge-invariant. For $(\boldsymbol{C_6})^6$, we have $(\boldsymbol{C_6})^6 = \qty( [T_1 C_6^2 ]^3 ) \otimes \qty( [T_1 C_6^2 ]^3 )$. Therefore, let's examine its action on sublattice $A$ first. Since $C_6^2 T_1 = C_6 \cdot [ \alpha_2 T_2 C_6] = \alpha_2 \alpha_3 \cdot T_1^{-1} T_2 C_6^2$, and similarly, $C_6^2 T_1^{-1} = \alpha_2 \alpha_3 \cdot T_2^{-1} T_1 C_6^2$, we have
\begin{eqnarray}
    \qty(T_1 C_6^2)^3  && = (\alpha_2 \alpha_3) \cdot T_2 C_6^4 T_1 C_6^2 = \nonumber \\
    && T_2 C_6^2 T_1^{-1} T_2  C_6^4 =  (\alpha_2 \alpha_3) \cdot T_1 C_6^2 T_2 C_6^4 = \nonumber \\
    && \alpha_3 \alpha_3 \cdot T_1 T_2^{-1} T_1^{-1} T_2 C_6^6 = \eta_1 \eta_3 \alpha_3 
\end{eqnarray}
where we used a relation $C_6^2 T_2 = C_6 [ \alpha_3 T_1^{-1} T_2 C_6] = \alpha_2 \alpha_3 T_2^{-1} C_6 T_2 C_6 = \alpha_2 T_2^{-1} T_1^{-1} T_2 C_6^2$ from second line to third line. Unlike other cases, we can see that when $(C_6)^6$ acts on a single spinon, we would get a gauge-dependent factor $\eta_1 \eta_3 \alpha_3 $. For a bound state, we get
\begin{eqnarray}
    (\boldsymbol{C_6})^6 &&= \qty(T_1 C_6^2)^3 \otimes \qty(T_1 C_6^2)^3 \nonumber \\
    &&= (\eta_1 \eta_3 \alpha_3^A) \otimes (\eta_1 \eta_3 \alpha^B_3) = \alpha_3^A \alpha_3^B  = 1
\end{eqnarray}
The last equality comes from the fact that the existence of ${\cal S}^2 = 1$ fixes gauge choice such that $\alpha_3^A = \alpha_3^B = 1$. Finally, for $\boldsymbol{\sigma}\boldsymbol{C}_6 \boldsymbol{\sigma} \boldsymbol{C}_6$, it is very straightforward to show that it is equivalent to $\sigma^2 \otimes \sigma^2 = \qty(\eta_2)^2 =  1$. Thus, we showed that all symmetry fractionalization class characterizing the bound state should be trivial for this gauge choice, and thus belong to the trivial class. 
\\

\section{Group Cohomology Calculation of Symmetry Fractionalization Pattern}\label{AppendixTwisted}

In this appendix, we elaborate on the second group cohomology of a symmetry group and how to generalize it to calculate fractionalization classes when symmetry operators can permute anyons with trivial mutual statistics between them. (For example, $e_A$ and $e_B$ have a trivial mutual statistics and they are permuted to each other under translation in the bipartite lattice construction.) In the section.~\ref{sec:review}, we gave the intuitive explanation and the lemma on how to characterize a fractionalization class by phase factors associated with symmetry relations of a group $G$. The lemma works similarly regardless of whether a symmetry permutes anyons or not, but we need extra care to define a coefficient group ${\cal A}$, i.e. a set of allowed phase factors.

First, let us summarize how the second cohomology group is defined in a physical context when group elements do not permute anyons. $\Gamma$ is called a linear representation if $\forall g_1,g_2 \in G$, it satisfies
\begin{equation}
    \Gamma(g_1) \Gamma(g_2) = \Gamma(g_1 g_2)
\end{equation}
On the other hand, $\bar{\Gamma}$ is called a projective representation extended by a coefficient group ${\cal A}$ if ${\cal A}$ is an abelian group (a set of $U(1)$ phase factors) and there exists a map $\omega: G\times G \mapsto {\cal A}$, called a factor set, such that: 
\begin{equation}
    \bar{\Gamma}(g_1) \bar{\Gamma}(g_2) = \omega(g_1,g_2) \bar{\Gamma}(g_1g_2)
\end{equation}
where $\omega(g_1,g_2)$ is the additional phase factor we discussed in the section.~\ref{sec:review}. The factor set should satisfy a consistency equation for a projective representation to be associative:
\begin{equation}\label{eq:cocycle_1}
    \omega(g_1,g_2) \omega(g_1g_2,g_3) = \omega(g_1,g_2g_3)  \, {}^{g_1}\omega(g_2,g_3) 
\end{equation}
where $^{g_1}\omega(g_2,g_3)$ is the image of $\omega(g_2,g_3)$ under a certain transformation depending on $g_1$ (the ``action" of $g_1$). The action of $g_1$ is often trivial.  However for a time-reversal symmetry, which acts anti-unitarily on complex numbers, $^{\cal T} \!\omega(g_2,g_3) = \omega(g_2,g_3)^{-1}$.  
The Eq.~\ref{eq:cocycle_1} is called a 2-cocycle condition, by analogy with a similar formula in topology. Thus, a projective representation is characterized by its factor set $\omega$. Projective representations can be redefined by a gauge transformation $\Gamma'(g) = \lambda(g) \Gamma(g)$ where $\lambda: G \mapsto {\cal A}$ is a map (which does not have to be a homomorphism) from $G$ to ${\cal A}$. If two projective representations can be related by a gauge transformation, they are called equivalent, and the phase factors are related by,
\begin{equation}\label{eq:coboundary_1}
    \omega'(g_1,g_2) = \frac{\lambda(g_1) \,  {}^{g_1}\!\lambda(g_2) }{\lambda(g_1 g_2)} \omega(g_1,g_2)
\end{equation}
which we denote as $\omega \sim \omega'$. For a given $G$ and $\cal A$, projective representations ($\omega$) form an abelian group of equivalence classes, called the second cohomology group of $G$, $H^2(G,{\cal A})$. Although we did not write down $\bar{\Gamma}$ when we related symmetry relations and phase factors in the main text, if $g_1^2 = 1$ is a symmetry relation and the action of $g_1^2$ on an anyon gives some phase factor, we should write it as $\qty[\bar{\Gamma}(g_1)]^2 = e^{i \theta}$. However, for a notational convenience, we often omit writing down $\bar{\Gamma}$.

So far, we have avoided a rigorous discussion on the definition of ${\cal A}$. Let ${\cal I}$ be an algebra of abelian anyons for a given topological order. For example, in a familiar $\mathbb{Z}_2$ topological order, ${\cal I} = \{1, e, m, em \,| \dots \}$, where $\dots$ are fusion rules. Then, we define a coefficient group ${\cal A}$ is defined as $\text{Hom}({\cal I}, U(1))$. Physically, the definition is motivated by the observation that there arises a gauge redundancy associated with fusion rules for a given topological order, $\bar{\Gamma}|_a \rightarrow f(a) \bar{\Gamma}|_a$ which can depend on the anyon type $a$. For example, when a topological excitation $a = b \times c$, it means that an anyon $a$ can fractionalize into $b$ and $c$. While doing so, operators for $a$,$b$ and $c$ can be redefined by phase factors consistent with this fusion rule. For any given fusion rule $a \times b = c$, 
$f$ must satisfy
\begin{equation} \label{eq:factorset_hom}
    f(a) \cdot f(b) = f(c) \qquad 
\end{equation}
Thus, $\cal A$ is not a set of phase factors (when the anyons are all considered simultaneously), but rather a set of homomorphisms from anyons to phase factors. By taking into account this fact, we can represent a factor set as a function $\omega: G \times G \times {\cal I} \rightarrow U(1)$ and a phase factor as $\lambda: G \times {\cal I} \rightarrow U(1)$. 

Now if the symmetries include crystal symmetries, $\omega(g_1,g_2,a)$ is not precisely a homomorphism.  The symmetry fractionalization patterns can be intertwined with the statistics because when two anyons are fused and then transformed, they may become braided with one another. In this situation, the assumption on the coefficient group that $\omega(g,h) \in \text{Hom}({\cal I}, U(1))$ is wrong. Instead, a factor set $\omega(g,h;a)$ would satisfy the following equation for a given fusion rule $a \times b = c$:
\begin{equation} 
\omega(g,h;a) \cdot \omega(g,h;b) = \Omega_{a,b}^c(g,h) \cdot \omega(g,h;c)
\end{equation}
where the phase factor $\Omega_{a,b}^c$ is called ``twist factor''\cite{EssinHermele_Z2_2013,YML2014,Zaletel_TwistFactor}, that can be determined from the mutual statistics of the anyons\cite{YML2014}. For many cases, such as $\mathbb{Z}_2$ order, we do not need to consider this complication since we can find a minimal set of anyons each of which has zero topological spin.  As long as these anyons are not permuted, each anyon in the minimal set of anyons generating the topological order carries an independent fractionalization class, and classification of a given topological phase would be given as a product of fractionalization classes of all independent anyons. When anyons are permuted, we factor the group of anyons ${\cal I}$ into group of anyons
that are exchanged with one another and that have trivial mutual statistics, and if this is possible, we do not need to consider topological data. (For example, we can consider the $e_A/e_B$
excitations and the $m_A/m_B$ excitations separately in the toric code model.)  Finally,
in the case of the double-semion states, the translation and time-reversal do not seem to have any interplay with braiding, which still allows us to use Eq.~\ref{eq:factorset_hom}.

Returning to our general discussion, we remark that a coefficient group ${\cal A}$ is called a $G$-module, which means that a group element $g \in G$ acts as a function $g: {\cal A} \mapsto {\cal A}$, either by a left-action or by a right-action. Previously discussed cases are when $g$ acts as an identity mapping in $\cal A$, since they do not act on a phase factor (when it is a time-reversal). However, when it comes to the case of anyon permuting symmetries, group elements can act on ${\cal A}$ by changing anyon arguments of $f \in \text{Hom}({\cal I}, U(1))$. In this case, 2-cocycle condition for factor sets should be modified as:
\begin{eqnarray}\label{C_eq2}
    &&\omega(g_1,g_2;{\bf g_3.a}) \cdot \omega(g_1 g_2, g_3;{\bf a})  \nonumber \\
    &&= \omega(g_1, g_2 g_3;{\bf a}) \cdot \, {}^{g_1}\omega(g_2,g_3;{\bf a}) ]\quad 
\end{eqnarray}
for $g_1,g_2,g_3 \in G$ where ${\bf g_3.a}$ is an anyon transformed from $\bf a$ by a symmetry $k$. Such a complication arises since the symmetry operator can act on ${\cal A} = \text{Hom}({\cal I}, U(1))$ in two different ways: the anyon part $\cal I$ and the phase factor part $U(1)$. Formally, allowing these two possibilities is equivalent to allow group elements to have both left and right action on elements of ${\cal A}$. Similarly, 2-coboundary condition is modified as following:
\begin{equation}\label{C_eq3}
    \omega'(g_1,g_2;{\bf a}) = \frac{\lambda(g_1;{\bf g_2.a})\, {}^{g_1}\!\lambda(g_2;{\bf a})}{\lambda(g_1 g_2; {\bf a})} \omega(g_1,g_2;{\bf a})
\end{equation}
An abelian group formed by equivalence classes of this $\omega$ in this case is called a \emph{twisted}-second group cohomology $H^2_t(G,{\cal A})$, where twisted means that elements of $G$ can act non-trivially on ${\cal A}$. Thus, when $G$ includes time-reversal operator ${\cal T}$, it should be called a twisted-group cohomology in a rigorous sense. However, in the case of $\mathbb{Z}_2$ topological order, phase factors are $\pm 1$, and ${\cal T}$ acts trivially on a coefficient group.

\begin{table}[t]
    \begin{center}
    \scalebox{0.96}{
      \begin{tabular}{| c | c | c | c | c |}
    \hline
    $\,$ Group Relations $\,$ & $\,\,$ \#1 $\,\,$&$\,\,$ \#2 $\,\,$&$\,\,$ \#3 $\,\,$&$\,\,$ \#4 $\,\,$ \\ \hline\hline
    $T_1 T_2 T_1^{-1} T_2^{-1} $& $M_{a,\boldsymbol{1}}$ & $M_{a,\boldsymbol{1}}$ & $M_{a,\boldsymbol{m_A m_B}}$  & $M_{a,\boldsymbol{m_A m_B}}$  \\ 
    \hline
    $\sigma^2 $& $M_{a,\boldsymbol{1}}$ & $M_{a,\boldsymbol{m_A m_B}}$ &$M_{a,\boldsymbol{1}}$ & $M_{a,\boldsymbol{m_A m_B}}$ \\
    \hline  
    $(C_6)^6 $ & $M_{a,\boldsymbol{1}}$ & $M_{a,\boldsymbol{1}}$ & $M_{a,\boldsymbol{1}}$ & $M_{a,\boldsymbol{1}}$ \\
    \hline
    $R^2 = (\sigma C_6)^2$ & $M_{a,\boldsymbol{1}}$ & $M_{a,\boldsymbol{m_A m_B}}$ & $M_{a,\boldsymbol{1}}$ & $M_{a,\boldsymbol{m_A m_B}}$ \\
    \hline
  \end{tabular}
  }
  \caption{     \label{HAP} \small Representations for four distinct equivalence classes in twisted-second group cohomology calculation $H^2_t(G_{p6m},\mathbb{Z}_2 \times \mathbb{Z}_2)$. Here we just consider $e$-particles, and full classificaiton should be direct product of this and group cohomology for $m$-particles, which has the same structure. In fact, third and fourth rows, phase factors for $(C_6)^6 $ and $R^2$ do not contribute to the classification because they are gauge-dependent. Here, $\alpha$ and $\beta$ are generators of $\mathbb{Z}_2 \times \mathbb{Z}_2 = \{1,M_{a,\boldsymbol{m_A}} \} \times \{1,M_{a,\boldsymbol{m_B}}\}$. As discussed, these generators are function from anyon category ${\cal I}$ to $U(1)$ phase factors. $M_{a,\boldsymbol{m_A}}$ spits out $-1$ for input of $e_A$, and $M_{a,\boldsymbol{m_B}}$ spits out $-1$ for $e_B$. Thus, $M_{a,\boldsymbol{m_A}} \cdot M_{a,\boldsymbol{m_B}} = M_{a,\boldsymbol{m_Am_B}}$ spits of $+1$ for $e_A e_B$, implying that the bound state $e_A e_B$ have a totally trivial fractionalization class for all cases. }
\end{center}
\vspace{-0.3in}
\end{table}

Equation.~\ref{C_eq2} defines allowed fractionalization classes, and Equation.~\ref{C_eq3} defines equivalence relations between fractionalization classes. Thus, if we give a rule for how $G$ acts on ${\cal A}$, in principle, above two equations define the problem of our interest. Although twisted group cohomology is very involved to analytically calculate, the calculation can be done computationally using a Homological Algebra Programming (HAP) package of the software called Group Algebra Programming (GAP)\cite{GAP} for various groups including space groups.

There is one subtlety. Since the coefficient group ${\cal A}$ is a group of homomorphisms from ${\cal I} = \expval{e_A,e_B}$ to $U(1)$ (which should be $\{1,-1 \}$ due to the fusion rules), we need a way to represent coefficients and associated group actions in a simple way. This can be done by representing element of ${\cal A}$ as a function giving a braiding statistics of a input particle with specific anyons. For example, let ${\cal I} = \expval{1,e,m,\epsilon=em}$. If $\gamma \in {\cal A}$ gives $\gamma({\bf 1}) = 1$, $\gamma(\boldsymbol{e}) = 1$, $\gamma(\boldsymbol{m}) = -1$, $\gamma(\boldsymbol{\epsilon}) = -1$, then we can represent
\begin{equation}
    \gamma(a) = M_{a,\boldsymbol{e}}
\end{equation}
where $M_{a,\boldsymbol{e}}$ represents a braiding statistics between an input anyon $a$ and $e$. Since braiding statistics satisfies group properties under multiplication, this indeed represents ${\cal A} = \text{Hom}({\cal I}, U(1))$ properly.\cite{Maissam_SymFrac_2014}

Now, let us consider the system of honeycomb lattice with $\mathbb{Z}_2 \times \mathbb{Z}_2$ topological order. $\text{Hom}({\cal I}, U(1))$ can be represented by generators, $\expval{M_{a,\boldsymbol{e_A}}, M_{a,\boldsymbol{e_B}}, M_{a,\boldsymbol{m_A}},M_{a,\boldsymbol{m_B}} }$. As $C_6$ rotational symmetry exchange two sublattices, $C_6$ acts on the coefficient group as
\begin{equation}
    C_6: \,\,M_{a,\boldsymbol{m_A}} \leftrightarrow M_{a,\boldsymbol{m_B}} \qquad M_{a,\boldsymbol{m_A}} \leftrightarrow M_{a,\boldsymbol{m_B}}
\end{equation}
while the other symmetry operators $T_1$, $T_2$, and $\sigma$ act trivially on coefficients. Consider only $e$-particles for now as $e$ and $m$ anyons can be treated independently. Then, ${\cal A}_e = \mathbb{Z}_2 \times \mathbb{Z}_2 = \{1,M_{a,\boldsymbol{m_A}} \} \times \{1,M_{a,\boldsymbol{m_B}}\}$. With all this information, we can finally calculate $H^2_t(G_{p6m},{\cal A}_e)$ using GAP. The resulting twisted-second group cohomology is isomorphic to $\mathbb{Z}_2 \times \mathbb{Z}_2$ (TABLE~\ref{HAP}). We remark that $(C_6)^6$ and $R^2$ in the table do not contribute to the classification since they are gauge-dependent. Under the gauge transformation ${\Gamma'}(C_6) = M_{a,\mathbf{e_A}} {\Gamma}(C_6)$, the action of $(C_6)^6$ and $R^2$ are multiplied by $M_{a,\boldsymbol{m_A m_B}}$. For example,
\begin{eqnarray}
    \qty[ {\Gamma'}(C_6) ]^6 &=& \qty[ M_{a,\mathbf{m_A}} \Gamma(C_6) ]^6 =  \qty( M_{a,\mathbf{m_A}} M_{a,\mathbf{m_B}} )^3 \,\,\, \nonumber \\
      && \times \,\, \qty[ \Gamma(C_6) ]^6 = M_{a,\mathbf{m_A m_B}} \qty[ \Gamma(C_6) ]^6 
\end{eqnarray}
since $\Gamma(C_6) M_{a,\mathbf{m_A}} = M_{a,\mathbf{m_B}} \Gamma(C_6)$ and also
\begin{eqnarray}
    \qty[ {\Gamma'}(C_6) \Gamma'(\sigma) ]^2 &=& \qty[ M_{a,\mathbf{m_A}} {\Gamma}(C_6) \Gamma(\sigma) ]^2 \nonumber \\
    &=& M_{a,\mathbf{m_A}} M_{a,\mathbf{m_B}} \qty[ {\Gamma}(C_6) \Gamma(\sigma) ]^2 \,\,\, \nonumber \\
    &=& M_{a,\mathbf{m_A m_B}} \qty[ {\Gamma}(C_6) \Gamma(\sigma) ]^2
\end{eqnarray}
Therefore, if the anyon of our interest is $a = e_A$, the phase factor is not well-defined since it can be $1$ or $-1$ depending on a gauge choice for these two relations. However, for the bound state $e_A e_B$, since $M_{\mathbf{e_A e_B},\mathbf{m_A m_B}} = 1$, phase factors for $e_A e_B$ are invariant.

In all four possible symmetry fractionalization classes for $e_A$, $e_B$ anyons, the bound state $e_A e_B$ have trivial phase factors under symmetry relations, implying that $e_A e_B$ can be a singlet of all possible symmetries. For an anyon $e_A$ (or $e_B$), a phase factor for $\boldsymbol{T}_1 \boldsymbol{T}_2 \boldsymbol{T}_1^{-1} \boldsymbol{T}^{-1}_2$ can be $\pm 1$. A phase factor for $\boldsymbol{\sigma}^2$ can be $\pm 1$. Such results agree with what we obtained in more direct approach in the main part of the paper. A phase factor for $(\boldsymbol{C}_6)^6$ has a gauge choice where it is trivial, which means that the phase factor can be always gauge transformed to be one. 

The resulting four symmetry fractionalization classes exactly agree with symmetry fractionalizations we can construct out of triangular sublattices in Equation.~\ref{symFrac_H}, which is labeled by $(\eta_1, \eta_2, \eta_1 \eta_3 \alpha_3, \eta_2)$, a set of phase factors for a single $e_A$ or $e_B$ anyon. Here $\eta_1$ and $\eta_2$ are determined by which SET phase is realized on triangular sublattices,  and the one involving $\alpha_3$ is gauge-dependent and does not contribute to the classification. The result further confirms the validity of our construction and tells us that symmetry fractionalization classes we obtained from bipartite construction are robust.

Although here we only discussed symmetry fractionalization class of $e$-particles, since coefficient group ${\cal A}$ factorizes into $e$-particle and $m$-particle components, full classification can be obtained by direct-product of individual cohomology group
\begin{equation}
    H^2_t(G_{p6m},{\cal A}) =  H^2_t(G_{p6m},{\cal A}_e) \times H^2_t(G_{p6m},{\cal A}_m)
\end{equation}
where in fact two cohomology groups $H^2_t(G_{p6m},{\cal A}_e)$ and $H^2_t(G_{p6m},{\cal A}_m)$ are isomorphic because a mathematical structure does not change under $e \leftrightarrow m$. Since $\mathbb{Z}_2$ SL we used in the construction can realize only one fractionalization class for visons as in TABLE~\ref{Triangle}, we do not have a freedom to realize all four possible classes for $m_A$, $m_B$ anyons in this case. Thus, the vison part would only realize the fractionalization class \#3 in TABLE~\ref{HAP}.

It seems interesting that in this calculation we took into account only how the anyons are acted on by the symmetries; we did not assume the state is made out of disconnected lattices. Still, possible SET phases we obtained through this general method is same as what we obtained through the bipartite lattice construction.  
In principle, one may also expect different SET phases where the anyons permute the same way, but it never happens actually, even for more complicated lattices.

Consider a lattice made from $k$ sublattices, $L_1,\dots L_k$ that are mapped to one another by symmetry.  Suppose that the symmetries on the sublattice $L_1$ form a group $H$, and let them be fractionalized: $\Gamma(h_1)\big|_a \Gamma(h_2)\big|_a=\omega_1(h_1,h_2;a)\Gamma(h_1 h_2)\big|_a$ for any $h_1,h_2\in H$ for a given anyon $a$. Assume that any element $h \in H$ does not not permute anyons. Can we extend this symmetry fractionalization pattern of the sublattice to the full system?  Suppose that for each anyon $a$, there are $n$ distinct copies of them denoted by $C_i(a)$ such that $C_i(a)$ and $C_j(a)$ have no mutual phases if $i \neq j$, and that they are permuted the same way as the sublattices, i.e., $g\qty(C_i(a))=C_j(a)$ if $g:L_i\rightarrow L_j$.  Then, we can find all the SETs for this system using cohomology to solve Eq.~\ref{C_eq2}---it turns out there is only one that extends $\omega_1$ of the sublattice.  We can see this by classifying the projective representation of the full lattice group $G$.  

Let ${\cal S}_i$ be a symmetry of the full system that maps $L_1\rightarrow L_i$, and choose some projective representations of $H$, restricted to the anyons in sublattice $L_1$. 
Now this projective representation can be extended to a general symmetry and anyons in other sublattices. For any $g\in G$ that maps $L_i$ to $L_j$, we must define $\Gamma(g)$'s phase for each sublattice $L_i$. Them, the relation $g={\cal S}_j({\cal S}_j^{-1}g{\cal S}_i){\cal S}_i^{-1}$ is convenient because the symmetry in the parentheses is an element of $H$ (it maps $L_1$ to itself). Hence $\Gamma(g)$ restricted to anyons in $L_1$ is already defined for this symmetry.

One can notice that $\Gamma(g)$ restricted to $C_i(a)$ can be expressed as $\Gamma ({\cal S}_j)|_{C_1(a)}\Gamma({\cal S}_j^{-1}g{\cal S}_i)|_{C_1(a)}\Gamma({\cal S}_i)^{-1}|_{C_i(a)}$ upto a phase factor $\omega(S_j, S_j^{-1} g S_i; C_1(a)) \cdot \omega(g S_i, S_i^{-1}; C_i(a))$, which can be set to be $1$ by a proper gauge choice for $\Gamma(g)$. Thus, one can express
\begin{eqnarray}\label{eq:sym_decomposition}
&&\Gamma(g)\big|_{C_i(a)}=\Gamma ({\cal S}_j)\big|_{C_1(a)}\Gamma({\cal S}_j^{-1}g{\cal S}_i)\big|_{C_1(a)}\Gamma({\cal S}_i)^{-1}\big|_{C_i(a)}, \nonumber \\
&&\mbox{}
\end{eqnarray}
Then one can work out the product of $\Gamma(g_1)$ and $\Gamma(g_2)$ for the anyon $C_i(a)$ to find $\omega\qty(g_1,g_2;L_i(a))$ because the factors of $\Gamma({\cal S})$'s cancel for all identity relations relevant to symmetry fractionalizations. Therefore, the answer is determined just by the symmetries within $L_1$.

For example, for the pg-group we discussed in the main text, we have two sublattices $A$ and $B$.  There are two types of symmetries: translational and orientation-reversing symmetries. The first
type preserves the sublattice and the second type exchanges them.  Suppose we fix the phases of $\Gamma(h)$ for a translation $h$ and for any anyon in the $A$-sublattice.
Then we can define the phases of all the other symmetries in the way we described---the Table \ref{table:square} summarizes how the action of any symmetry operator can be expressed in the form of Eq.~\ref{eq:sym_decomposition} given the choice of ${\cal S}_1 = I$, ${\cal S}_2 = g_{\text{glide}}$. (Here, $h$ is some properly chosen translation operator.) For example, if
$g$ is a symmetry that exchanges $A$ and $B$, one can either express it as $g_{\text{glide}}h_1$ or as $h_2 g_{\text{glide}}^{-1}$.
To use the Eq.~\ref{eq:sym_decomposition}, one should use the first expression if the symmetry is to be applied to an $A$-anyon and the second if it is to be applied to a $B$-anyon, and then replace each element by the projective representation $\Gamma$ corresponding to it. 
This choice affects only the phase of the symmetry because $\Gamma$ preserves relationships (such as $g=g_{\text{glide}} h_1$) up to a phase.  It is allowed to choose the phase differently for an $A$ and a $B$-anyon.
One can easily check that when symmetries are expressed in this form and multiplied together, the form is preserved, and one can extend symmetry fractionalization of the sublattice into the full system.

\begin{table}
\begin{tabular}{l|c|c}
Type of symmetry& A-anyons & B-anyons\\
\hline
Lattice-Fixing & $\Gamma(h)$ & $\Gamma(g_{\text{glide}})\Gamma(h)\Gamma(g_{\text{glide}})^{-1}$\\
Lattice-Exchanging &$\Gamma(g_{\text{glide}})\Gamma(h)$ & $\Gamma(h)\Gamma(g_{\text{glide}})^{-1}$
\end{tabular}
\caption{\label{table:square} Any symmetry operation can be expressed in the above form. For instance, the gilde-reflection symmetry in the main text would be expressed as $\Gamma(g_{\text{glide}})$ for $A$-anyons and $\Gamma(T_x) \Gamma(g_{\text{glide}})^{-1}$ for $B$-anyons, where $T_x$ is translation in $x$-direction. 
}
\end{table}

\section{Bosonic System with Magnetic Translation Symmetry}\label{Magnetic}
When we map a spin model into a bosonic model with a constraint ($n_b \leq 2S$) by Holstein-Primakoff transformation, magnetic translation symmetry changes as the following:

\begin{equation}\label{eq:mag_translation}
    T_1 T_2 T_1^{-1} T_2^{-1} = \prod_r e^{i S^z_r \Phi_0} = \prod_r e^{i (n_r-S) \Phi_0}
\end{equation}

However, when we think about a bosonic tight-binding model with  magnetic flux, it is more reasonable to write it as the following:

\begin{equation}
    T_1 T_2 T_1^{-1} T_2^{-1} = \prod_r e^{i n_r \Phi_0}
\end{equation}

There comes the difference by the constant phase factor $\prod_r e^{iS\Phi_0}= e^{iNS\Phi_0}$. Since this is the operator equation, the phase factor cannot be removed. One may wonder whether the mapping can be even well-defined when we have a magnetic field. In fact, if we require the condition $NS\Phi_0 \equiv 0 \mod 2\pi$, then it is possible to properly define the corresponding bosonic model under Holstein-Primakoff transformation. Now one can ask: what does spin-flip symmetry correspond to in the bosonic model? Under ${\cal P}: S_z \mapsto -S_z$, we see that ${\cal P}: n_b \mapsto 2S - n_b$. Thus, we see that

\begin{equation}\label{eq:appendix_map}
{\cal P} \prod_r e^{i n_r \Phi_0} {\cal P}^{-1} = \prod_r e^{i (2S - n_r) \Phi_0} = e^{2iSN\Phi_0} \prod_r e^{-i n_r \Phi_0}
\end{equation}

Because of the above condition $NS\Phi_0 \equiv 0 \mod 2\pi$, the prefactor in the Eq.~\ref{eq:appendix_map} disappears. Thus, during the discussion of magnetic translation symmetry, regardless of whether it is a spin or boson model, we can safely treat the action of ${\cal P}$ as flipping the $U(1)$ charge. Of course, when we think about particle density $\expval{n_b}$, we should keep in mind that ${\cal P} n_b {\cal P} = 2S - n_b$. Thus, for a particle-hole symmetric system, we must have a half-filling $\expval{n_b} = S$, where $0\leq n_b \leq 2S$. In the main text, we ignored this subtlety. For a generic boson model, ${\cal P}: n_b \mapsto N - n_b$, where $N$ is an arbitrary integer that one has a freedom to choose. Thus, for $n_b$ fractional, there is no particle-hole symmetry. 



\section{Constraints on Topological Orders}\label{appendix:Fluxon}

In the main text, we explained why certain SETs cannot be realized for a given symmetry setting. To do so, we showed that the flux seen by each anyon with a certain quantum number does not transform properly under the spin-flip/time-reversal symmetry. 

In fact, it can be shown in a more compact description of a symmetry fractionalization of magnetic translation algebra. Assume that anyon does not permute under magnetic translation symmetries. Our analysis relies on the following formula given in Ref.~\onlinecite{YML2017}: 
\begin{equation} \label{eq:fractionalized_magnetic_algebra}
	\Omega_{\tilde{T}_1} 	\Omega_{\tilde{T}_2} 	\Omega_{\tilde{T}_1}^{-1} 	\Omega_{\tilde{T}_2}^{-1} = \lambda_a \Omega_{U(\phi)}
\end{equation}
where $\Omega_{\cal O}$ represents a fractionalized action of a symmetry operator $\cal O$, $\phi$ represents magnetic flux per unit cell, and $\lambda_a$ represents anyon flux of $a$ per unit cell. For a given localized (anyonic) excitation, Eq.~\ref{eq:fractionalized_magnetic_algebra} gives a phase factor acquired when we move the excitation around one unit cell. Because this phase factor is a physical observable, it should be invariant under the on-site symmetry action which commutes with magnetic translations. From twisted-group cohomology condition, this condition can be also shown directly. For $\phi=\pi$ flux case, we know that spin-flip and time-reversal symmetry commutes with magnetic translations. Moreover, $\lambda_a \Omega_{U(\phi)}$ transforms in the following way under each symmetry action:
\begin{eqnarray} \label{eq:anyon_transform}
&&{\cal P}: \,\,\, \lambda_a \Omega_{U(\pi)} \mapsto \lambda_{a^{\cal P}} \Omega_{U(-\pi)} = \lambda_{a^{\cal P} \cdot  \bar{F}} \Omega_{U(\pi)} \nonumber \\
&&{\cal T}_1: \,\,\, \lambda_a \Omega_{U(\pi)} \mapsto  \lambda_{a^{{\cal T}_1}} \Omega_{U(-\pi)} = \lambda_{a^{{\cal T}_1} \cdot \bar{F}} \Omega_{U(\pi)} \nonumber \\
&&{\cal T}_2 = {\cal P} \cdot {\cal T}_1: \,\,\, \lambda_a \Omega_{U(\pi)} \mapsto \lambda_{a^{{\cal T}_2}} \Omega_{U(\pi)}
\end{eqnarray}
where we used the fact that the action of $U(-2\pi)$ is equivalent to the braiding with anti-fluxon $\bar{F}$. For $\mathbb{Z}_2$ toric code order, anyon does not transform but fluxon must be equivalent to a vison. For $\mathbb{Z}_2$ double-semion order, $s \leftrightarrow \bar{s}$ under time-reversal symmetry, and fluxon must be equivalent to a bosonic anyon $b=s \bar{s}$. Using Eq.~\ref{eq:anyon_transform}, one can check whether the RHS of Eq.~\ref{eq:fractionalized_magnetic_algebra} is invariant under symmetry actions, and determine whether the symmetry fractionalization pattern is consistent. For the generalization to smaller fractions for the flux or filling, we can apply this analysis in the exact same manner to $\mathbb{Z}_n$ topological orders, knowing that how anyons transform under the symmetries.




\bibliographystyle{aipnum4-1}
\bibliography{Archive_JongYeon.bib}

\end{document}